%% file: ms_final.tex
\documentclass[numberedappendix]{emulateapj} 
\usepackage{graphics}
\usepackage{amsmath}
\usepackage{color}
\usepackage{rotating}
\usepackage{natbib}
\newcommand{\degree}{\hbox{$^\circ$}}

\newcommand{\ltsimeq}{\la}
\newcommand{\gtsimeq}{\ga}

\newcommand{\msun}{M$_{\odot}$}

\newcommand{\HI}{H{\sc i}}

\shortauthors{McQuinn et al. (2016)}
\shorttitle{DUSTiNGS: Radial Distribution of Intermediate-age and Old Stars in Dwarfs}

\begin{document}
\title{DUSTiNGS III: Distribution of Intermediate-Age and Old Stellar Populations in Disks and Outer Extremities of Dwarf Galaxies}
\author{Kristen B.~W.~McQuinn\altaffilmark{1,2}, 
Martha L.~Boyer\altaffilmark{3,4},
Mallory B.~Mitchell\altaffilmark{2},
Evan D.~Skillman\altaffilmark{2},
R.~D.~Gehrz\altaffilmark{2},
Martin A.~T.~Groenewegen\altaffilmark{5},
Iain McDonald\altaffilmark{6},
G.~C.~Sloan\altaffilmark{7,8,9}
Jacco Th.~van Loon\altaffilmark{10},
Patricia A.~Whitelock\altaffilmark{11,12},
Albert A.~Zijlstra\altaffilmark{6}
}

\altaffiltext{1}{University of Texas at Austin, McDonald Observatory, 2515 Speedway, Stop C1402, Austin, Texas 78712; \ {\it kmcquinn@astro.as.utexas.edu}}
\altaffiltext{2}{Minnesota Institute for Astrophysics, School of Physics and
Astronomy, 116 Church Street, S.E., University of Minnesota,
Minneapolis, MN 55455; \ {\it mitch925@umn.edu, skillman@astro.umn.edu, gerhz@astro.umn.edu}} 
\altaffiltext{3}{Observational Cosmology Lab, Code 664, NASA Goddard Space Flight Center, Greenbelt, MD 20771, USA; \ {\it martha.boyer@nasa.gov}}
\altaffiltext{4}{Department of Astronomy, University of Maryland, College Park, MD 20742 USA}
\altaffiltext{5}{Koninklijke Sterrenwacht van Belgi\"e, Ringlaan 3, B-1180 Brussels, Belgium; \ \email{martin.groenewegen@oma.be}} 
\altaffiltext{6}{Jodrell Bank Centre for Astrophysics, School of Physics and Astronomy, University of Manchester, Manchester, M13 9PL, UK\ {\it Iain.Mcdonald-2@manchester.ac.uk}}
\altaffiltext{7}{Cornell Center for Astrophysics \& Planetary Science,
  Cornell Univ., Ithaca, NY 14853-6801, USA, sloan@isc.astro.cornell.edu}
\altaffiltext{8}{Department of Physics and Astronomy, University of North
  Carolina, Chapel Hill, NC 27599-3255, USA}
\altaffiltext{9}{Space Telescope Science Institute, 3700 San Martin Dr.,
  Baltimore, MD 21218, USA}
\altaffiltext{10}{Astrophysics Group, Lennard-Jones Laboratories, Keele University, Staffordshire ST5 5BG, UK; \ {\it j.t.van.loon@keele.ac.uk}}
\altaffiltext{11}{Astronomy Department, University of Cape Town, 7701 Rondebosch, South Africa}
\altaffiltext{12}{South African Astronomical Observatory (SAAO), PO Box 9, 7935 Observatory, South Africa}

\begin{abstract}
We have traced the spatial distributions of intermediate-age and old stars in nine dwarf galaxies in the distant parts of the Local Group, using multi-epoch 3.6 and 4.5 $\micron$ data from the DUST in Nearby Galaxies with Spitzer (DUSTiNGS) survey. Using complementary optical imaging from the \emph{Hubble Space Telescope}, we identify the tip of the red giant branch (TRGB) in the 3.6 $\micron$ photometry, separating thermally-pulsating asymptotic giant branch (TP-AGB) stars from the larger red giant branch (RGB) populations. Unlike the constant TRGB in the I-band, at 3.6 $\micron$ the TRGB magnitude varies by $\sim$0.7 mag, making it unreliable as a distance indicator. The intermediate-age and old stars are well mixed in two-thirds of the sample with no evidence of a gradient in the ratio of the intermediate-age to old stellar populations outside the central $\sim1-2$\arcmin. Variable AGB stars are detected in the outer extremities of the galaxies, indicating that chemical enrichment from these dust-producing stars may occur in the outer regions of galaxies with some frequency. Theories of structure formation in dwarf galaxies must account for the lack of radial gradients in intermediate-age populations and the presence of these stars in the outer extremities of dwarfs. Finally, we identify unique features in individual galaxies, such as extended tidal features in Sex~A and Sag~DIG and a central concentration of AGB stars in the inner regions of NGC~185 and NGC~147.
\end{abstract} 

\keywords{galaxies:\ dwarf -- galaxies:\ photometry -- galaxies:\ stellar content -- galaxies:\ fundamental parameters -- galaxies:\ evolution -- galaxies:\ Local Group}

\section{Stellar Population Gradients in Dwarf Galaxies}\label{sec:intro}
Stellar population age gradients trace the combined effects of galaxy formation and evolution. Young stellar population gradients are universal in dwarf galaxies, found in both dwarf irregular (dI) and dwarf spheroidal (dSph) morphological types. Early studies with ground-based optical telescopes revealed that dI galaxies in particular have strong age gradients where the youngest stars are centrally concentrated and the oldest stars have the largest physical extents \citep[e.g.,][and references therein]{Aparicio2000a}.  

The \emph{Hubble Space Telescope (HST)} allowed resolved star photometry to significantly fainter magnitudes and with higher precision than possible from the ground, which allowed a separation of the red giant branch (RGB) stars from the red helium burning (HeB) stars and of the main sequence (MS) from the blue HeB stars \citep[e.g.,][]{Dohm-Palmer1997, Dohm-Palmer1998, Gallagher1998, Tolstoy1998, McQuinn2012a}.  As a result, the spatial distributions of these different stellar populations could be studied with greater precision, showing the degree of central concentration decreases with the mean age of the populations of high mass MS stars, blue and red HeB stars, and RGB stars. For dIs, this leads to the question of whether the outer stars represent a true halo structure or just an extension of the exponential disk \citep{Aparicio2000b}.

Simulations of dwarf galaxies have shown that internal processes (i.e., supernova feedback and stellar migration driven by, for example, a changing potential) may play an important role in shaping radial age gradients \citep[e.g.,][]{Stinson2009, El-Badry2016}. On the one hand, stellar migration may help to erase radial gradients by mixing stars of different ages over time. On the other hand, if stellar feedback and migration primarily drive stellar populations outward, these internal processes may actually help to create gradients in the sense that the oldest stars will be pulled repeatedly over longer timescales and reach larger radii than younger stars. These processes may be particularly relevant at describing the re-distribution of stars in the outer extremities of galaxies where little in-situ star formation is expected. 

External tidal disturbances and mergers may also play a role in shaping radial age gradients in dwarf galaxies. In the outer regions and halos of massive spiral galaxies, the stellar mass assembly of a predominantly ancient stellar structure is thought to be driven by hierarchical merging \citep[e.g.,][]{Searle1978}. In dwarf galaxies, such a process would require the acquisition of increasingly low-mass haloes. While such a scenario is unlikely, tidal disturbances and interactions do impact the stellar distributions of a dwarf galaxy. 

The relative importance of internal versus external processes is unclear. A key test of the different theories of evolution and structure formation in dwarfs is whether or not stellar populations extending to the outer extremities of galaxies are purely old populations (supporting in-situ star formation in a contracting gaseous disk) or a mix of intermediate-age and old stars (supporting a combination of in-situ star formation, internal drivers of extended structure such as stellar feedback and migration, and possibly tidal disturbances). 

Here, using the photometric catalogues from the DUST in Nearby Galaxies with {\it Spitzer} (DUSTiNGS) survey \citep[][hereafter Paper I]{Boyer2015a}, we study the radial distributions of the intermediate-age stellar population in nine dwarf galaxies (traced by thermally-pulsating asymptotic giant branch (TP-AGB) stars) and compare their distributions with that of an older stellar population (traced by RGB stars). We separate the majority of the TP-AGB population from the RGB populations by uniquely identifying the tip of the red giant branch (TRGB) at 3.6 $\micron$. We also identify many of the largest dust producing, AGB stars by applying photometric variability criteria.  The three populations of evolved stars $-$ TP-AGB and RGB stars, and AGB stars identified through variability analysis $-$ allow us to not only examine the radial structure and distribution of intermediate-age and old stars in the interiors of the galaxies, but also, importantly, to identify whether intermediate-age populations exist in the outer extremities of dwarfs.

The paper is organized as follows. Section~\ref{sec:data} describes the general properties of AGB stars, the DUSTiNGS survey, and the data reduction. In Section~\ref{sec:trgb}, we identify the TRGB in the 3.6 $\micron$ photometry by matching point sources with HST optical imaging. In Section~\ref{sec:geometry} we measure the elliptical parameters of the galaxies from the infrared data and compare these structural parameters with those derived from optical data in the literature. Using the measured geometry, in Section~\ref{sec:profiles} we create radial profiles of the resolved AGB and RGB stellar populations and measure the effective radius of each galaxy. In Section~\ref{sec:galaxies} we discuss each galaxy in detail; Section~\ref{sec:comparison} presents an inter-sample comparison. In Section~\ref{sec:scenarios} we discuss the implications of the distribution of AGB stars in the context of galaxy evolution scenarios. Our conclusions are summarized in Section~\ref{sec:conclusion}.

\section{AGB Stars and The DUSTiNGS Survey}\label{sec:data}
\subsection{Summary of AGB Properties}
AGB stars are intermediate-mass (1 \msun\ $\ltsimeq M \ltsimeq$ 8 \msun) evolved stars that are characterized by periods of mass-loss. AGB stars go through two basic phases: the early (E-)AGB and the TP-AGB phases. E-AGB stars are characterized by shell He-burning; they increase in luminosity (the second ascent of the red giant branch) as the degenerate C/O core increases in mass. When the He-burning shell runs out of fuel, the H-shell is reignited and what follows is the TP-AGB phase. During a thermal pulse, a thermally-unstable, thin He layer undergoes a shell He flash. This triggers convection into the intershell region, which dredges up newly synthesized material (the third dredge up). The dust produced helps drive stellar winds that are chemically enriched with dredged-up material. TP-AGB stars are more luminous than E-AGB stars, with most E-AGB stars remaining fainter than the TRGB, while most TP-AGB stars are brighter than this limit \citep[e.g.,][]{Rosenfield2014}. 

\input{tab1}

The AGB stage of evolution is short-lived \citep[t$\ltsimeq0.5-10$ Myr depending on the mass of the star;][]{Vassiliadis1993, Girardi2007, Rosenfield2014}, making these stars fairly rare in low-mass galaxies compared with longer-lived RGB stars. AGB stars exhibit variability, with periods and amplitudes generally increasing as the star evolves. TP-AGB stars are characterized by increasing mass loss that partially or totally obscures up to $30-40$\% of the stars at optical wavelengths \citep{Jackson2007a, Jackson2007b, Boyer2009}, making it difficult to catalog a complete census of them from optical imaging alone. 

Dust production from mass-loss in AGB stars increases as a star reaches the end of its evolution, so the dustiest AGB stars tend to have the longest periods and largest amplitudes \citep[e.g.,][among others]{Whitelock1987, vanLoon1999, Ita2011, Javadi2013, McDonald2016}. These highly evolved stars are in the superwind phase and most easily detected in the infrared. They are sometimes referred to as ``extreme'' AGB stars (or x-AGB)\footnote{The x-AGB class is an empirical classification based solely on the infrared photometry from {\it Spitzer} \citep[see ][]{Blum2006, Boyer2011}. It includes the dustiest AGB stars, regardless of chemical composition of the circumstellar envelope (C/O). Most x-AGB stars are in the superwind phase at the end of their evolution where dust drives a strong wind that causes the star to lose mass at a faster rate than nuclear consumption \citep{vanLoon1999}.}. The majority of the reddest AGB stars are carbon rich, with masses $\gtsimeq1.5 $\msun\ \citep{Boyer2015c, Dell'Agli2015a, Dell'Agli2015b}. Thus, AGB stars with the red infrared colors characteristic of x-AGB stars are representative of the bulk intermediate-age AGB population. 

\subsection{Overview of DUSTiNGS}
The DUSTiNGS survey was designed to create a near-complete census of the most luminous AGB population in the local universe by imaging 50 nearby dwarf galaxies with the {\it Spitzer Space Telescope} \citep{Werner2004, Gehrz2007} Infrared Array Camera \citep[IRAC; ][]{Fazio2004} at 3.6 and 4.5 $\micron$ over 2 epochs. The full sample includes all dwarf galaxies within 1.5 Mpc that were known at the time of the observations and that lacked sufficient existing coverage with {\it Spitzer}. The next nearest galaxy ($d = 1.7$ Mpc) is beyond {\it Spitzer's} ability to spatially resolve stars. One advantage of the DUSTiNGS survey is that in the infrared, dust-enshrouded AGB stars are recovered that are easily missed at shorter wavelengths.

The DUSTiNGS epochs were separated by approximately 6 months to aid in identifying variable AGB stars at infrared wavelengths \citep{LeBetre1992, LeBetre1993, McQuinn2007, Vijh2009}, thereby eliminating confusion with other infrared-bright point sources. In total, DUSTiNGS identified 710 variable AGB star candidates including 526 variable x-AGB star candidates \citep[][hereafter Paper II]{Boyer2015b}. Twelve of these x-AGB star candidates are in galaxies with [Fe/H]$<-$2.0, making them the most metal-poor dust-producing AGB stars known. Less dusty AGB stars have typical amplitudes too small to be detected given the sensitivity limit of DUSTiNGS and, therefore, are not identified as variable.

\subsection{Using AGB Stars to Trace Intermediate-Age Gradients}
AGB stars can be a unique tracer of an intermediate-age population in a galaxy. While AGB stars have lifetimes spanning from  tens of Myr to $>$10 Gyr, more massive AGB stars have typical stellar lifetimes of only 0.1 to 3 Gyr, depending on their initial mass and evolution \citep{Marigo2013}. In addition to the initial mass of an AGB stars correlating positively with the length of time a star spends above the TRGB, these massive TP-AGB stars have higher luminosities towards the end of their lives than the lower mass TP-AGB stars. Thus, for typical star-forming dwarfs, the maximum contribution of AGB stars to the 3.6$\micron$ luminosity of a galaxy is comprised of stars with lifetimes of $\sim1$ Gyr \citep{Mouhcine2002}, making the bright TP-AGB stars in DUSTiNGS excellent tracers of an intermediate-age population. Confirming the intermediate-age nature of TP-AGB stars, \citet{Feast2006} found a mean age for Galactic carbon stars in the solar neighborhood of $1.8\pm0.4$ Gyr and \citet{Held2010} find bright carbon-rich AGB stars are generally less than 2 Gyr old with a smaller percentage as old as 3 Gyr.

Observationally, unambiguously cataloguing a sufficient number of AGB stars in dwarfs for a radial study can be challenging. Yet, with carefully chosen observing strategies, AGB and RGB stars can be separated in galaxies. In a series of papers using wide-field imaging in the near-infrared, carbon-rich AGB stars were photometrically separated from the RGB population in a few Local Group dwarf galaxies \citep{Albert2000, Battinelli2000, Demers2002, Letarte2002, Demers2004, Battinelli2004a, Battinelli2004b, Battinelli2004c}.  Based on comparisons of the AGB and RGB radial profiles, these studies report the presence of both intermediate-age and older populations in the outer regions of a sub-sample of the galaxies studied.

In this study, we selected the nine nearby dwarf galaxies that host $>90$\% of the identified variable AGB star candidates from the larger DUSTiNGS sample. The variable AGB star candidates are a powerful tracer of stars at greater radii in the galaxies as the identification of the variable AGB stars is independent of photometric contamination. Thus, membership of an individual point source to the galaxy is more secure than for point sources selected based solely on photometric criteria. We use the distribution of variable AGB stars to probe the outer reaches, or ``extremities'', of the galaxies. We have chosen not to use the term ``halo'' which is typically used to describe the spherical distribution of ancient stellar populations detected around massive galaxies. It is unclear whether dwarf galaxies have such halo structures and we do not have a sufficient sample of variable AGB stars to model their 3-D distribution. As described below (see Section~\ref{sec:profiles}), we define an approximate boundary between the interior of a galaxy and the outer extremities at $\sim3 \times$ the effective radius ($R_e$). The outer extremities probed by our data are within $6\times R_e$, well within the expected tidal radii of the galaxies \citep[see e.g.,][and reference therein]{Sanchez-Salcedo2007, Sanna2010, Bellazzini2014}.

We use the DUSTiNGS photometric catalogs to measure the radial distribution of intermediate-age stars traced by TP-AGB stars brighter than the TRGB, and older stars traced by the RGB population. The study of the AGB and RGB stars' spatial extents is confined to the interior of the galaxies. For ease of notation, we refer to the interior of the galaxies as the stellar disks. In the outer extremities of the galaxies, the TP-AGB and RGB stars become indistinguishable in the DUSTiNGS data from background and foreground contamination.

\subsection{The Sample, Observations, and Data Reduction}
The full DUSTiNGS sample includes 37 dSphs, 8 dIs, and 5 dTrans. For this study, we are interested in exploring the radial profiles of intermediate-age and old stellar populations in the disks and outer extremities of galaxies. Thus, from the full DUSTiNGS sample, we selected the nine galaxies with at least 6 variable AGB star candidates including six dI (IC~1613; IC~10; Sextans~A; Sextans~B; WLM; Sag~DIG), two dSphs (NGC~185; NGC~147), and one dTrans (DDO~216). The majority of the galaxies have a sufficient number of variable AGB star candidates ($>25$) to build a radial profile. For two galaxies (DDO~216; Sag~DIG), the smaller number of stars precludes building a variable AGB radial profile. In these cases, we construct stellar radial profiles from the RGB stars and TP-AGB stars, and identify where the variable AGB stars are located. Table~\ref{tab:properties} lists the nine galaxies that make up our sample as well as their basic properties and observational details.

Here, we briefly summarize the observing strategy and data reduction; we refer the interested reader to Paper~I for a full description. Each galaxy was imaged simultaneously at 3.6 and 4.5 $\micron$ in two epochs. The mosaic images created from these two epochs map each galaxy to at least twice its half-light radius to assist in the statistical subtraction of background and foreground contamination. Point-spread function (PSF) photometry was performed on the images using DAOphot II and ALLSTAR \citep{Stetson1987}, with corrections recommended by the Spitzer Science Center applied to the final photometry. We use the photometry from the DUSTiNGS ``good-source'' catalogs (GSC), which has been culled to include only high-confidence point sources. 

Individual variable AGB star candidates were identified from the full GSCs using the variability index defined by \citet{Vijh2009}. Archival observations for three of our selected sample (IC~1613, DDO~216, Sextans~A, WLM) were also available, which were processed in a uniform manner and provided an additional epoch (Epoch 0) to the variability analysis. The colors and magnitudes of individual variable AGB stars were then used to separate x-AGB stars from the less dusty AGB stars. A full description of the method used to both identify variable AGB stars and characterize their dust production is given in Paper~II. 

\begin{figure*}
\includegraphics[width=\textwidth]{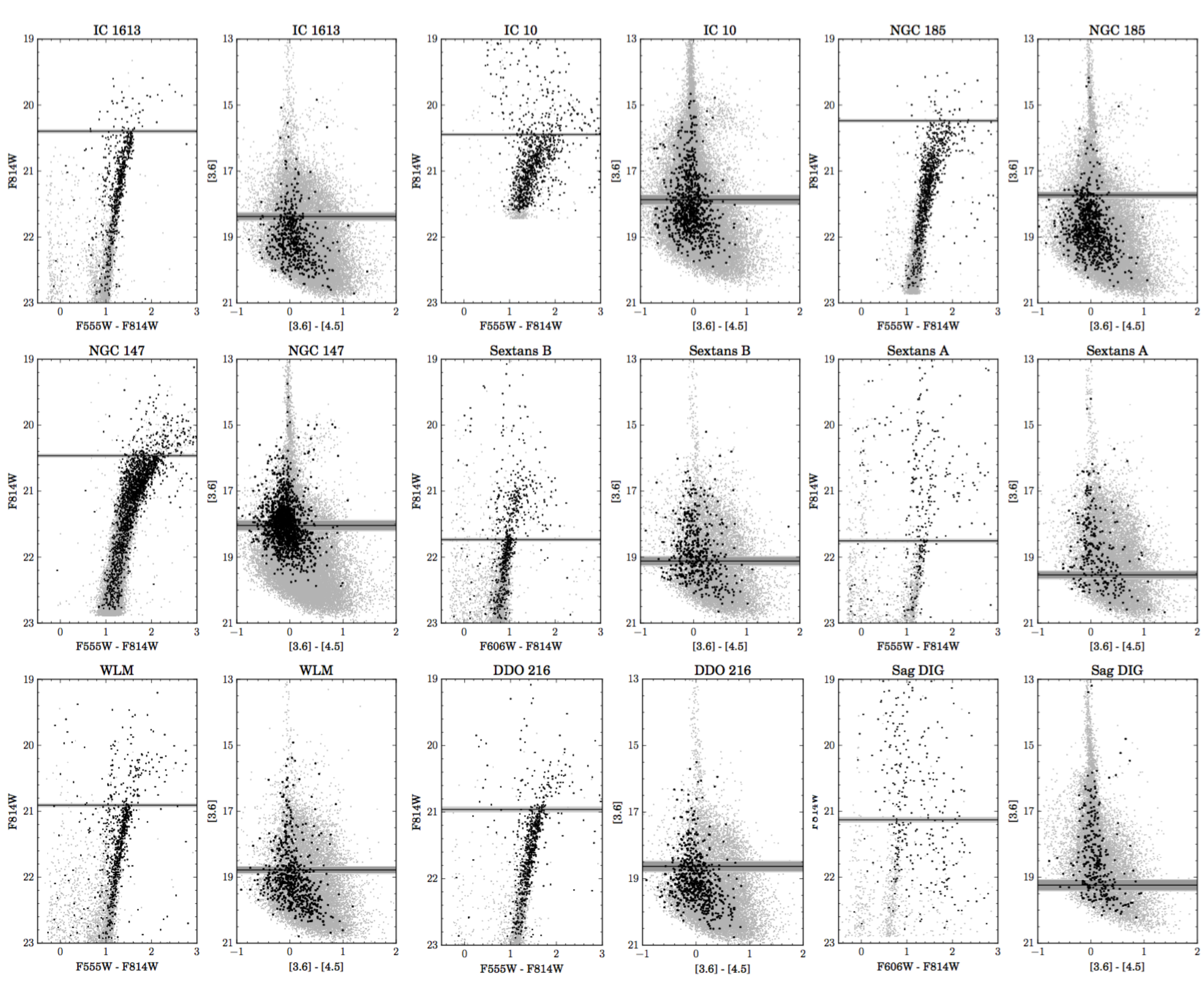}
\caption{Optical HST CMDs side-by-side with the infrared {\it Spitzer} CMDs for each galaxy in the sample. Gray points mark everything detected in each respective survey. The black points are sources in common between the HST and {\it Spitzer} catalogs. The TRGBs are clearly distinguishable in the optical HST CMDs, but not in the {\it Spitzer} data. The horizontal black lines mark the TRGB we identify in the optical CMDs and the TRGB luminosity we identify in the infrared CMDs using both data sets; gray shading denotes the $1\sigma$ uncertainty. \label{fig:cmd}}
\end{figure*}

\section{The TRGB at 3.6 $\micron$}\label{sec:trgb}
\subsection{Identifying the TRGB}
The majority of the TP-AGB population not identified through our variability analysis are brighter than the TRGB luminosity (corresponding to the luminosity of low-mass stars just prior to the helium flash). Thus, by identifying the TRGB in the 3.6 $\micron$ photometry, the TP-AGB and RGB stars can be separated by applying simple photometric cuts to the data. The difficulty is that the TRGB luminosity is not well-established in the {\it Spitzer} infrared filters. At 3.6 $\micron$, the TRGB is estimated to be $\sim-6$ mag, but is seen to vary by a few tenths of a magnitude \citep{Jackson2007a, Jackson2007b, Boyer2009}.  Furthermore, the DUSTiNGS photometry is incomplete in the range of the expected TRGB, which results in false TRGB detections. This is especially problematic in the most distant and/or crowded DUSTiNGS galaxies.

To identify the TRGB in our 3.6 $\micron$ photometry, we use HST WFPC2 optical photometry in the F814W filter for our sample as a guide. The TRGB luminosity in the F814W filter (approximately the $I$-band) is known to be relatively constant with a modest, but predictable, dependence on metallicity \citep[e.g.,][]{Rizzi2007a}. For all galaxies except IC\,10, HST photometry is from \citet{Holtzman2006}, and can be found online.\footnote{astronomy.nmsu.edu/holtz/archival/html/lg.html} This archive includes 23 DUSTiNGS galaxies with 1$-$4 WFPC2 and/or ACS fields, including the 9 galaxies that are the focus of this paper. The fields used here are listed in Table~\ref{tab:trgb}. We selected fields that avoided the most crowded center regions of the galaxies. The region in IC\,10 included in \citet{Holtzman2006} suffers from severe crowding, so we instead use data from HST GO program 10242, which includes WFPC2 images in an outer region of the galaxy ($d \approx 4.\arcmin5$). We performed photometry on this field with DOLPHOT \citep{Dolphin2000}, using AstroDrizzle 2.0 to create the reference image.

To estimate the 3.6 $\micron$ TRGB, we start by locating the TRGB in F814W, which allows us to identify RGB and AGB star candidates. The HST fields are small ($\approx$$2\arcmin \times 2\arcmin$) compared to the {\it Spitzer} fields ($>$$10\arcmin \times 5\arcmin$), but by identifying RGB and AGB stars in the overlapping region, we can minimize contaminating sources and confusion from large photometric uncertainties at the faint end of the {\it Spitzer} luminosity function, where the TRGB is located. 

The TRGB is typically located by applying an edge-detection filter to the luminosity function of the stars in the region of the RGB. Bayesian maximum likelihood (ML) techniques have also been used to measure TRGB luminosities and can result in lower statistical uncertainties of order 0.1 mag or less \citep[e.g.,][]{Makarov2006, Conn2012, McQuinn2016a}. For our purposes of identifying the TRGB in order to separate the RGB and AGB populations, the use of an edge-detection filter provides sufficient precision of $\ltsimeq0.2$ mag. To find the optical TRGB, we use the strategy described in detail by \citet{Mendez2002} and employed for similar data by \citet{Dalcanton2012}. We start by correcting for extinction using $A_V$ from \citet{Weisz2014a}, measured using the same HST data \citep[$A_{\rm F814W}/A_{\rm V} = 0.57$;][]{Girardi2008}. We then select stars belonging to the RGB as those within 3$-$5\,$\sigma$ of a line fit to the RGB in a color-magnitude diagram (CMD) of F555W$-$F814W or F606W$-$F814W vs.\ F814W. The stricter sigma cut is used on the blue side of the RGB to minimize contamination from the main sequence and red supergiants. The resulting F814W luminosity function is then Gaussian-smoothed to avoid issues with discrete binning and passed through a Sobel edge-detection filter. The HST observations are sensitive to well below the TRGB, so the Sobel filter response is not affected by incomplete photometry. 

The uncertainty on the TRGB in the F814W filter is determined with 1000 Monte Carlo Bootstrap resampling trials, each with a conservative, additional 4\,$\sigma$ Gaussian random errors and with random variations on the bin size and starting magnitude of the luminosity function. We then fit a Gaussian function to the returned TRGB estimates, setting the mean as the final TRGB estimate. The F814W TRGBs were measured in this way in each of the fields selected (see Appendix~\ref{sec:append_trgb} for individual measurements); for galaxies with two observed fields, the final measurements are the weighted average of the measurements. The resulting F814W TRGBs, marked in the HST CMDs shown in Figure~\ref{fig:cmd} and listed in Table~\ref{tab:trgb}, are consistent with those measured in the same fields by \citet{Weisz2014a}. We use the empirically calibrated relation specific to the WFPC2 filters from \citet{Rizzi2007a} with a color-based metallicity correction to transform the F814W TRGB magnitudes to a distance modulus.

Next, we match the stellar positions from the HST photometry filtered for RGB stars to those of the {\it Spitzer} photometry, which is complicated by the significant difference in wavelength, photometric uncertainty, and angular resolution between HST and {\it Spitzer}. We use the DAOphot routines DAOmatch$+$DAOmaster \citep{Stetson1987} which iteratively solve the transformation coefficients between cross-identified stellar coordinates as the user decreases the allowable separation distance. We start with a 5-pixel (3$\arcsec$) separation and decrease to 1-pixel (0\farcs6), excluding faint sources from the HST catalog that are unlikely to be detected by {\it Spitzer}. 

Figure~\ref{fig:cmd} shows the CMDs from the HST and {\it Spitzer} photometry side-by-side for each galaxy, with the matched sources overplotted in each panel. The TRGB, marked in each CMD, is clearly identifiable in the optical data but not in the infrared data. We use the combined HST$+${\it Spitzer} photometry to estimate the 3.6~\micron\ TRGB by applying four strategies, detailed in Appendix~\ref{sec:append_trgb}. In uncrowded, nearby fields (e.g., IC\,1613; see Fig.~\ref{fig:lfunc} in Appendix~\ref{sec:append_trgb}), all four estimates agree, albeit with different uncertainties. For these, our final estimate of the TRGB is a weighted mean of the four results; uncertainties are the standard error on the weighted mean. For more difficult fields that have higher levels of incompleteness due to their farther distance and/or greater amount of stellar crowding, we take the weighted mean of a subset of the measurements as described in Appendix~\ref{sec:append_trgb}. The TRGB measurements in the {\it Spitzer} data calculated in this way are likely biased to slightly brighter magnitudes because of a combination of random noise and completeness issues. Based on artificial star tests in Paper~I (see their Figure~7), the recovered stars in the range of the TRGB magnitudes are $0.05-0.1$ mag brighter than the input stars as random noise on individual sources at the limit of detection can scatter sources brighter. Thus, we add this additional uncertainty to the upper error on the weighted mean for the TRGB magnitudes. Final estimates are listed in Table~\ref{tab:trgb}, including the corresponding 3.6$\micron$ absolute magnitudes based on the distances and uncertainties measured from the HST F814W imaging. 

\input{tab2}

We then use the TRGB luminosity at 3.6 $\micron$ to separate the TP-AGB from RGB stars in the DUSTiNGS data. Hereafter, we refer to this population of AGB stars simply as AGB stars, while the stars identified via variability analysis in Paper~II are referred to as variable AGB stars. Note that the RGB population still contains less evolved E-AGB stars. Based on galaxy simulations, \citet{Rosenfield2014} estimates E-AGB stars constitute $\sim$20\% of the stars below the TRGB in the optical and near infared (while TP-AGB stars constitute $<5$\%), depending on the magnitude range considered. These AGB stars are inseparable photometrically in our data, but they also represent a small fraction of the stars below the TRGB. Thus, for the remainder of the paper, we will refer to the population of stars below the TRGB luminosity as the RGB population and representative of a generally older stellar population. In Table~\ref{tab:photometry} we list the total number of RGB stars, AGB stars, variable AGB stars, and the sub-sample of variable x-AGB candidates identified in each galaxy. 

\begin{figure}
\includegraphics[width=\columnwidth]{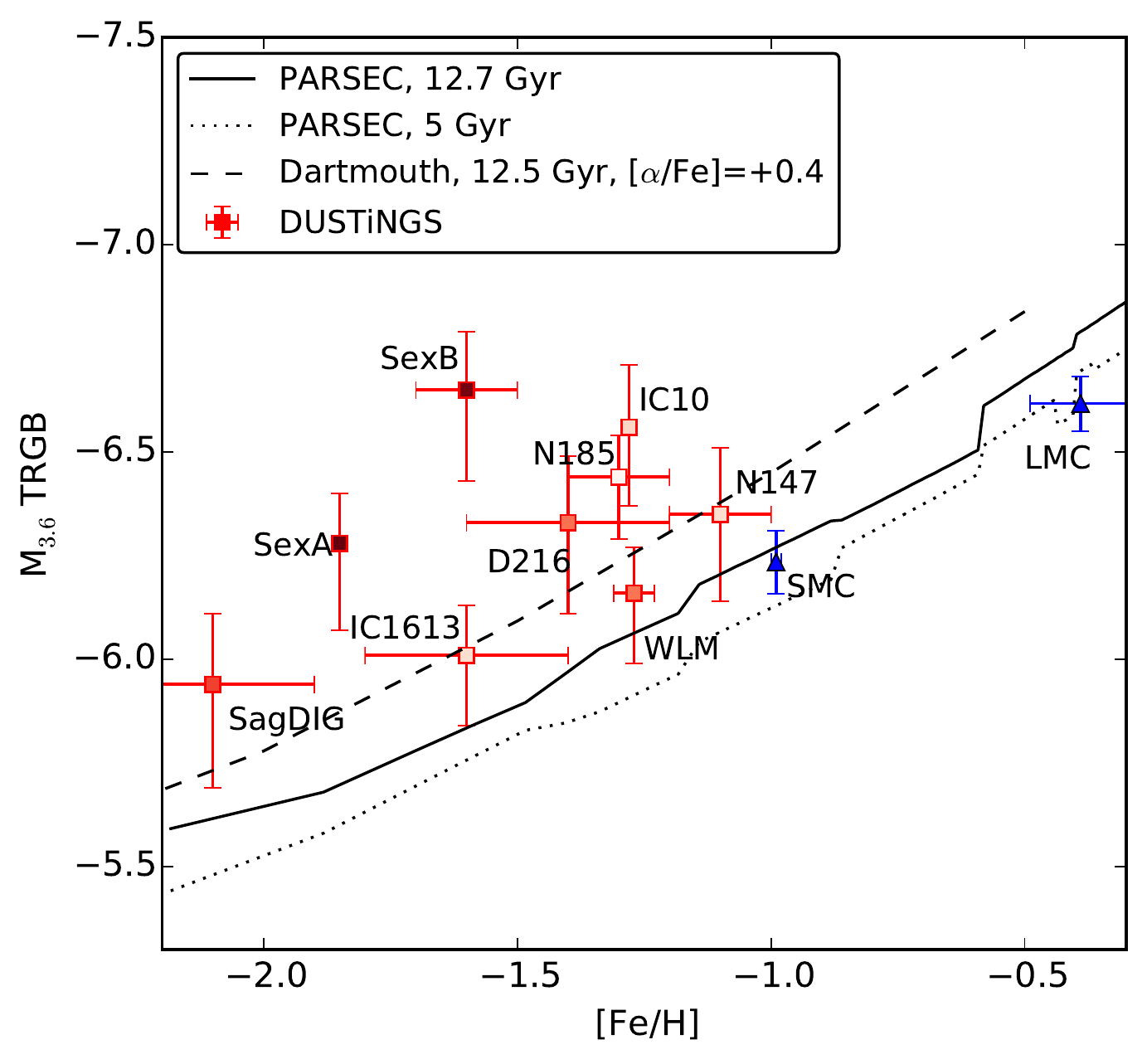}
\caption{The metallicity of the sample compared with the 3.6~\micron\ TRGB. The points are color-coded by distance with closer galaxies plotted in light red and more distant galaxies plotted with darker red. Uncertainties in M$_{3.6}$ include the uncertainties in the TRGB distance from the F814W data. Theoretical predictions are shown from the PARSEC stellar evolution models \citep{Bressan2012} for a 12.7~Gyr population (solid black line), for a 5 Gyr population (dashed black line), and from the Dartmouth stellar evolution models for a 12.5 Gyr population with an alpha-element enhancement of $+0.4$ \citep{Dotter2008}. Similarly derived values for the SMC and LMC are also shown. The DUSTiNGS galaxies approximately follow the predicted metallicity trend, but are offset to brighter magnitudes. The difference may be at least partially caused by the presence of a younger stellar population, metallicity changes, and/or completeness effects in the more distant galaxies. \label{fig:z_trend}}
\end{figure}

\subsection{The Non-Universality of the 3.6~\micron\ TRGB}
In the galaxies studied, the TRGB luminosity is not constant at 3.6~\micron\ within 1\,$\sigma$, varying from $-5.94^{+0.25}_{0.17}$ to $-6.65^{+0.22}_{0.14}$ mag. Without a well-calibrated correction, the luminosity of the TRGB at 3.6~\micron\ does not appear to be a predictable distance indicator. 

\citet{Salaris2005} examined the stability of the TRGB luminosities in the $I$ and $K$-bands using synthetic stellar populations with varying metallicity and age. The I-band showed the known, modest dependence on metallicity, which is typically included in TRGB calibration relations \citep[e.g., see][]{Rizzi2007a}, while the $K$-band showed a stronger dependence ($\sim1$ mag) with both metallicity and age of the RGB stars. 

Similarly, we examine whether the TRGB luminosities at 3.6 $\micron$ show a dependence on metallicity or the average age of the stellar populations. Figure~\ref{fig:z_trend} compares the TRGB luminosities at 3.6 $\micron$ for our sample as a function of metallicity. Also plotted are the TRGB luminosities at 3.6 $\micron$ for the Small and Large Magellanic Clouds (SMC and LMC respectively) calculated in the same manner as described above using data from the {\it Spitzer} Surveying the Agents of Galaxy Evolution program \citep[SAGE;][]{Meixner2006, Gordon2011}. Stellar metallicities for the SMC and LMC are from \citet{Dobbie2014} and \citet{Choudhury2016} respectively. Theoretical predictions for the TRGB luminosity at 3.6$\micron$ as a function of metallicity are shown from the PARSEC models for a 5 Gyr and 12.7 Gyr population \citep{Bressan2012}, and from the Dartmouth models for a 12.5 Gyr population with an alpha-element enhancement of $+0.4$ \citep{Dotter2008}. Stellar spectroscopic measurements of RGB stars show alpha enhancements of order $0.1-0.3$ in satellites of the Milky Way and M31, including the two dSph NGC~147 and NGC~185 studied here \citep[][and references therein]{Vargas2014}. Thus, we chose an alpha enhancement of $+0.4$ to bracket the predicted maximum increase to the TRGB luminosity. Experiments with different values of the efficiency factor $\eta$ ranging from 0.0$-$0.5 in the PARSEC models, which scales the mass-loss by stellar winds in the RGB phase of stellar evolution \citep{Reimers1975}, showed a negligible change on the TRGB 3.6$\micron$ luminosity. 

In Figure~\ref{fig:z_trend}, four galaxies (SMC, NGC 147, WLM, and IC 1613) are in agreement with the PARSEC stellar evolution models for older stars. Despite the range in models shown in Figure~\ref{fig:z_trend}, the TRGB luminosities for the remainder of the DUSTiNGS galaxies are not a clear function of [Fe/H]. It is possible that the stellar metallicities reported for our sample are preferentially measuring older, more metal-poor stars. If the upper RGBs are instead populated with a significant number of younger, more metal-rich stars, then this could partially explain why the TRGB luminosities are predominantly brighter than the models predict. We investigated whether the galaxies with brighter TRGB luminosities had a higher fraction of their stars formed more recently using star-formation histories derived from resolved stellar populations \citep{Weisz2014a}. We did not find a correlation between the TRGB luminosity at 3.6 $\micron$ and the star formation in the galaxies over timescales of 1, 2.5, 5, or 10 Gyr ago. Thus, while differences in metallicity and age may both play a part in the scatter and offset we measure in the 3.6 $\micron$ TRGB luminosities, they are unlikely to account for the full variability and offset seen in Figure~\ref{fig:z_trend}.

The differences between the measured TRGB luminosities at 3.6$\micron$ and the theoretical predictions may be related to the incompleteness or crowding in the data. The farthest galaxies (Sex~A, Sex~B) are two of the more discrepant measurements from the theoretical predictions, suggesting that photometric confusion or blending may be impacting the measurements in these two systems. Similarly, while we used fields in the outer regions of the galaxies for the measurements, it is possible that stellar crowding in some galaxies may result in measuring a brighter TRGB luminosity. Deeper, high-resolution imaging from the James Webb Space Telescope at similar wavelengths could reduce these uncertainties, providing improved constraints on the TRGB luminosities in the infrared. Regardless, this first comparison for a sample of galaxies with theoretical predictions shows variability in the TRGB luminosity at 3.6$\micron$ that is not well characterized. Without further constraints, the 3.6 $\micron$ TRGB luminosity does not appear to be a predictable distance indicator. 

\input{tab3}

\section{Constructing Radial Profiles from the Resolved Stellar Populations}\label{sec:geometry}
We created radial profiles from both the AGB stars and the RGB population following several steps. First, elliptical parameters were determined from the 3.6 $\micron$ images, which provided the appropriate geometry for defining radial annuli. Second, the stars were binned into annuli, with corrections made for outer annuli that were only partially included in the field of view. Third, stellar counts were corrected for photometric incompleteness as a function of radius based on artificial star tests. Lastly, stellar counts were corrected for foreground and background contamination. 

\subsection{Measuring Elliptical Parameters from the {\it Spitzer} Images}

The ellipticities and position angles in all nine galaxies have previously been measured using surface-brightness isophotal fits to optical images. However, at optical wavelengths, young star-forming complexes can twist or distort isophotes as a function of radius, impacting the robustness of the derived parameters. Because the surface-brightness in the infrared is not as biased by young star-forming regions, measuring the elliptical parameters from surface-brightness isophotal fits to the $Spitzer$ images provides a consistency check on previous measurements. Furthermore, the wide-field DUSTiNGS maps are ideal for fitting isophotes to larger radii.

\begin{figure}
\includegraphics[width=0.47\textwidth]{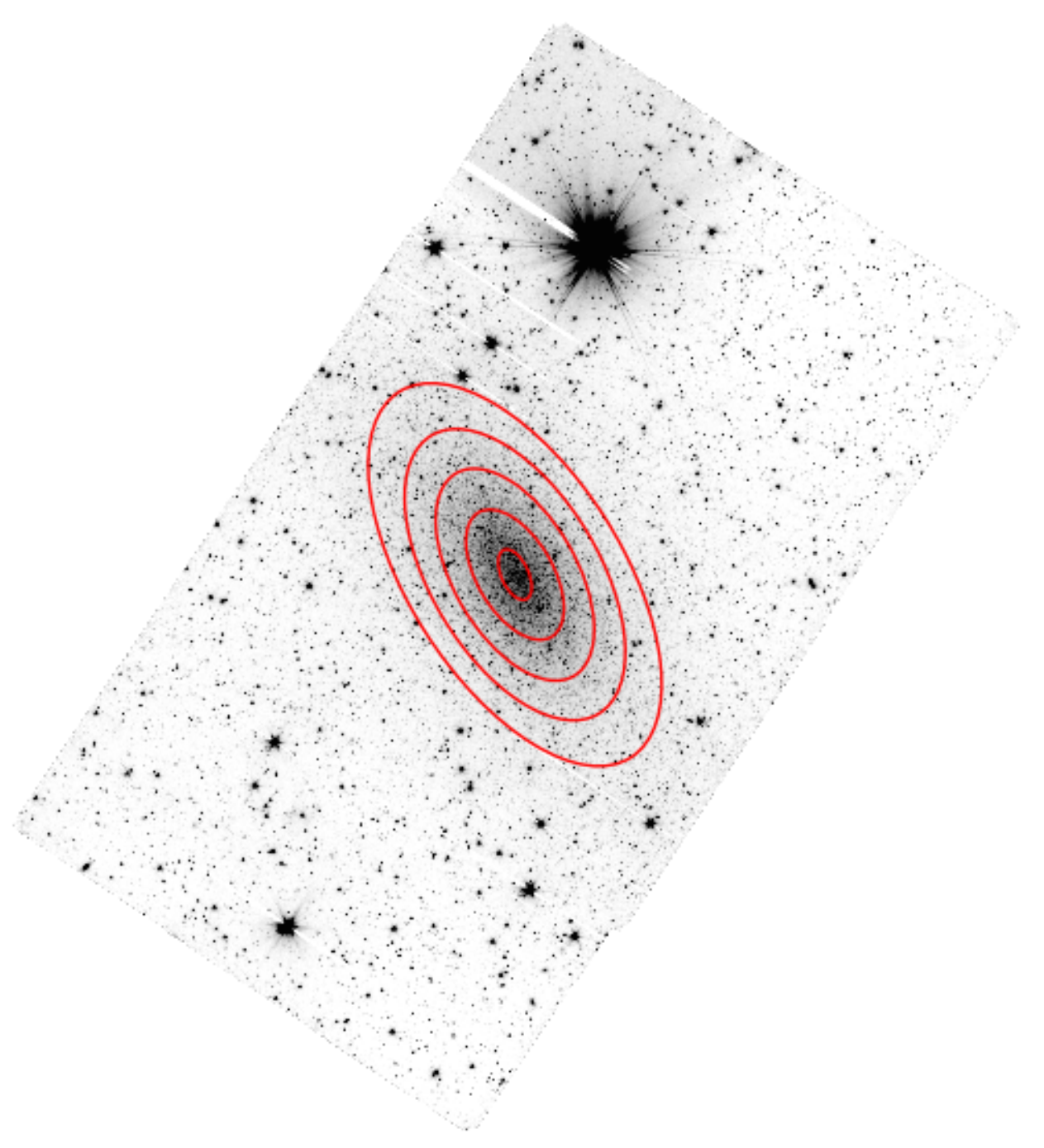}
\caption{3.6 $\micron$ image of NGC~147 with the best-fitting isophotes from the IRAF task {\sc ellipse}. The ellipticity and position angle are consistent over the optical disk of the galaxy. The best-fitting elliptical parameters for seven galaxies in the sample are listed in Table~\ref{tab:structure_params}. Consistent sets of concentric isophotes did not fit the disks of the remaining two galaxies; elliptical parameters from the literature are adopted as described in the text.}
\vspace{10pt}
\label{fig:isophotes}
\end{figure}

Figure~\ref{fig:isophotes} presents an example DUSTiNGS 3.6 $\micron$ mosaic of NGC~147 with ellipses fit to the surface-brightness isophotes. The wide field of view is typical of the observations. We measured the elliptical parameters from the 3.6 $\micron$ images using the Imaging Reduction and Analysis Facility (IRAF)\footnote{IRAF is distributed by the National Optical Astronomy Observatories, which are operated by the Association of Universities for Research in Astronomy, Inc., under cooperative agreement with the National Science Foundation.}  task {\sc ellipse}. The few, significant foreground stars close to and within the galaxy were removed from the images prior to fitting elliptical contours using the IRAF task {\sc imedit}, with interpolation from neighboring pixels. 

Table~\ref{tab:structure_params} lists the ellipticity and position angle derived from the isophotal fits as well as the values previously derived by other studies. For two galaxies, IC~1613 and Sag~DIG, the {\sc ellipse} task was unable to converge to stable parameters due to the more dispersed stellar populations and lower surface-brightness of these galaxies. For these two galaxies, we adopt the elliptical parameters from \citet{Bernard2007} and \citet{Higgs2016} respectively. For the rest of the sample, our measurements of the ellipticity and position angle agree with previous values and measured uncertainties with two exceptions. The first is Sex~A which has low ellipticity allowing for multiple fits with small changes in axial ratios at different orientations. The second is NGC~185, where we find a higher value of ellipticity with a larger position angle than previous optical studies \citep{Geha2010, McConnachie2012, Crnojevic2014}.

\subsection{Dividing into Radial Bins and Correcting for Incompleteness} 
Using the elliptical parameters in Table~\ref{tab:structure_params}, the stellar counts were separated into 1\arcmin\ annular regions. The stars within each annulus were counted and normalized by area. As not all outer annuli were fully covered in the image footprints, we subtracted the missing areas of the elliptical sectors from a given annulus. Details on calculating the area of the elliptical sectors are given in Appendix~\ref{sec:append_sectors}.

\input{tab4}

\begin{figure*}
\includegraphics[width=\textwidth]{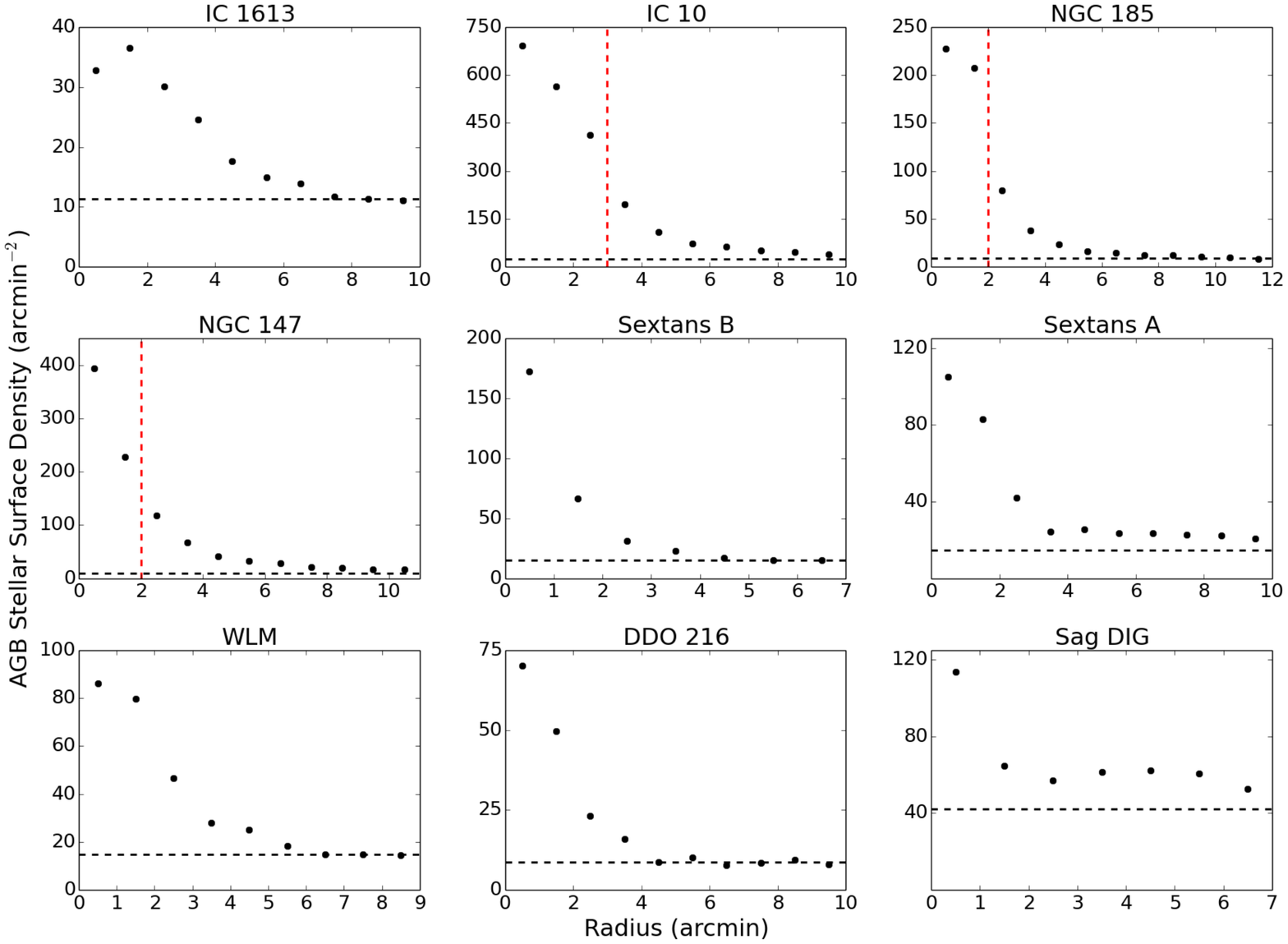}
\vspace{-35pt}
\caption{Completeness-corrected AGB stellar surface-density profiles for the sample. The horizontal dashed line marks the density at which the profiles flatten, providing an estimate of the contamination surface density in the magnitude range of the AGB stars. The vertical lines for IC~10, NGC~147, and NGC~185 mark the transition radius between the inner regions significantly affected by stellar crowding and the less crowded outer regions.}
\label{fig:initial_agb}
\end{figure*}

\begin{figure*}
\includegraphics[width=\textwidth]{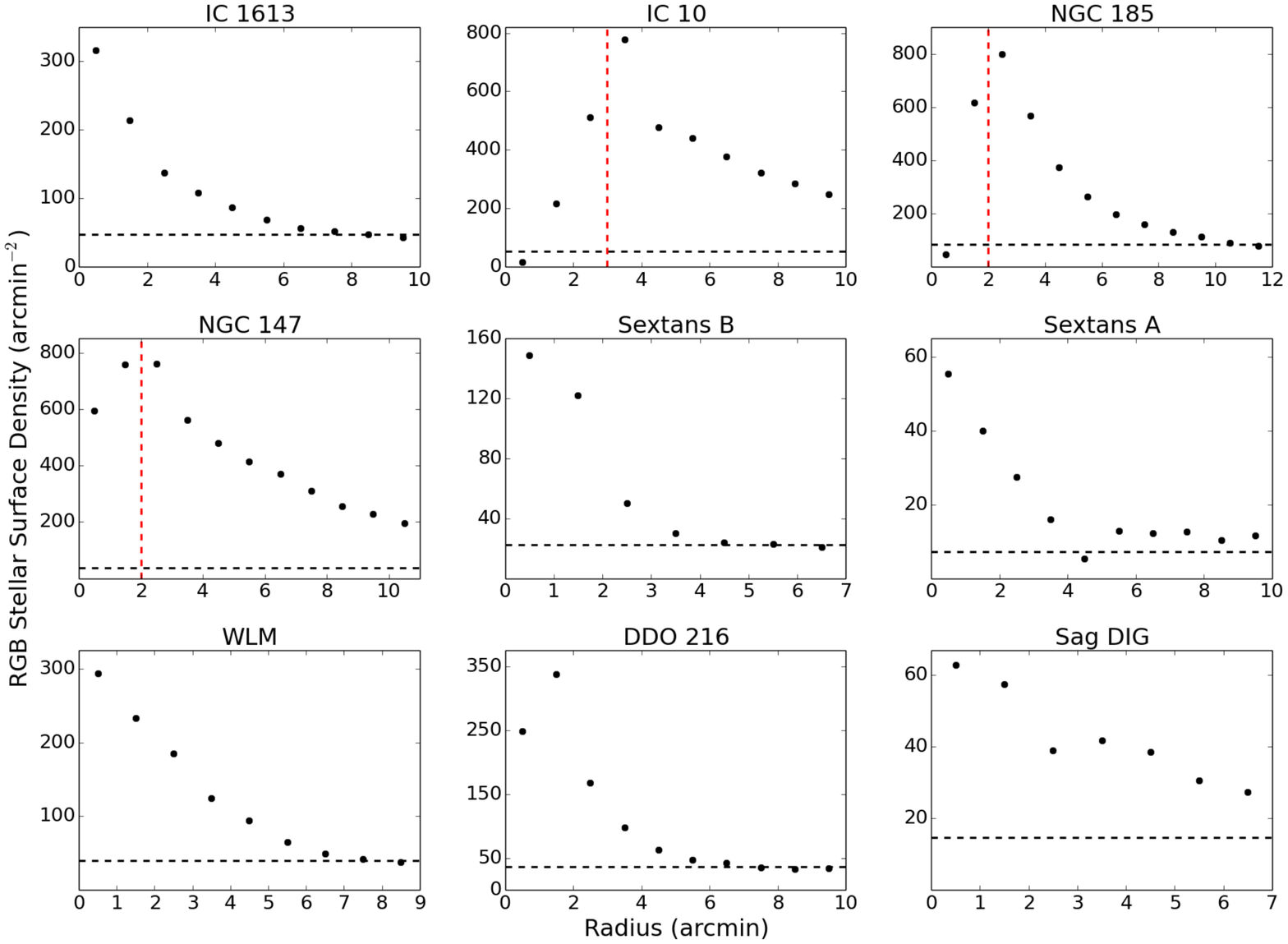}
\vspace{-35pt}
\caption{Completeness-corrected RGB stellar surface-density profiles for the sample. The horizontal dashed line marks the surface density of the estimated contamination in the magnitude range of the RGB stars. The vertical lines for IC~10, NGC~147, and NGC~185 mark the transition radius between the inner regions significantly affected by stellar crowding and the less crowded outer regions.}
\label{fig:initial_rgb}
\end{figure*}
 
The DUSTiNGS survey was designed to be $\sim$75\% complete at $M_{3.6} = -6$ mag in each epoch which is the approximate faint limit for the TRGB \citep{Jackson2007a, Boyer2009, Boyer2011}. Based on the total epoch 1 exposure times of 1080 s, the mean 75\% completeness limits were measured from artificial star tests to be [3.6] $= 19.1\pm0.1$ and [4.5] $= 18.7\pm0.2$, as reported in Paper~I. These values measure the global completeness limits of each data set. 

To obtain higher accuracy for the radial stellar profiles, we applied completeness corrections on each galaxy as a function of radius using artificial star tests as described in Paper~I. The individual corrections were estimated by binning the artificial star tests by both magnitude and radius. The completeness functions were then calculated within each radial bin. Finally, the radially-binned stellar counts were divided into magnitude bins and scaled up according to the magnitude dependent completeness functions calculated for each annulus.

Figures~\ref{fig:initial_agb} \& \ref{fig:initial_rgb} plot the completeness-corrected, AGB and RGB surface-density profiles as a function of radius. Each profile shows higher stellar densities at smaller radii followed by a decrease, which eventually flattens to a constant value. This floor to the stellar density provides an estimate of the foreground and background contamination in each field for the magnitude range of the AGB and the RGB stars. In a few cases, the profiles were reduced but not flat at the largest annulus around the galaxy that fit in the field of view. We estimate the contamination for these galaxies using regions at larger distances off the minor-axes as described in the next section. Three galaxies (IC~10, NGC~185, NGC~147) have higher levels of incompleteness from stellar crowding in their central regions and consequently show declining RGB stellar densities in the innermost radii. No additional corrections were made to these inner profiles. 

\begin{figure*}
\includegraphics[width=\textwidth]{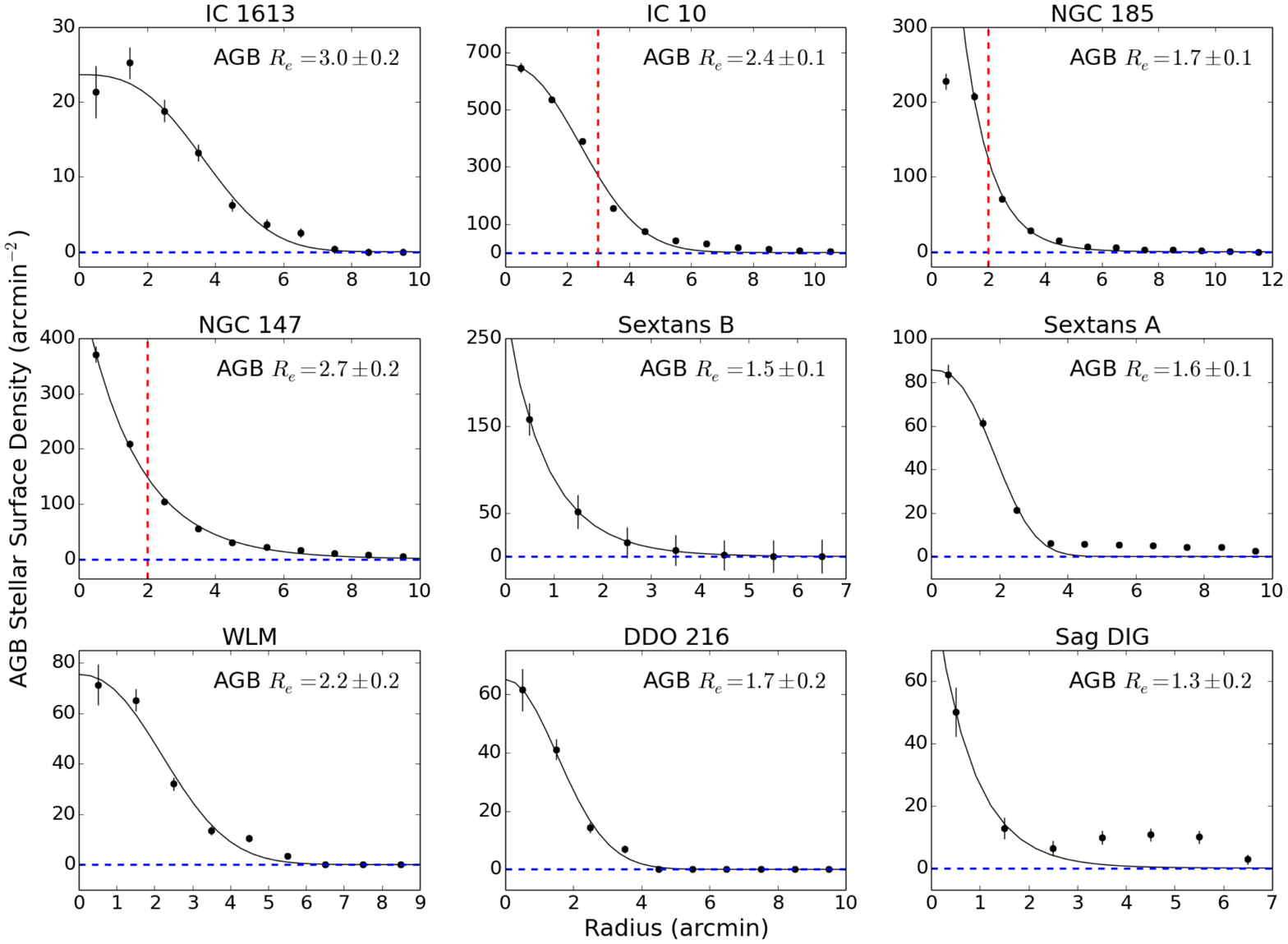}
\vspace{-35pt}
\caption{Final stellar surface-density profiles of AGB stars for the sample with the best-fitting S\'ersic profiles overlaid. Uncertainties represent the Poisson statistics on the stellar counts. Vertical lines separate the inner regions in three galaxies significantly affected by stellar crowding from their outer regions; only the outer profiles were used in the profile fits for these systems. Effective radii of the AGB stars (AGB $R_{e}$) based on the best-fitting profiles are noted in the upper right corner of each panel in units of arcmin and are listed in Table~\ref{tab:radial_extent} along with the S\'ersic indices.}
\label{fig:sersic_agb}
\end{figure*}

\begin{figure*}
\includegraphics[width=\textwidth]{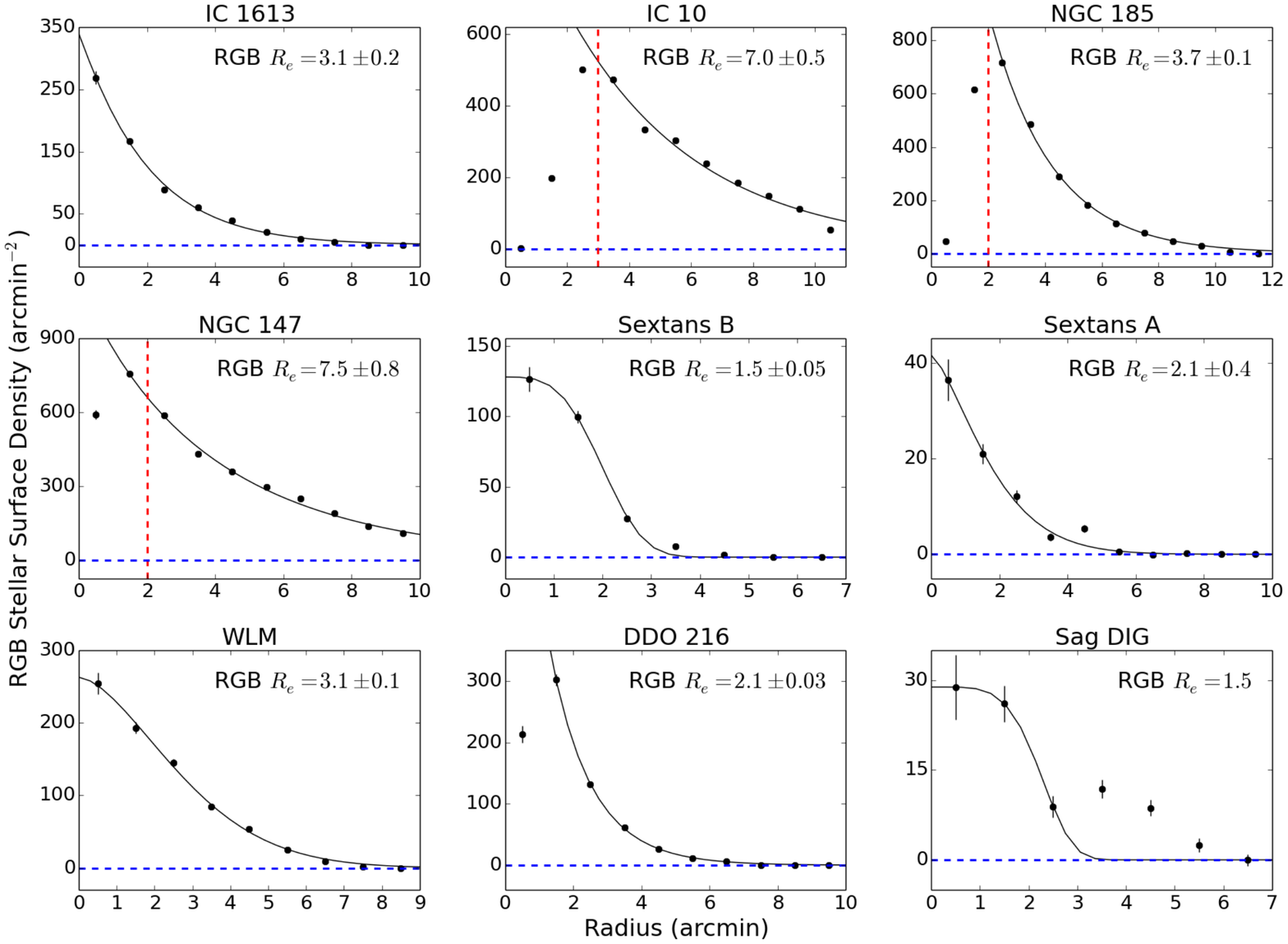}
\vspace{-35pt}
\caption{Final stellar surface-density profiles of RGB stars for the sample with the best-fitting S\'ersic profiles overlaid. Uncertainties represent the Poisson statistics on the stellar counts. Vertical lines separate the inner regions in three galaxies significantly affected by stellar crowding from their outer regions; only the outer profiles were used in the profile fits for these systems. Effective radii of the RGB stars (RGB $R_{e}$) based on the best-fitting profiles are noted in the upper right corner of each panel in units of arcmin and are listed in Table~\ref{tab:radial_extent} along with the S\'ersic indices.}
\label{fig:sersic_rgb}
\end{figure*}

\subsection{Statistically Subtracting Foreground and Background Contamination}
While the stellar catalogs have been culled to include point sources meeting a quality threshold, Galactic foreground stars and unresolved background galaxies are still present in the stellar counts. The fractional contributions of bona fide stars in a galaxy, foreground stars, and background galaxies change as a function of radius. In the inner regions of galaxies with high stellar crowding, fewer faint background sources are photometrically recovered. Thus, the surface density is predominantly comprised of stars in the host galaxy and brighter foreground contamination. Over radial annuli without significant stellar crowding, all three components contribute. At larger radii, where the contribution from the galaxy is small, the surface density is comprised predominantly of foreground and background contamination. The contamination is also a function of magnitude. Thus, estimates of the contamination affecting the RGB and the AGB populations are calculated separately. 

For the three galaxies (IC~10, NGC~185, and NGC~147) with higher levels of stellar crowding in their centers, only foreground sources are subtracted in the inner regions. As seen in the initial RGB radial profiles in Figure~\ref{fig:initial_rgb}, the surface densities at the center of these galaxies are diminished compared with the peak values. The radius at which stellar crowding impacts the radial profiles is readily measured from Figure~\ref{fig:initial_rgb}. We confirmed this radius by plotting the luminosity functions for these galaxies as a function of radius. Progressing from larger annuli inward, the radius where there were fewer numbers of bright stars than that identified in the prior outer radius (i.e., the radius at which the bright end of the luminosity function turned over) agreed with the radius estimated from the radial profiles. This radius, demarcated with a red dashed line in Figures~\ref{fig:initial_agb} \& \ref{fig:initial_rgb}, defines where foreground contamination alone is subtracted from the profiles.

IC~1613 and DDO~216 show slightly lower surface densities in the first bin of the AGB and RGB radial profiles, respectively. However, in both cases, the completeness function calculated for each bin was consistent with that of the surrounding annulus. Furthermore, the luminosity functions did not turn over or indicate that the inner annulus was impacted by higher stellar crowding. 

We used {\sc trilegal}, a simulator of photometry for stellar fields in the Galaxy \citep{Girardi2012}, to estimate the number of Galactic sources within the appropriate area. Table~\ref{tab:photometry} lists both the inner radii most impacted by stellar crowding and the surface density of foreground sources estimated from {\sc trilegal}. The simulated photometry is separated by magnitude; sources above the apparent TRGB 3.6$\micron$ magnitude (see Table~\ref{tab:trgb}) for the individual galaxies were subtracted from the AGB inner radial profiles and those below were subtracted from the inner RGB radial profiles. 

The DUSTiNGS observing strategy was designed such that the fields of view are larger than the half-light radii of the galaxies. The extended areal coverage provides a means to statistically estimate the combined contamination from foreground and unresolved background sources for the RGB and the AGB populations, as demonstrated in Figures~\ref{fig:initial_agb} \& \ref{fig:initial_rgb}. We subtracted the average surface densities calculated from the flattened density profiles at larger radii, adjusted by area, from the radial profiles. Similarly to the foreground corrections, the surface density of sources above and below the apparent TRGB magnitude for each galaxy was subtracted from the AGB and RGB stellar counts, respectively. The correction was applied to all stellar counts except those within the inner radii of the three galaxies with high levels of crowding. In two cases (IC~10 and NGC~147), the combination of the larger angular extent of the stellar populations and the orientation of the stellar disks within the field of view resulted in profiles that did not flatten completely. Additionally, the outer radii in Sex~A showed a slight increase after initially flattening, consistent with the detection of an extended stellar tidal structure previously identified \citep{Bellazzini2014}. Finally, the outer radii in Sag~DIG show significant variability in the stellar surface densities. For these four systems (IC~10, NGC~147, Sex~A, Sag~DIG), we estimated the background and foreground contamination by measuring the density of sources in a blank area of sky off the minor axes. The surface densities of contamination measured for Sex~A and Sag~DIG are lower than the stellar densities at larger radii in Figures~\ref{fig:initial_agb} \& \ref{fig:initial_rgb}, supporting the assumption that the slightly elevated densities in the extended profiles are due to stars which belong to the galaxies. The final stellar density profiles are presented in Figure~\ref{fig:sersic_agb} \& \ref{fig:sersic_rgb} and discussed in more detail below.

\section{The Radial Extent of the AGB stars, the RGB stars, and the variable AGB stars.}\label{sec:profiles}
The radial extent of the AGB stars and the RGB stars are proxies for exploring the relative distribution of an intermediate-age population with an older-age population inside the disks, while the variable AGB stars trace the distribution of intermediate-age stars inside the disks {\em and} the outer extremities. There are a number of different approaches to measuring and comparing the extent of the stellar population distributions. Here, we explore three different visualizations of the data. First, we measure the scale length of the AGB and RGB using S\'ersic profiles fit to the linear radial profiles after correcting for incompleteness and contamination. We do not attempt to fit the variable AGB distributions as the majority of galaxies do not have sufficient statistics for a robust fit. Second, we explore the cumulative distributions of the AGB stars, the RGB stars, and the variable AGB stars. Finally, we present the radial distribution of the ratio of AGB to RGB stars and of variable AGB to RGB stars.

\subsection{Radial AGB and RGB S\'ersic Profiles}
First, we fit S\'ersic profiles \citep{Sersic1963} to the AGB and RGB surface-density radial profiles. The S\'ersic function is a generalized exponential profile more often used to fit surface-brightness profiles. However, it has also been applied to stellar surface-density profiles with consistent results \citep[e.g.,][]{McConnachie2007, Battinelli2007, Javadi2011}.

We use a S\'ersic function with the following form for the AGB and RGB surface-density profiles:

\begin{equation}
\rho = \rho_{e}\exp\Big(-b_n\Big(\big(\tfrac{r}{R_{e}}\big)^{1/n}-1\Big)\Big)
\label{eq:sersic}
\end{equation}

\input{tab5}

\noindent where $b_n = 1.992n - 0.3271$ \citep{Capaccioli1989}. This profile is conveniently parametrized such that the free parameter is a radial scale length. We refer to these  measured scale lengths as effective radii ($R_{e}$), which provides a means to compare the radial extent of the different aged populations within the sample. For comparison with future studies, we note that this effective radius is analogous to a half-light radius found from fitting surface-brightness profiles, assuming a constant stellar mass-to-light ratio, a reasonable assumption in the infrared. Our calculated values of  $R_{e}$ from the RGB stars generally agree with previously reported half-light radii measured from surface-brightness fits where available \citep{McConnachie2012}, confirming our assumption. The exceptions are IC~10 and NGC~147 for which we measure larger effective radii than previously reported.

Figure~\ref{fig:sersic_agb} presents the radial AGB surface-density profiles with S\'ersic fits for the sample; Figure~\ref{fig:sersic_rgb} presents the same for the RGB population. Uncertainties represent the Poisson statistics on the stellar counts. For the galaxies where crowding impacts the inner profiles, we excluded the central regions from the fits. We also excluded the extended stellar feature in the outer regions of Sex~A and Sag~DIG (see Section~\ref{sec:galaxies}). The best-fitting profiles are overlaid with effective radii noted in each panel; Table~\ref{tab:radial_extent} lists the S\'ersic indices and effective radii. The S\'ersic index is close to unity for five galaxies, indicating nearly exponential profiles. The remaining four galaxies have indices less than one. The effective radii found for the RGB stars were comparable to or larger than the effective radii found for the AGB stars as discussed in Sections~\ref{sec:galaxies} $-$ \ref{sec:scenarios}. 

The RGB radial profiles provide a measure of the extent of the main stellar disk. The radius at which the stellar counts drop to background levels has previously been established as the approximate boundary between the main stellar disk and the extended stellar envelope of a dwarf galaxy as described by \citet{Minniti1996} for WLM (although they use ``halo'' to describe the extended distribution of older stars versus the more centrally concentrated young stars). With the exception of IC~10 and NGC~147, the RGB surface densities consistently reach a floor at $\sim3\times R_{e}$, where the stellar counts become indistinguishable from the background surface density. Thus, we refer to the regions farther than $\sim3\times R_{e}$ of the RGB stars as the ``outer extremities'' of the galaxies.

\subsection{Cumulative Radial Distributions of AGB, RGB, and Variable AGB Stars}
\begin{figure*}
\includegraphics[width=\textwidth]{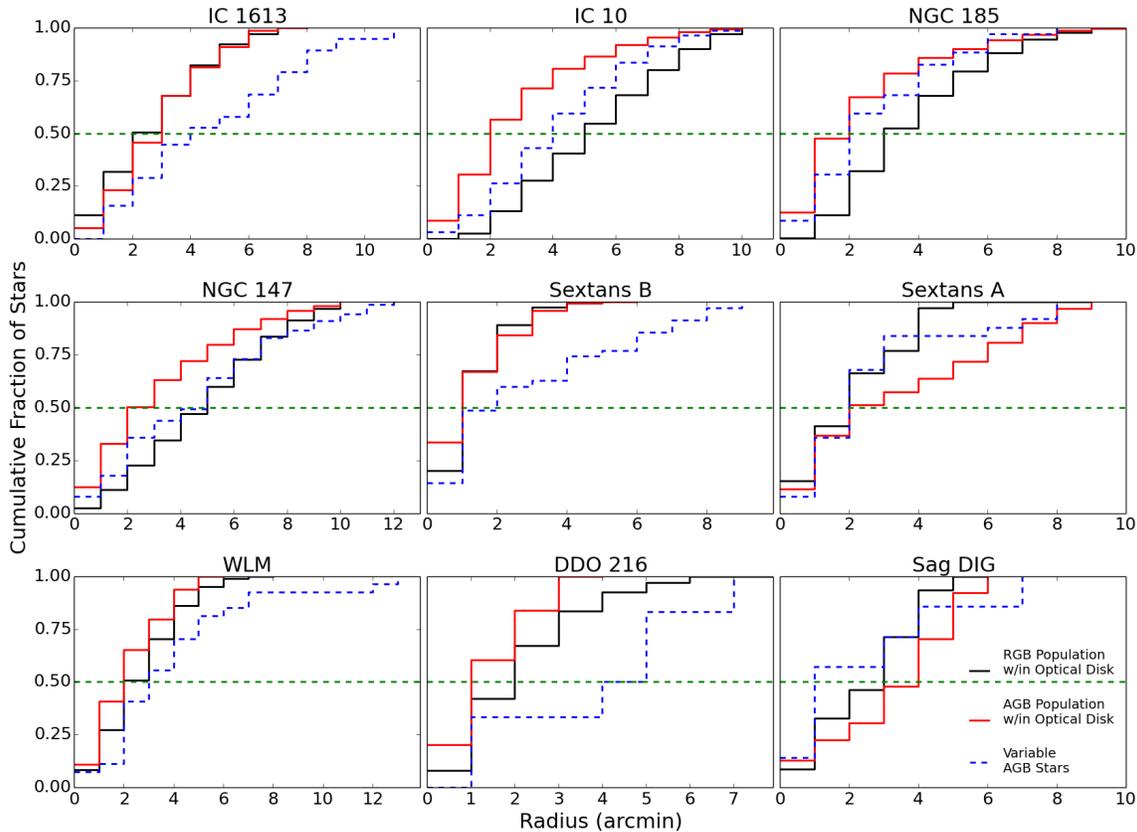}
\vspace{-25pt}
\caption{The cumulative distribution of the AGB stars (red), the RGB stars (black), and the variable AGB stars (dashed blue). The AGB and RGB stellar distributions are plotted out to $\sim3 \times R_e$, while the variable AGB stellar distribution reaches into the extremity of the galaxies in all cases with extended areal coverage. The 50\% fraction is marked with a dashed green horizontal line. While coarsely binned, the cumulative distributions show whether the intermediate-age AGB stars follow similar radial distributions as the older RGB stars.}
\label{fig:cumulative}
\end{figure*}

Figure~\ref{fig:cumulative} presents the cumulative fraction of the AGB stars, the RGB stars, and the variable AGB stars. The histograms are calculated using the number of sources in each 1$\arcmin$ bin normalized by the total number of sources for each stellar type. These cumulative plots allow a direct comparison of the distribution of the intermediate-age AGB stars and the older RGB stars within the disks of each galaxy, as well as the distribution of variable AGB stars in the disks and extremities. The radius which encloses 50\% of the RGB stars is a second measure of a scale length of a galaxy; the precision of this method is dominated by the width of the 1\arcmin\ bin size. As listed in Table~\ref{tab:radial_extent}, the scale lengths of the RGB stars measured with this approach agree with the effective radii determined from the S\'ersic profiles (RGB $R_{e}$) within the uncertainties, providing a consistency check on our measurements. We adopt the $R_e$ calculated from the S\'ersic profiles as the scale length of the galaxies and use these in the subsequent analysis.

Figure~\ref{fig:cumulative} highlights the similar distribution and scale length of the intermediate-age and old stellar populations in four galaxies (IC~1613, Sex~B, WLM, DDO~216). These systems lack evidence of a strong intermediate-age radial gradient in the sense that the intermediate-age stars have a similar distribution relative to the old stars. Three galaxies (IC~10, NGC~185, NGC~147) show a somewhat more centrally concentrated intermediate-age population. In two galaxies, Sex~A and Sag~DIG, the AGB stars have more extended spatial distribution than the RGB stars. The galaxies with areal coverage past $3\times R_e$ show extended distributions of the variable AGB stars. Sections~\ref{sec:galaxies}$-$\ref{sec:scenarios} discuss this further. Note that the extended distribution of variable AGB stars past $3\times R_e$ cannot be directly compared to the AGB and RGB as their identification is based on photometry alone and confined to the disk of the galaxies. 

\subsection{Ratio of AGB and Variable AGB to RGB Stars as a Function of Effective Radius}
\begin{figure*}
\includegraphics[width=\textwidth]{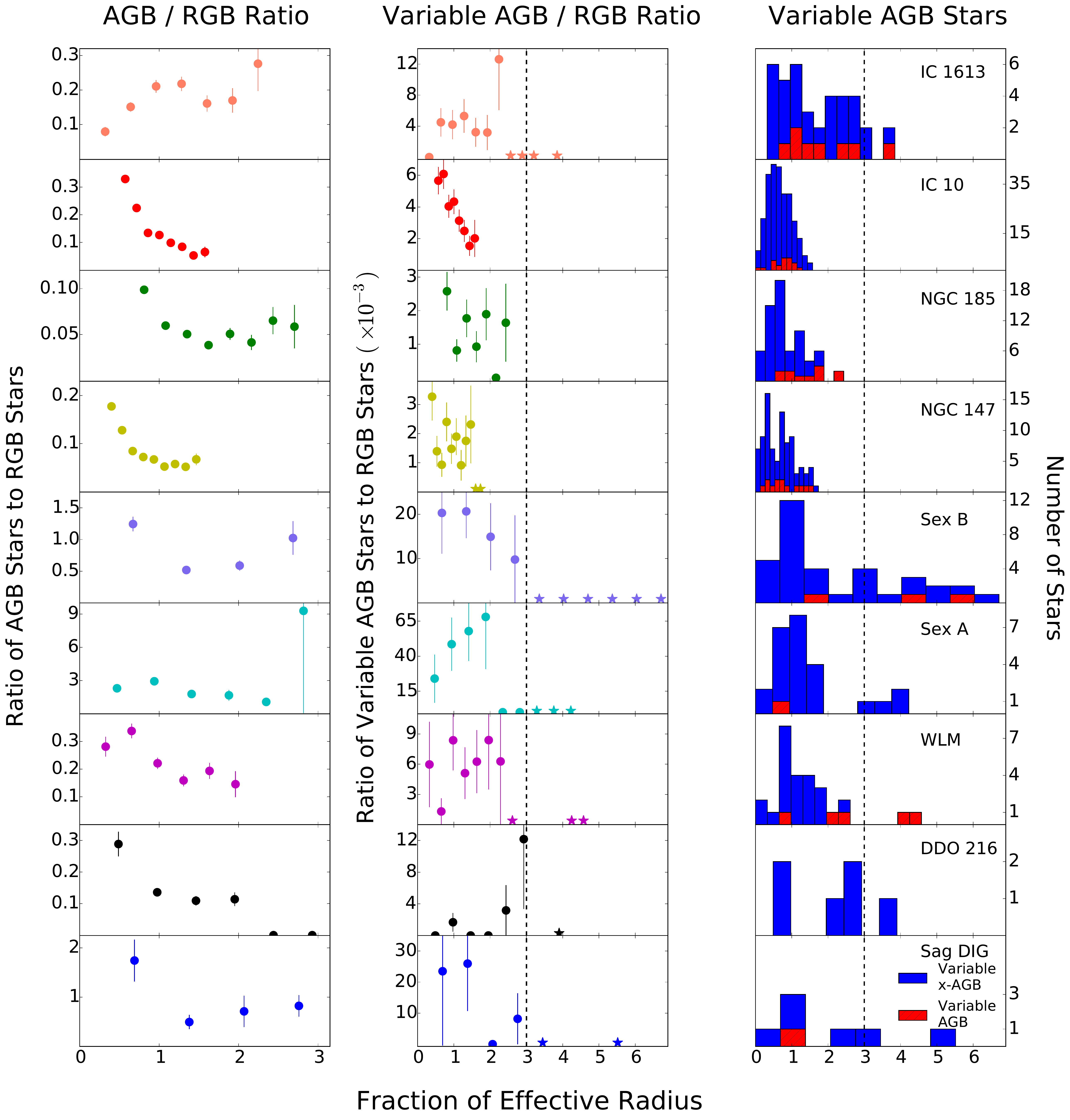}
\caption{The left panel shows the ratio of AGB to RGB stars as a function of effective radius; uncertainties are based on Poisson statistics from the radial profiles. The ratio of stars in the inner radii of the crowded galaxies are omitted due to the higher uncertainties (IC~10, NGC~147, NGC~185). The middle panel shows the ratio of variable AGB stars to RGB stars. The vertical dashed lines are set at $3\times R_e$, marking the approximate transition from the stellar disk to the outer extremity of the galaxy. A `star' symbol is used to denote annuli where we find variable AGB stars outside the stellar disk of the galaxy; all galaxies with extended areal coverage have AGB stars detected in their stellar extremities. The ratios trace relative changes in the populations as a function of radius; they do not represent the absolute values for the galaxies' stellar populations. The right panel shows the number of variable AGB and x-AGB stars detected in each radial annulus. }
\label{fig:ratios}
\end{figure*}

The left panel of Figure~\ref{fig:ratios} presents the ratio of AGB to RGB stars as a function of effective radius for the sample. The radial annuli have been converted from arcmin to $R_e$ based on the RGB scale lengths measured from the S\'ersic profiles to allow for an inter-sample comparison (see Table~\ref{tab:radial_extent}, column 8). We omit the ratio of the inner-most radii in the three galaxies affected by stellar crowding. The AGB / RGB ratios are plotted across the stellar disks to $\sim3 \times R_{e}$ in the cases for which there are sufficient data in both the AGB and RGB counts. 

The middle panel of Figure~\ref{fig:ratios} presents the ratio of variable AGB stars to RGB stars. The variable AGB / RGB ratios are plotted to a similar extent as the AGB / RGB ratios. However, we also mark radial bins past  $3 \times R_{e}$ with identified variable AGB stars; these are noted as star symbols along the x-axis of each subplot. For comparison, the right panel presents histograms of the number of variable AGB stars as a function of effective radius, with the variable x-AGB stars also noted. The transition between disk and outer extremity is marked with a vertical dashed line. We did not account for RGB stars below the photometric limit of the DUSTiNGS data (i.e., we made no adjustment for the star-formation history of the galaxies nor the initial mass function). Thus, the ratio values trace the changes of the two populations relative to one another; the values do not represent the absolute ratio of all AGB to RGB stars present in the galaxies. 

The ratios of AGB to RGB stars as a function of radius confirm the trends seen in Figures~\ref{fig:sersic_agb}$-$\ref{fig:cumulative} that the majority of the sample have well-mixed intermediate-age and old populations, with the exceptions of IC~10, and to a lesser degree, NGC~147 and NGC~185. 

\section{Notes on Individual Galaxies}\label{sec:galaxies}
Here, we describe the galaxy profiles individually and compare our results with previous findings from the literature. Many of the galaxies were included in a study that mapped the distributions of carbon AGB stars (hereafter carbon stars) and RGB stars; separate references from this study are listed below for the individual galaxies. Many of the previously identified carbon stars are detected in the DUSTiNGS stellar catalogs (see Paper~II), but not all of these stars meet our variability criteria given the cadence and sensitivity limits of our observations. 

\begin{figure}
\includegraphics[width=\columnwidth]{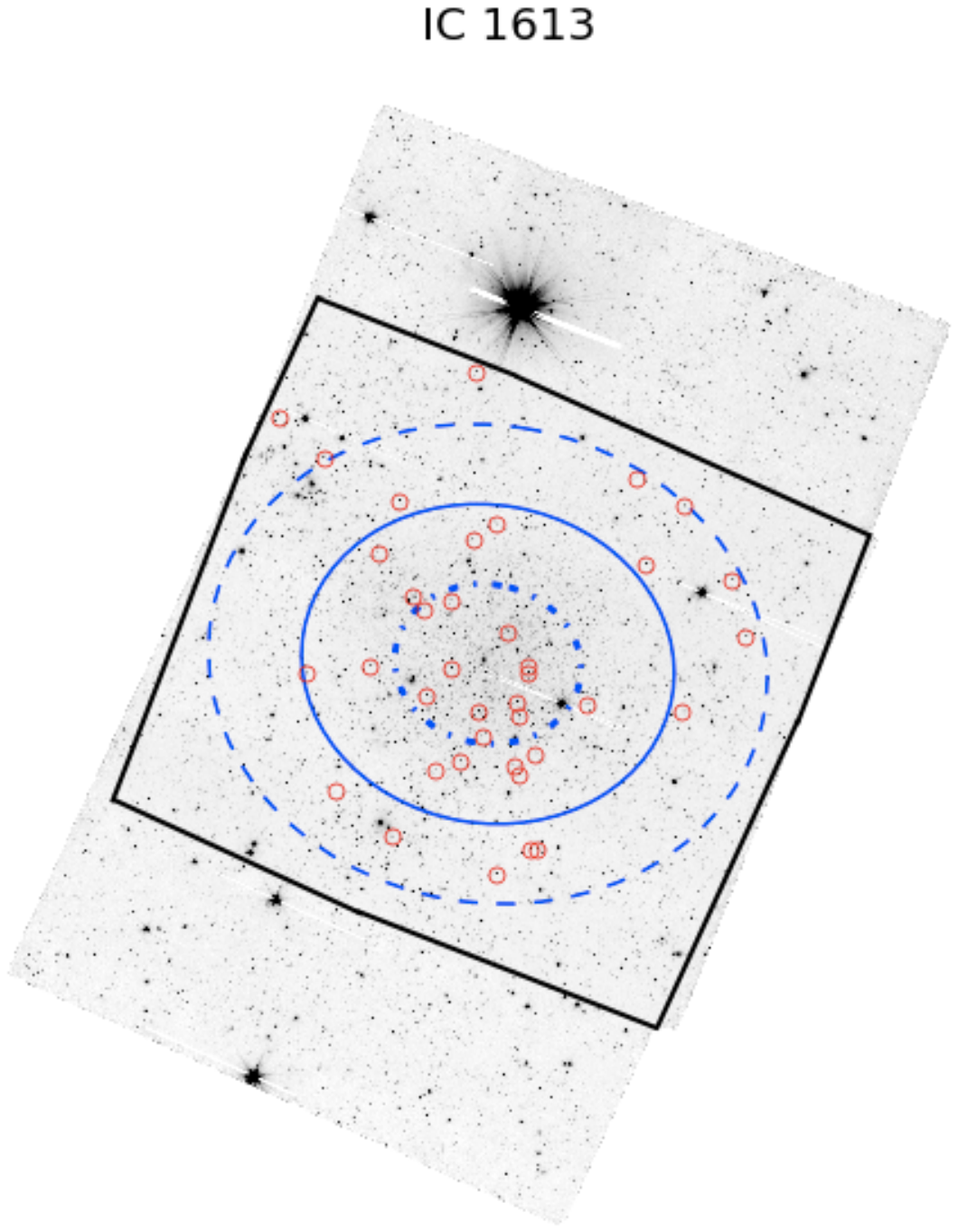}
\caption{3.6 $\micron$ map of IC~1613 with North up and East left. The black outline marks the area covered at 3.6 and 4.5 $\micron$ over two epochs, defining the region within which variable stars may be identified. The inner dot-dashed ellipse corresponds to 1 $R_e$ based on our adopted elliptical parameters. Each successive ellipse is spaced by 1 $R_e$; the dashed ellipse marks $3\times R_e$ which defines the boundary between the main stellar disk of the galaxy and the outer extremities. The ellipses correspond to those used in the radial profile analysis. Red circles mark the location of identified variable AGB star candidates. }
\label{fig:IC1613}
\end{figure}

\begin{figure}
\includegraphics[width=\columnwidth]{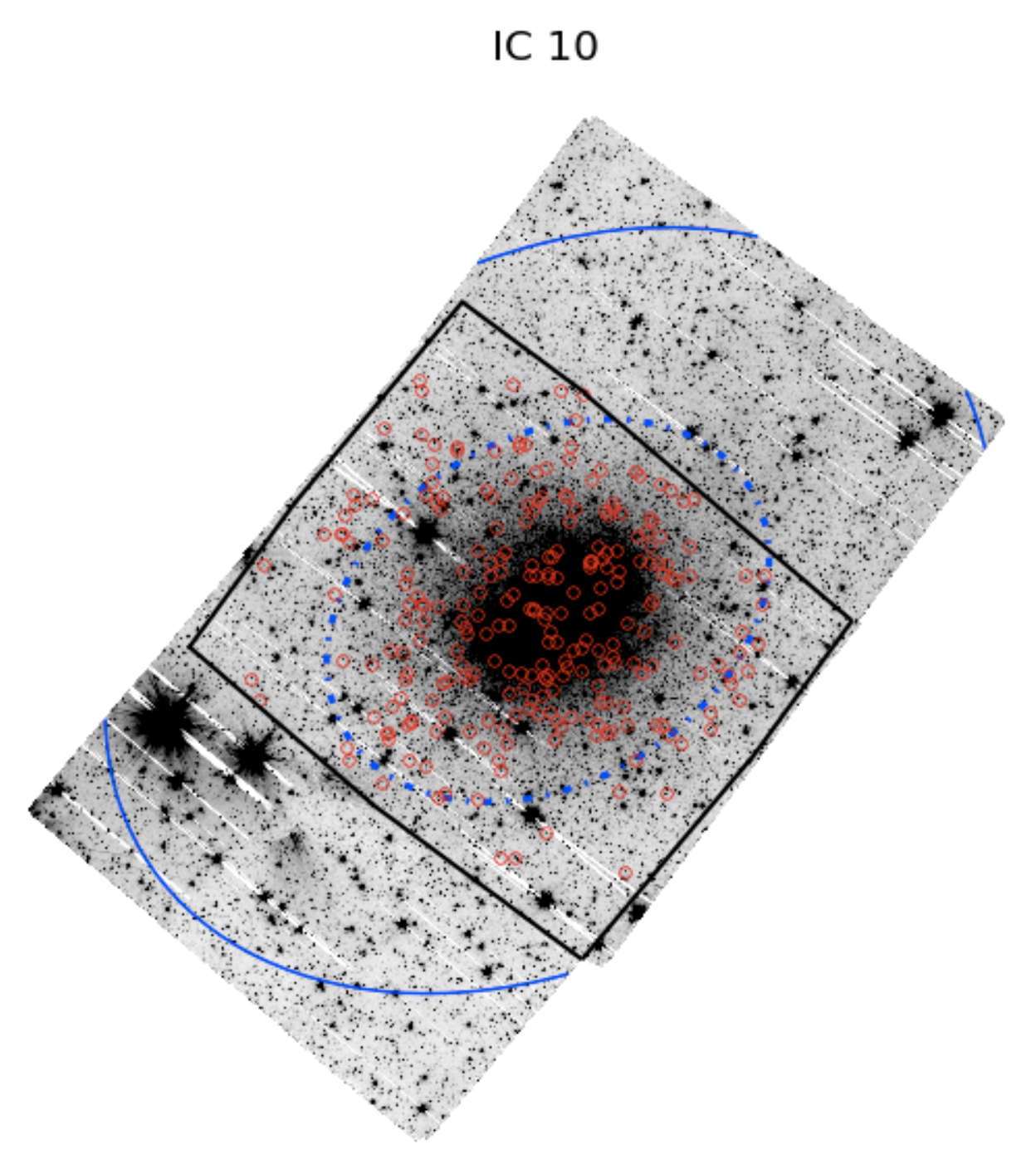}
\caption{3.6 $\micron$ map of IC~10 with overlapping footprint of 2 epochs in 3.6 and 4.5 $\micron$ outlined, 1 and 2 $\times R_e$ ellipses marked, and the locations of variable AGB stars. All colors and symbols are the same as those used in Figure~\ref{fig:IC1613}.}
\label{fig:IC10}
\end{figure}

In addition to a focused discussion on each galaxy, we also present maps of the sample in Figures~\ref{fig:IC1613}$-$\ref{fig:SagDIG} that highlight the overlapping footprints of our multi-wavelength, multi-epoch observations, ellipses based on the geometries in Table~\ref{tab:structure_params}, and locations of the variable AGB stars. Presented in this way, one can more easily compare the stellar disk with the outer extremities in each galaxy, and assess the spatial extent of the variable AGB population. Section~\ref{sec:comparison} compares the sample. 

\subsection{IC~1613}
IC~1613 is a gas-rich dI that is relatively isolated within the Local Group. Based on the resolved stellar populations in an inner and outer field, IC~1613 has had a constant star-formation history over its lifetime with higher overall star formation at smaller radii \citep{Skillman2003, Skillman2014}. Young and intermediate-age large amplitude AGB variables have been identified from $JHK_S$ photometry \citep{Menzies2015}. This study also derives a distance modulus of $24.37\pm0.08$ mag from C-rich Miras using the period$Ð$luminosity relation, in close agreement with our value of $24.39\pm0.03$. Previous ground-based, wide-field imaging found that the evolved stellar populations extend much farther than previously thought, reaching a galactocentric radius of 16.\arcmin5 \citep{Bernard2007}. Young stars were found to be located only in the inner 10\arcmin, with the majority inside a radius of $\sim7$\arcmin, confirming a strong young-to-old age gradient. A currently active star-forming region is off-center, at $2-4\arcmin$ from the center of the galaxy. Photometrically identified carbon stars have been found to extend to radii farther than the young stars, with 24 optically classified carbon stars past $\sim10\arcmin$ \citep{Albert2000} and a carbon star radial profile declining at approximately the same rate as the RGB stars out to $\sim18\arcmin$ \citep{Sibbons2015}. We find that the ratio of AGB to RGB stars increases at $\sim1\times R_e$ (see Figure~\ref{fig:ratios}), which corresponds to the radius of the off-center star-forming region where younger AGB stars likely contribute to the higher relative number of AGB stars. 

Figure~\ref{fig:IC1613} shows the 3.6 $\micron$ map of IC~1613 with the area mapped by both the 3.6 and 4.5 $\micron$ observations over 2 epochs. This is the region within which the variable AGB stars are identified. Also shown are ellipses with the geometry derived above (see Table~\ref{tab:structure_params}). Variable AGB stars are detected in the extremity of the galaxy out to $\sim4\times R_{e}$ (i.e., $\sim12$\arcmin), the extent of the multi-wavelength, multi-epoch observations.

\subsection{IC~10}
IC~10 is a unique system in the Local Group, considered to be a starburst galaxy with significant star formation over the last $\sim0.5$ Gyr \citep{Sanna2010}. The starburst may have resulted from a collision with an \HI\ cloud \citep[e.g.,][]{Huchtmeier1979, Sanna2010}. IC~10 has a larger angular size traced by an extended low surface-brightness population of RGB stars out to $\sim18-23\arcmin$ \citep{Sanna2010}. Because of the larger angular extent, our observations do not reach the edge of the stellar disk. The RGB population has been reported to be not only more spatially extended than the young population and the AGB stars, but offset from these two populations \citep{Gerbrandt2015}. This previous study measured the effective radius based on a S\'ersic profile of the RGB stars to be $5.\arcmin75\pm0.11$ and of the AGB stars to be $3.\arcmin22\pm0.30$. We derive an effective radius of the RGB stars to be $7.\arcmin$0 based on the best-fitting S\'ersic profile, but also note that the scale length measured by the cumulative distributions is $5.\arcmin5\pm0.5$, in close agreement with the value from \citet{Gerbrandt2015}. 

The radial profiles of IC~10 in Figures~\ref{fig:sersic_agb} $-$ \ref{fig:ratios} show that the AGB stars are more centrally concentrated than the RGB stars. Given the recent starburst in IC~10, the higher number of AGB stars at smaller radii could include fairly young AGB stars formed in the aftermath of this star-formation event. Yet, even though our areal coverage is limited to within $2\times R_e$, the ratio of AGB to RGB stars smoothly declines within this region and no AGB stars are identified in the outskirts of our observations. Figure~\ref{fig:IC10} presents the map of IC~10 which shows that few AGB stars are detected between $1-2 \times R_e$. The detection of such an gradient agrees with results from \citet{Gerbrandt2015}, but conflicts with previous findings that the carbon star population (with typical ages of $\sim1$ Gyr) is well mixed with a scale length comparable to that of the RGB population \citep{Demers2004}. This suggests that the AGB population we detect is not dominated by Carbon stars, providing supporting evidence of primarily younger AGB population. 

\begin{figure}
\includegraphics[width=\columnwidth]{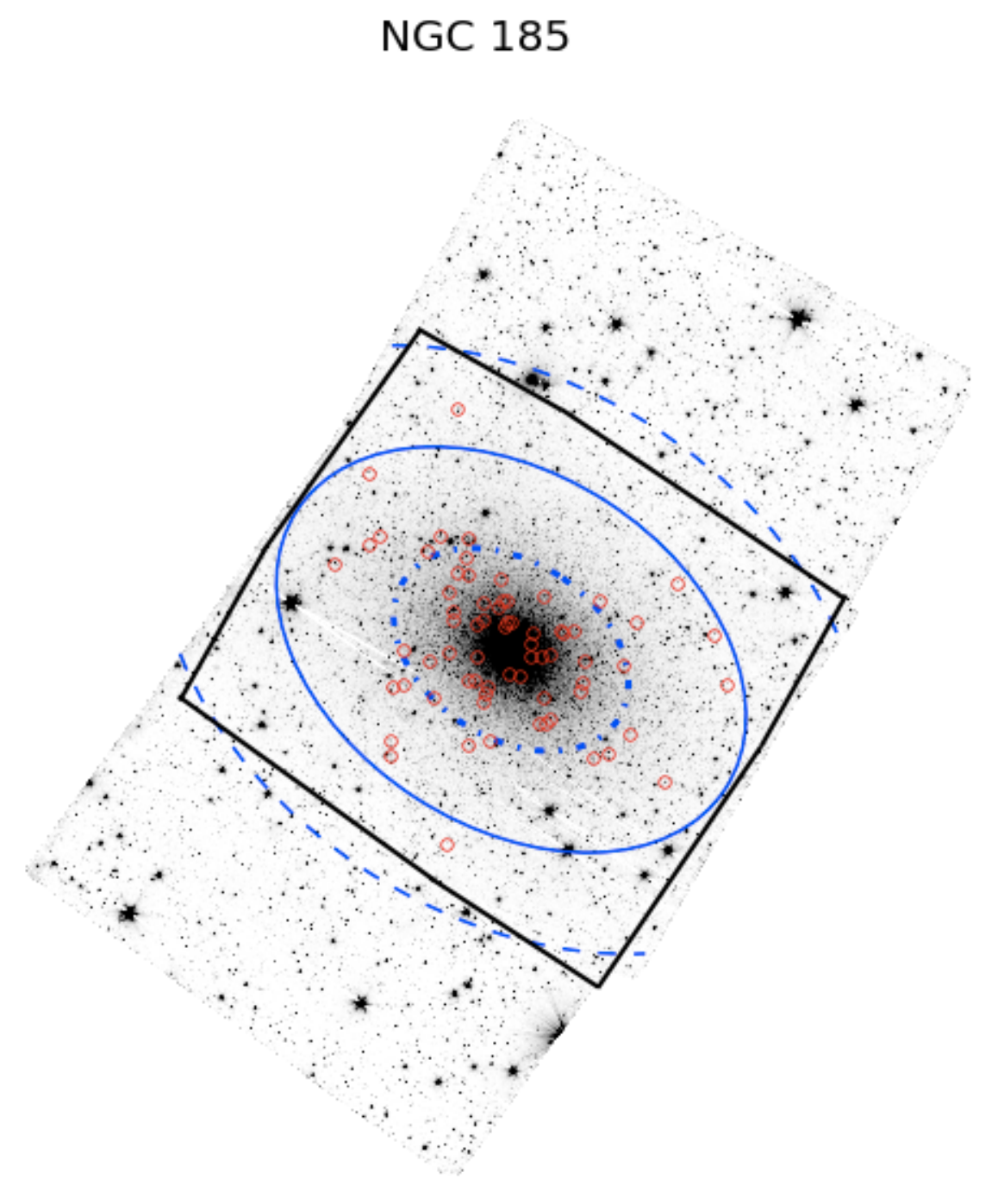}
\caption{3.6 $\micron$ map of NGC~185 with overlapping footprint of 2 epochs in 3.6 and 4.5 $\micron$ outlined, 1, 2, and  $3 \times R_e$ ellipses marked, and the locations of variable AGB stars. All colors and symbols are the same as those used in Figure~\ref{fig:IC1613}.}
\label{fig:NGC185}
\end{figure}

\begin{figure}
\includegraphics[width=\columnwidth]{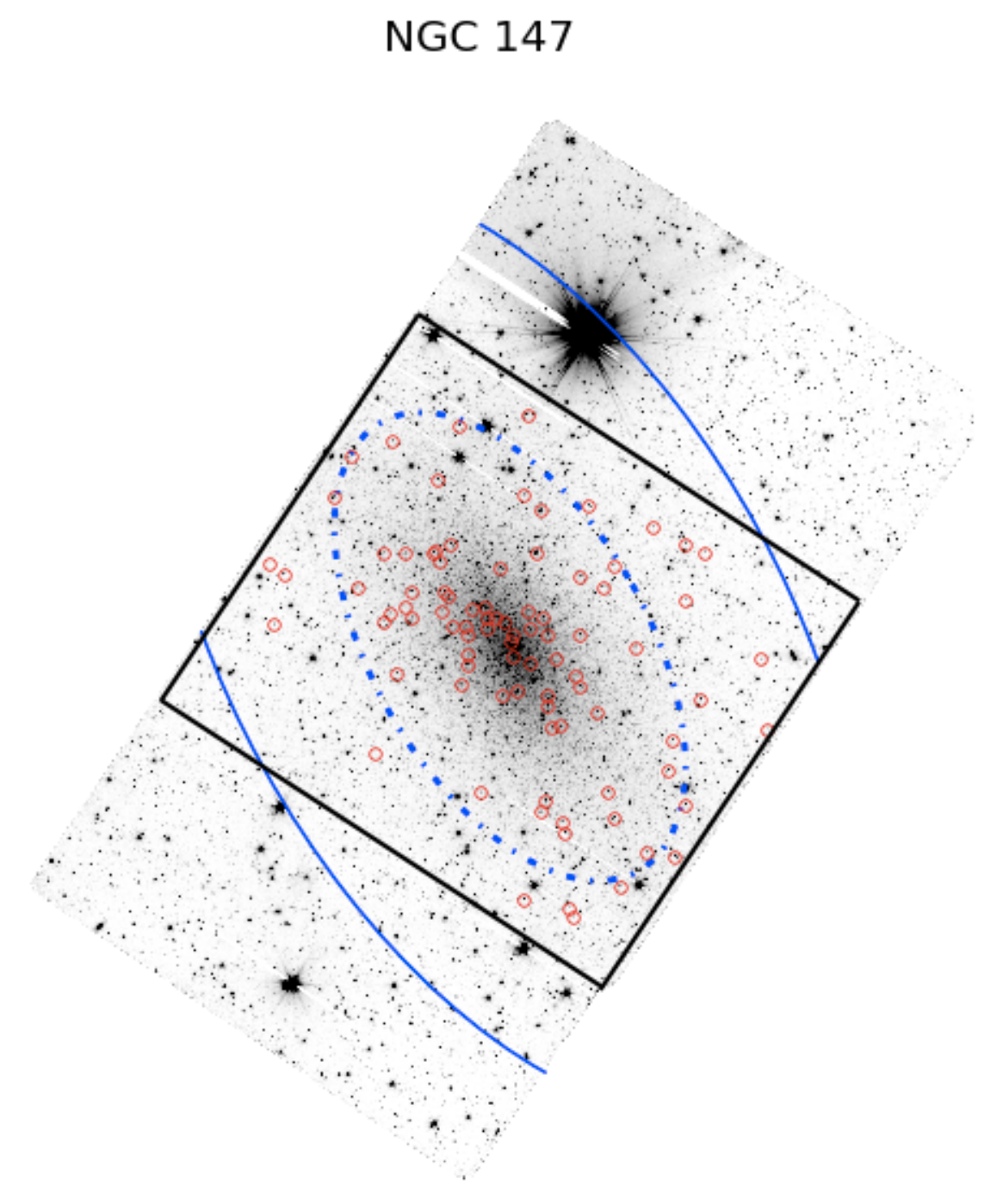}
\caption{3.6 $\micron$ map of NGC~147 with overlapping footprint of 2 epochs in 3.6 and 4.5 $\micron$ outlined, 1 and $2 \times R_e$ ellipses marked, and the locations of variable AGB stars. All colors and symbols are the same as those used in Figure~\ref{fig:IC1613}.}
\label{fig:NGC147}
\end{figure}

\subsection{NGC~185}
NGC~185 is one of two dSph in our sample. The galaxy has a gas content of $\sim3\times10^5$ \msun\ \citep{Marleau2010}, and hosts a population of stars younger than $\sim1$ Gyr \citep{Martinez-Delgado1999, Davidge2005}, and possibly as young as 100 Myr \citep[][; Golshan et al. submitted]{Davidge2005}. Stellar metallicity gradients have been reported based on observations of RGB stars both spectroscopically \citep{Vargas2014, Ho2015} and photometrically \citep{Crnojevic2014}.

From Figure~\ref{fig:ratios}, the ratio of AGB to RGB stars is higher in the inner region. Interestingly, outside the inner $\sim1\times R_e$, the ratio of AGB / RGB stars remains constant. Thus, the inner regions show evidence of an intermediate-age gradient, while at larger radii the intermediate-age and old stellar populations appear to be well-mixed. This indicates that we may be detecting a younger population of AGB stars centrally and is consistent with previous findings of a higher ratio of carbon stars in the central region \citep{Nowotny2003, Battinelli2004b}. As shown in Figure~\ref{fig:NGC185}, the detection of variable stars is confined to within $2 \times R_e$ due to the larger angular size of the galaxy and the orientation of the images. 

\subsection{NGC~147}
NGC~147 is the second dSph in our sample. The galaxy has little gas content with the last star-formation event occurring between $\sim300$ Myr (Golshan et al. submitted) and $\sim1$ Gyr \citep[e.g.,][]{Sage1998, Marleau2010}. \citet{Crnojevic2014} reported evidence of emerging symmetric tidal tails in the NW and SE directions, but at larger radii than covered by our observations. The tidal tails include an extended stellar stream and are possibly due to an interaction with M31 \citep{Geha2015, Arias2016}. Similar to IC~10, the best-fitting S\'ersic profile yields an effective radius ($7.\arcmin5$), while the cumulative distributions suggest a scale length of $5.\arcmin5\pm0.5$. The smaller value agrees with results from \citet{Crnojevic2014} who report an effective radius of $5.\arcmin23\pm0.03$ based on surface-brightness profiles. 

The AGB / RGB ratio declines with radius to $\sim1\times R_e$, indicating an intermediate-age gradient in the inner regions. Similar to NGC~185, the ratio of AGB to RGB stars flattens and remains relatively constant across the remainder of the disk covered in the DUSTiNGS data. The constant ratio at larger radii is in agreement with the previously reported extended intermediate-age population that is well-mixed with an old stellar population based on comparing carbon and RGB stars \citep{Battinelli2004c}. From Figure~\ref{fig:NGC147}, our multi-epoch observations are within $2 \times R_e$. Thus, we are not able to detect variable AGB stars in the outer extremity of the galaxy, even if a smaller effective radius of $5.\arcmin5$ is adopted. 

\begin{figure}
\includegraphics[width=\columnwidth]{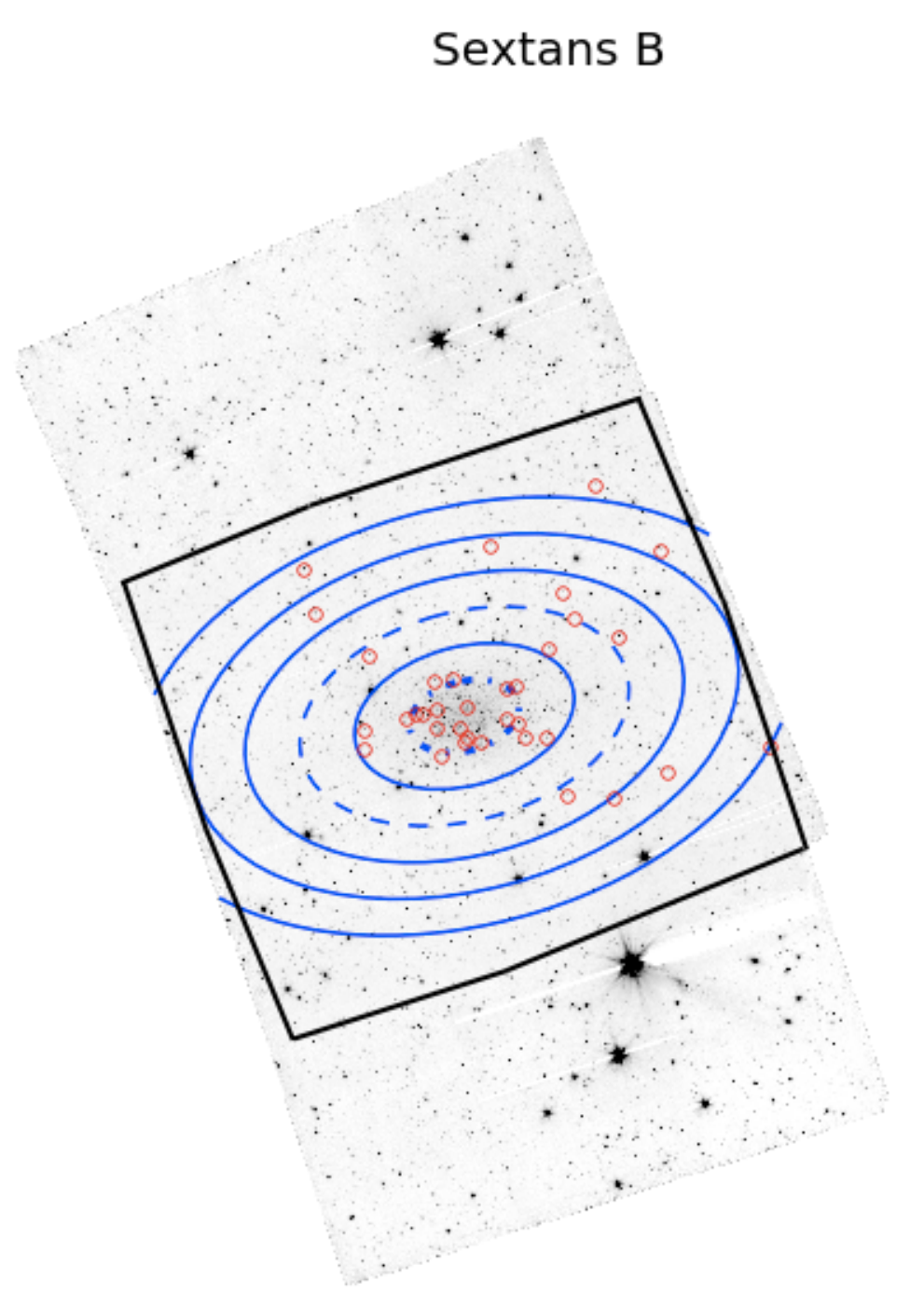}
\caption{3.6 $\micron$ map of Sextans~B with overlapping footprint of 2 epochs in 3.6 and 4.5 $\micron$ outlined, $1-6 \times R_e$ ellipses marked, and the locations of variable AGB stars. All colors and symbols are the same as those used in Figure~\ref{fig:IC1613}. The variable AGB stars extend to large radii, primarily along the minor axis and in the North and West directions.}
\label{fig:SextansB}
\end{figure}

\begin{figure}
\includegraphics[width=\columnwidth]{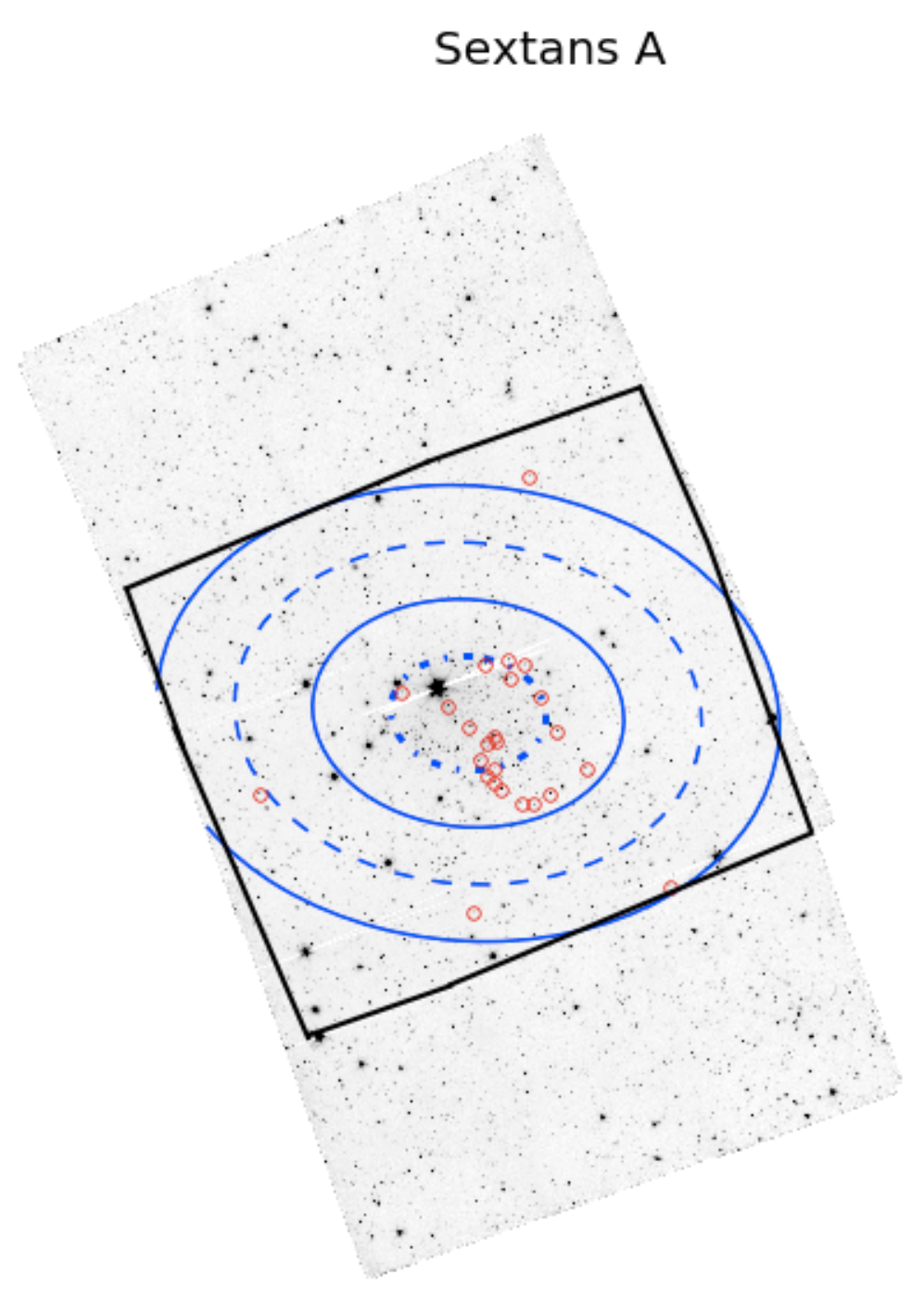}
\caption{3.6 $\micron$ map of Sextans~A with overlapping footprint of 2 epochs in 3.6 and 4.5 $\micron$ outlined, $1-4 \times R_e$ ellipses marked, and the locations of variable AGB stars. All colors and symbols are the same as those used in Figure~\ref{fig:IC1613}. Interestingly, the variable AGB stars within $\sim2\times R_e$ form a pattern similar to a spiral arm in the disk. }
\label{fig:SextansA}
\end{figure}

\subsection{Sextans~B}
Sex~B is a gas-rich dI that lies in the dwarf galaxy association $14+12$ \citep{Tully2006} which includes Sex~A, NGC~3109, Antlia dwarf, and Leo~P \citep{McQuinn2013}. Sex~B lies $\sim250$ kpc from its nearest neighbor Sex~A, and $\sim700$ kpc from the most massive dwarf in the association, NGC~3109. Simulations taking into account the velocities and possible orbits of Local Group galaxies suggest that Sex~B may have passed through the virial volume of the Milky Way at some time in the past \citep{Teyssier2012}. The distribution of the AGB and RGB stars closely trace each other, with a constant AGB / RGB ratio across the disk. From Figure~\ref{fig:SextansB}, the variable AGB stars are quite extended with stars detected in the outer extremities at $\sim6\times R_e$, primarily in the North and West directions. Many of these variable AGB stars at larger radii are located along the minor axis, consistent with a puffy, extended stellar distribution. 

\subsection{Sextans~A}
Similar to Sex~B, Sex~A is gas-rich dI found in the dwarf galaxy association $14+12$. The RGB surface-density profile shows a fairly flat profile at the innermost radii. The radial profile of the AGB stars declines until it reaches a floor at radii $\sim4\arcmin$ ($\sim2\times R_e$), which continues at a relatively constant surface density to $\sim9\arcmin$ ($\sim4\times R_e$). \citet{Bellazzini2014} noted a similarly extended, low-density and asymmetrical envelope across similar radii, which they suggest is a stellar tidal tail resulting from a previous interaction within the $14+12$ association and/or possibly with the Milky Way. Interestingly, \citet{Skillman1988} discovered a disturbed \HI\ velocity field which may be indicative of a recent interaction. While the stellar and gas tidal features are still of uncertain origin, the same simulations described above for Sex~B suggest Sex~A has had a previous passage through the virial volume of the Milky Way \citep{Teyssier2012}.  

Our measurements of the cumulative distributions show the AGB stars extend farther than the RGB stars with a longer scale length, and an AGB / RGB ratio that increases near $3\times R_e$ (see Figure~\ref{fig:ratios}). Because of the higher surface density of AGB stars, we are able to trace this extended feature past $3\times R_e$. Variable AGB stars are detected within the stellar disk of Sex~A, peaking outside the central annuli. From Figure~\ref{fig:SextansA}, we also detect variable AGB stars in the extended stellar population out to $\sim4\times R_e$ in a region coincident with the asymmetric feature. The larger spatial extent of the AGB stars compared with that of the older RGB stars and the distribution of variable AGB stars suggest that the extended stellar feature may have formed the stars in situ or have a separate origin from the RGB stars in the disk of Sex~A, providing supporting evidence of a tidal interaction in the last few Gyr. 

\subsection{WLM}
\begin{figure}
\includegraphics[width=\columnwidth]{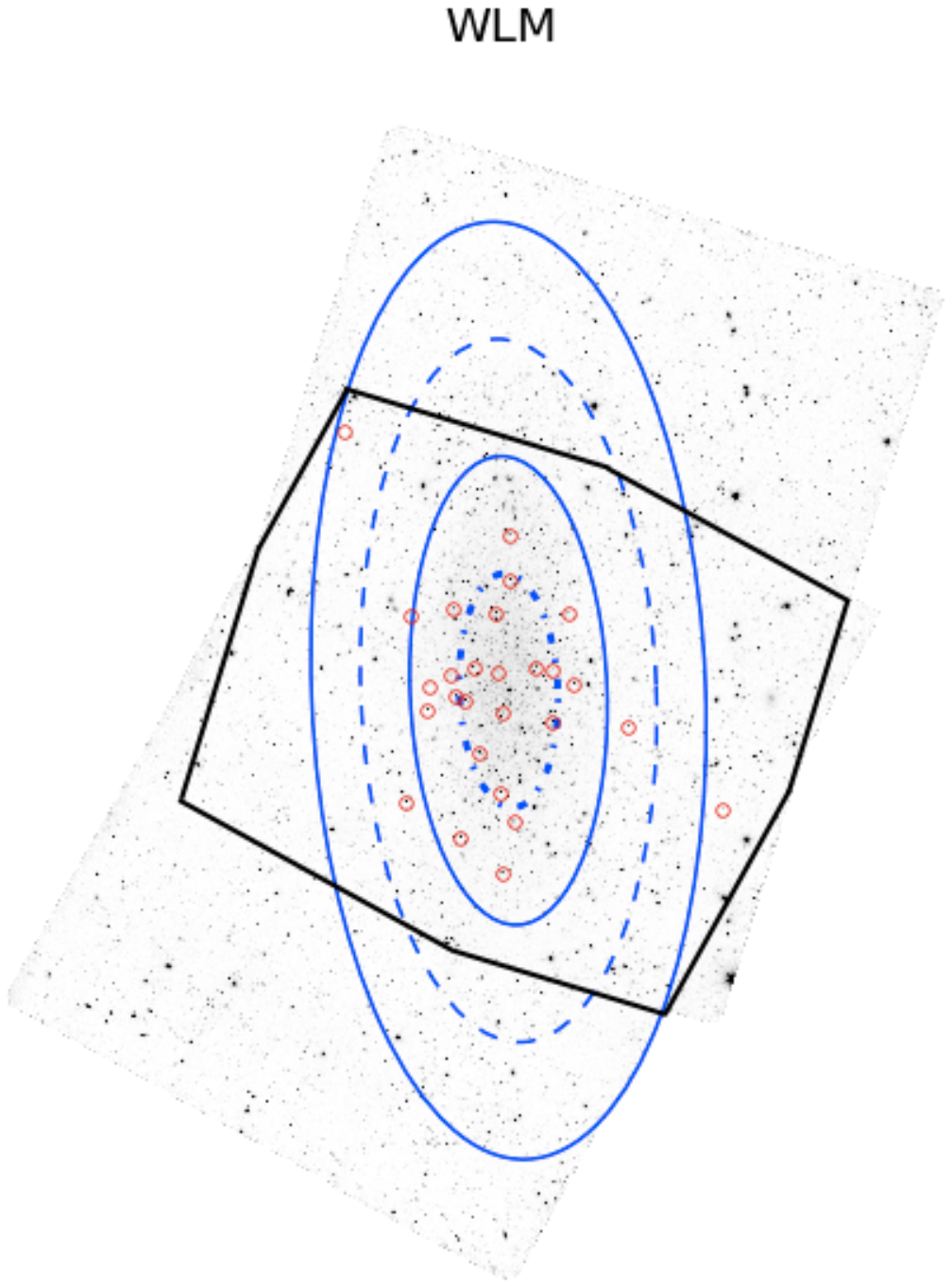}
\caption{3.6 $\micron$ map of WLM with overlapping footprint of 2 epochs in 3.6 and 4.5 $\micron$ outlined, $1-4 \times R_e$ ellipses marked, and the locations of variable AGB stars. All colors and symbols are the same as those used in Figure~\ref{fig:IC1613}.}
\label{fig:WLM}
\end{figure}

WLM is a gas-rich dwarf in the Local Group with a similar absolute magnitude as Sex~A and Sex~B, and a comparable number of detected variable AGB stars. The structure of the galaxy has been described as having a boxy inner component surrounded by an elongated halo, possibly tracing a thick disk seen nearly edge on \citep{Beccari2014}. A previous study of carbon stars in WLM found that these intermediate-age stars were well mixed with RGB stars within the stellar disk \citep{Battinelli2004a}. From Figure~\ref{fig:ratios}, we similarly find the ratio of AGB / RGB stars is fairly constant across the stellar disk. The ratio of variable AGB to RGB stars is also relatively constant, with more variation likely due to the small number of variable AGB stars identified. As shown in Figure~\ref{fig:WLM}, variable AGB stars are detected in the outer extremity of WLM, including 2 at $\sim4\times R_{e}$, corresponding to an angular distance of $\sim14\arcmin$. 

\begin{figure}
\includegraphics[width=\columnwidth]{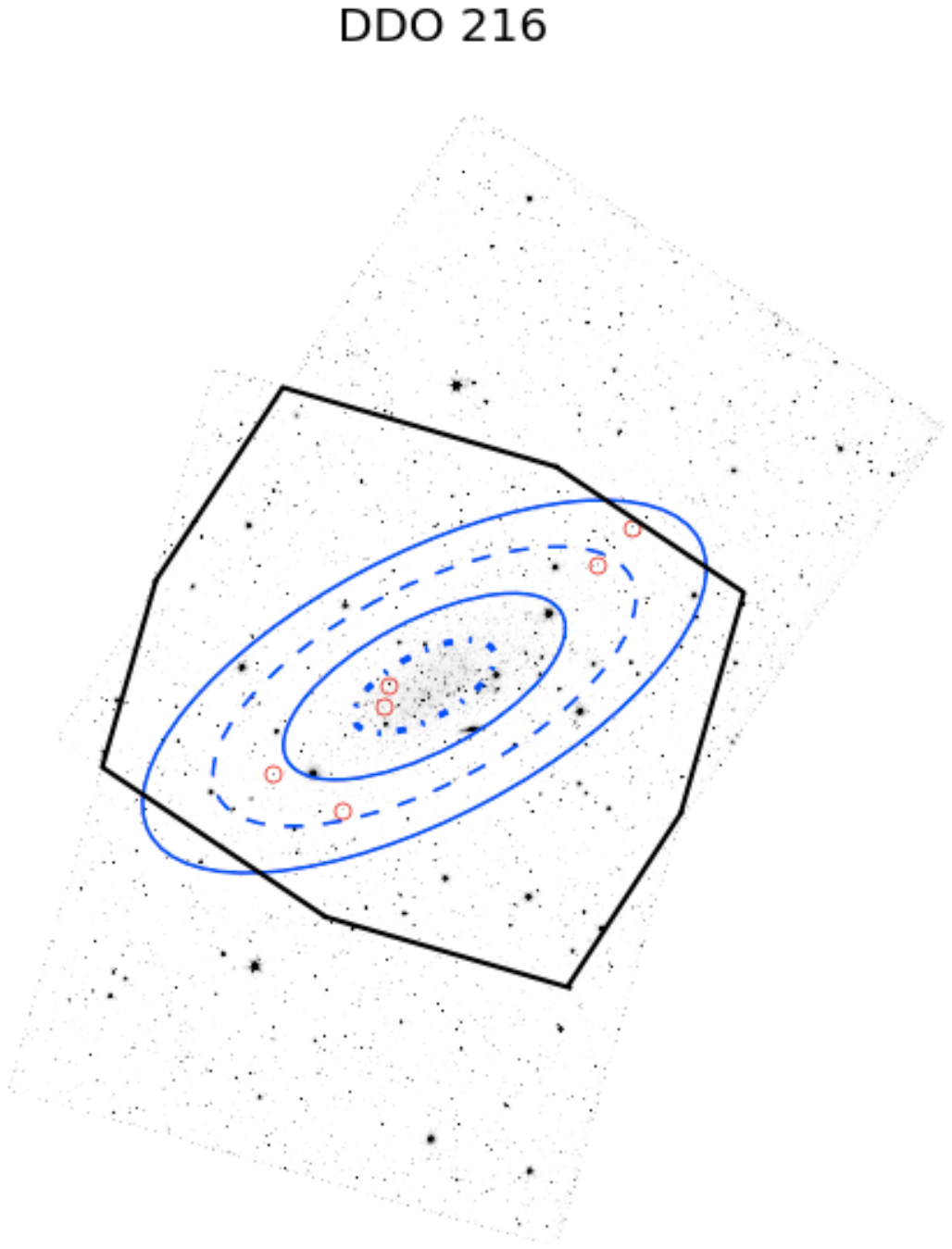}
\caption{3.6 $\micron$ map of DDO~216 with overlapping footprint of 2 epochs in 3.6 and 4.5 $\micron$ outlined, $1-4 \times R_e$ ellipses marked, and the locations of variable AGB stars. All colors and symbols are the same as those used in Figure~\ref{fig:IC1613}.}
\label{fig:DDO216}
\end{figure}

\subsection{DDO~216}
DDO~216 is a metal-poor, low-luminosity dI in the Local Group \citep{Skillman1997}. The RGB radial profile is measured to $\sim6$ \arcmin, or $\sim$1.6 kpc at the adopted distance of DDO~216 listed in Table~\ref{tab:trgb}. This is consistent with previous measurements of the main stellar disk of the galaxy \citep{deVaucouleurs1991}. A faint, extended stellar population has also been identified which reaches much farther to $\sim14$\arcmin, or $\sim4$ kpc, and consists of stars with estimated ages $>5$ Gyr \citep{Kniazev2009}. The profile is flattened in the inner $\sim2\arcmin$. The galaxy has a slightly higher ratio of AGB to RGB stars in the central region, with a constant ratio of AGB to RGB stars across the disk of the galaxy suggesting that the intermediate and older age stars are well mixed. The small number of identified variable AGB stars is insufficient to create a statistical profile. As seen in Figure~\ref{fig:DDO216}, variable AGB star candidates are found at different radii, including one x-AGB star candidate at $\sim3.5\times R_e$. This agrees with a previous study of the carbon star population in DDO~216 which found carbon stars located in the outer regions of the galaxy \citep{Battinelli2000}.

\subsection{Sag DIG}
Sag DIG is a metal-poor, low-luminosity dI in the Local Group. The \HI\ forms a complete ring with star formation occurring in a small number of gas clumps distributed asymmetrically around the ring \citep{Young1997}. \HI\ velocity maps show disturbed gas kinematics \citep{Hunter2012}. The radial extent of the RGB profiles reaches close to 6$\arcmin$ ($4\times R_e$), in agreement with deep optical imaging that revealed low surface-brightness structure ($\mu_V \sim  30.0$ mag arcsec$^{-2}$) out to $\sim5\arcmin$ \citep{Beccari2014} and recent results of RGB stars out to a similar radius \citep{Higgs2016}. 

\begin{figure}
\begin{center}
\includegraphics[width=0.7\columnwidth]{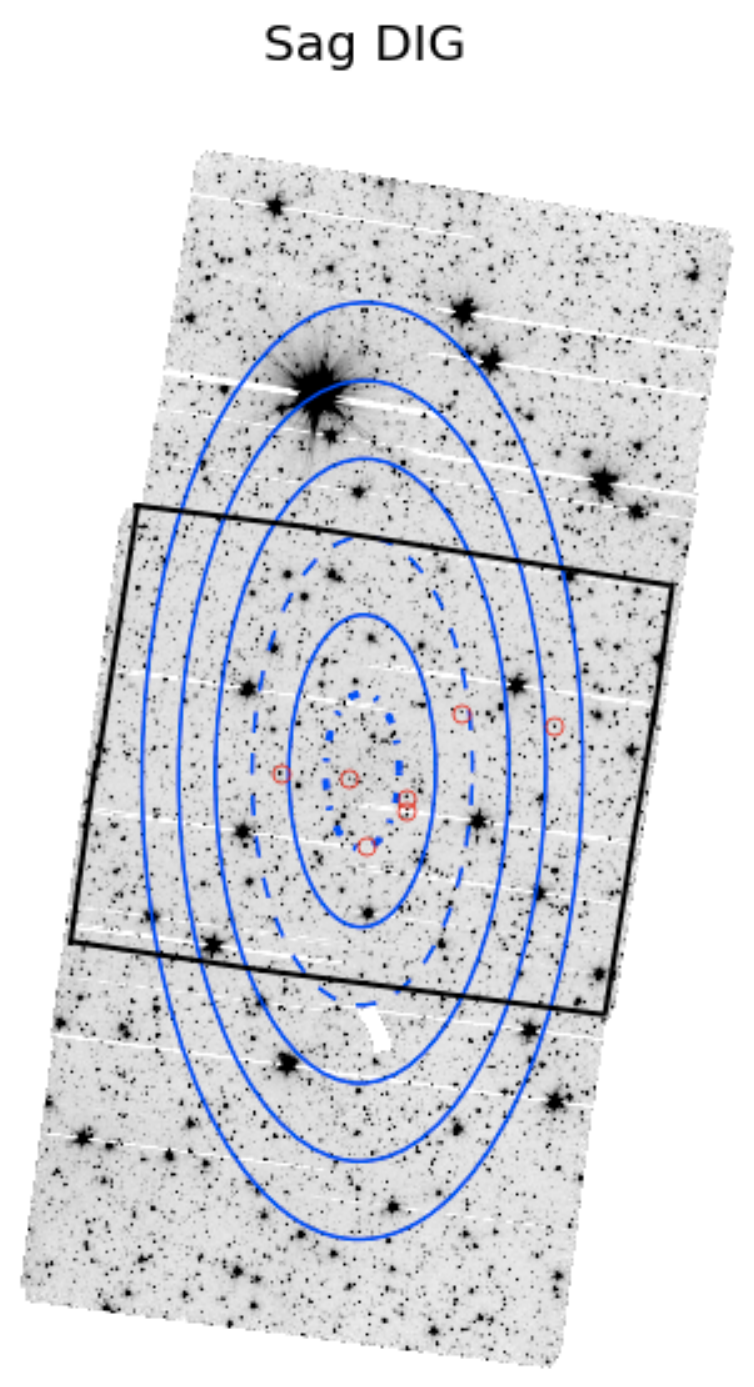}
\end{center}
\caption{3.6 $\micron$ map of Sag~DIG with overlapping footprint of 2 epochs in 3.6 and 4.5 $\micron$ outlined, $1-6 \times R_e$ ellipses marked, and the locations of variable AGB stars. All colors and symbols are the same as those used in Figure~\ref{fig:IC1613}.}
\label{fig:SagDIG}
\end{figure}

Breaks in both the AGB and the RGB surface densities at $\ltsimeq3\arcmin$ are detected. A similar change in surface density of the RGB stars was reported by \citet{Higgs2016} and attributed to marking the transition from stellar disk to stellar halo. However, we do not find a similar feature at the boundary between the stellar disks and outer extremities in the other galaxies in our sample. Moreover, the approximate extent of the \HI\ ring \citep{Young1997} corresponds to the break in the stellar surface-density profiles. The dip in stellar surface-density at $\ltsimeq3\arcmin$ coincides with a region of lower \HI\ column densities just outside the clumps in the \HI\ ring. The combined observations of disturbed gas kinematics, gas morphology, and break in the stellar surface-density profiles suggest that Sag~DIG has experienced an interaction despite its current relative isolation in the Local Group. As seen in Figure~\ref{fig:SagDIG}, Sag~DIG has 7 variable AGB star candidates detected in the DUSTiNGS data, one of which is detected in the outer extremity of the galaxy based on our $R_e$.

\section{An Inter-sample Comparison}\label{sec:comparison}
We have presented the radial profiles of intermediate-age and old stars, traced by AGB and RGB stars respectively, across the disks of 9 dwarf galaxies in and around the Local Group. We have also presented the distribution of variable AGB stars that can be detected at radii outside the stellar disks of the galaxies. The profiles suggest evidence of intermediate-age radial gradients in the sense that the intermediate-age stars are more centrally concentrated than the old stars in three of the nine systems: IC~10, NGC~185, and NGC~147. 

\begin{figure}
\includegraphics[width=0.49\textwidth]{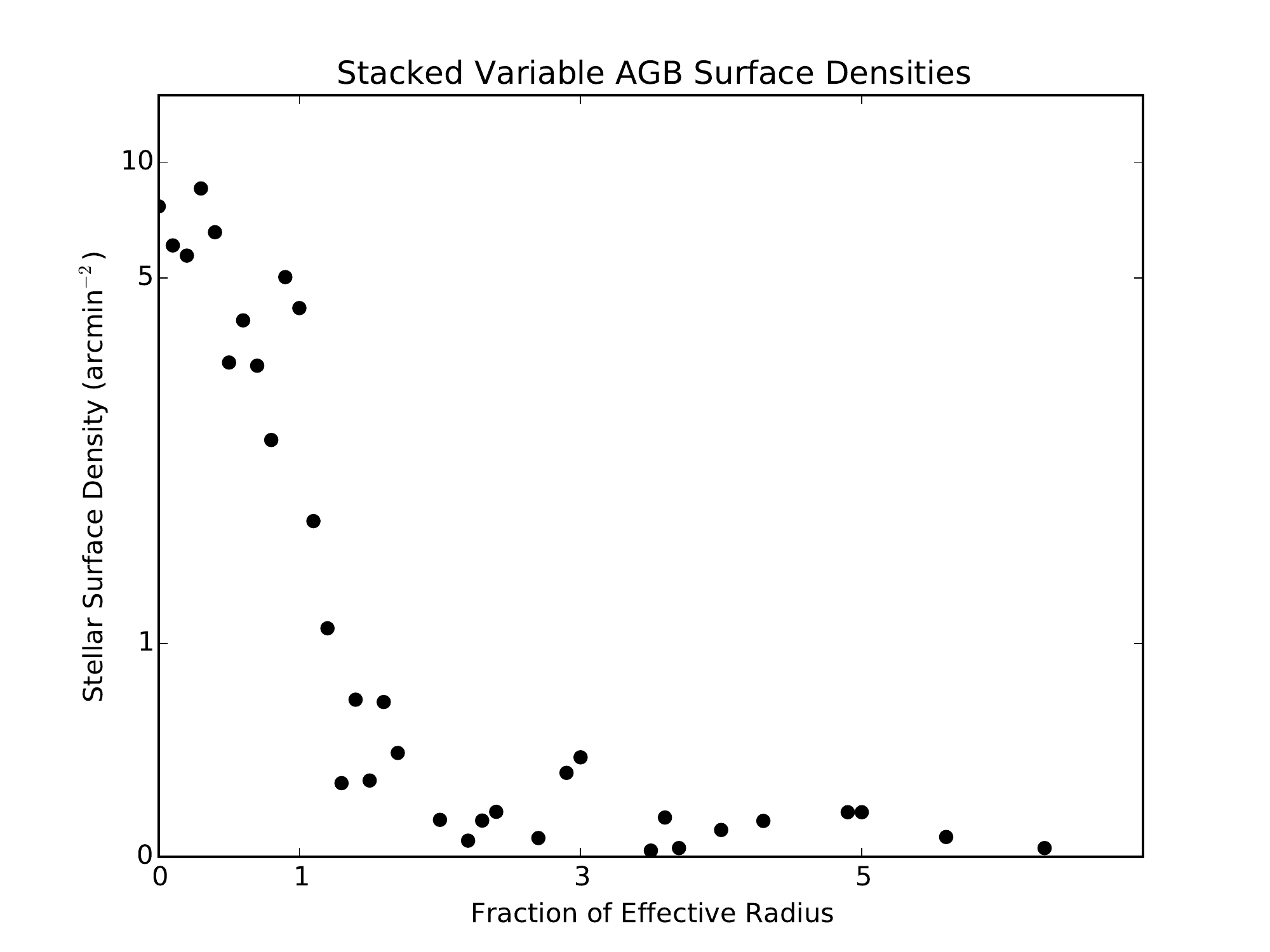}
\caption{The log surface density of variable AGB stars as a function of effective radius calculated by ÒstackingÓ the radial variable AGB counts from all nine galaxies in our sample. Despite the limited areal coverage for some galaxies, there is a clear detection of variable AGB stars in the extremities of the galaxies out to $6 \times R_e$.}
\label{fig:agb_stacked}
\end{figure}

From Figure~\ref{fig:ratios}, IC~10 shows a clear, positive age radial gradient with the TP-AGB stars more centrally concentrated than the old stars. Two galaxies, NGC~185 and NGC~147, show a higher ratio of AGB to RGB stars in the inner annuli, with a declining ratio that flattens at larger radii. The innermost radial bins affected by higher levels of stellar crowding (and higher, unquantified uncertainties) have already been omitted in Figure~\ref{fig:ratios} for these three systems. Thus, the variation in the AGB to RGB ratios at small radii is not in the region most affected by incompleteness.

The rest of the sample show fairly constant AGB to RGB ratios across their disks, with somewhat higher ratios in the inner measurement for WLM, DDO~216, and Sag~DIG. Most galaxies exhibit some variability in their ratios across the disks that is likely due to local asymmetries in the stellar distributions and variable star-formation histories. The generally constant AGB to RGB ratios are consistent with the similar scale lengths of the two populations calculated from the best-fitting S\'ersic profiles and the cumulative plots from Figures~\ref{fig:sersic_agb}$-$\ref{fig:cumulative}. 

Variable AGB stars are detected in the outer extremities as far as $6\times R_e$ in the galaxies with extended, multi-epoch areal coverage. The exceptions are IC~10, NGC~185, and NGC~147; these galaxies also have larger angular sizes, and thus, the DUSTiNGS areal coverage does not extend past $\sim3\times R_e$. Figure~\ref{fig:agb_stacked} shows the log of the surface density of all variable AGB stars from the sample based on stacking the variable AGB counts as a function of effective radius. The majority of stars are contained within $3 \times R_e$, with a tail of the distribution out to $6 \times R_e$. 
While there is some variability in the distribution due to binning the small number of stars, the radial gradient in the stellar surface-density indicates that most of the variable AGB star candidates are indeed galaxy members, rather than more uniformly distributed background or foreground objects. Our findings agree with results from the carbon-star survey \citep[e.g.,][]{Albert2000} which identified carbon stars photometrically at large radii in a number of the same galaxies in our sample, as noted in the discussion on individual galaxies in Section~\ref{sec:galaxies}. 

\begin{figure}
\includegraphics[width=0.49\textwidth]{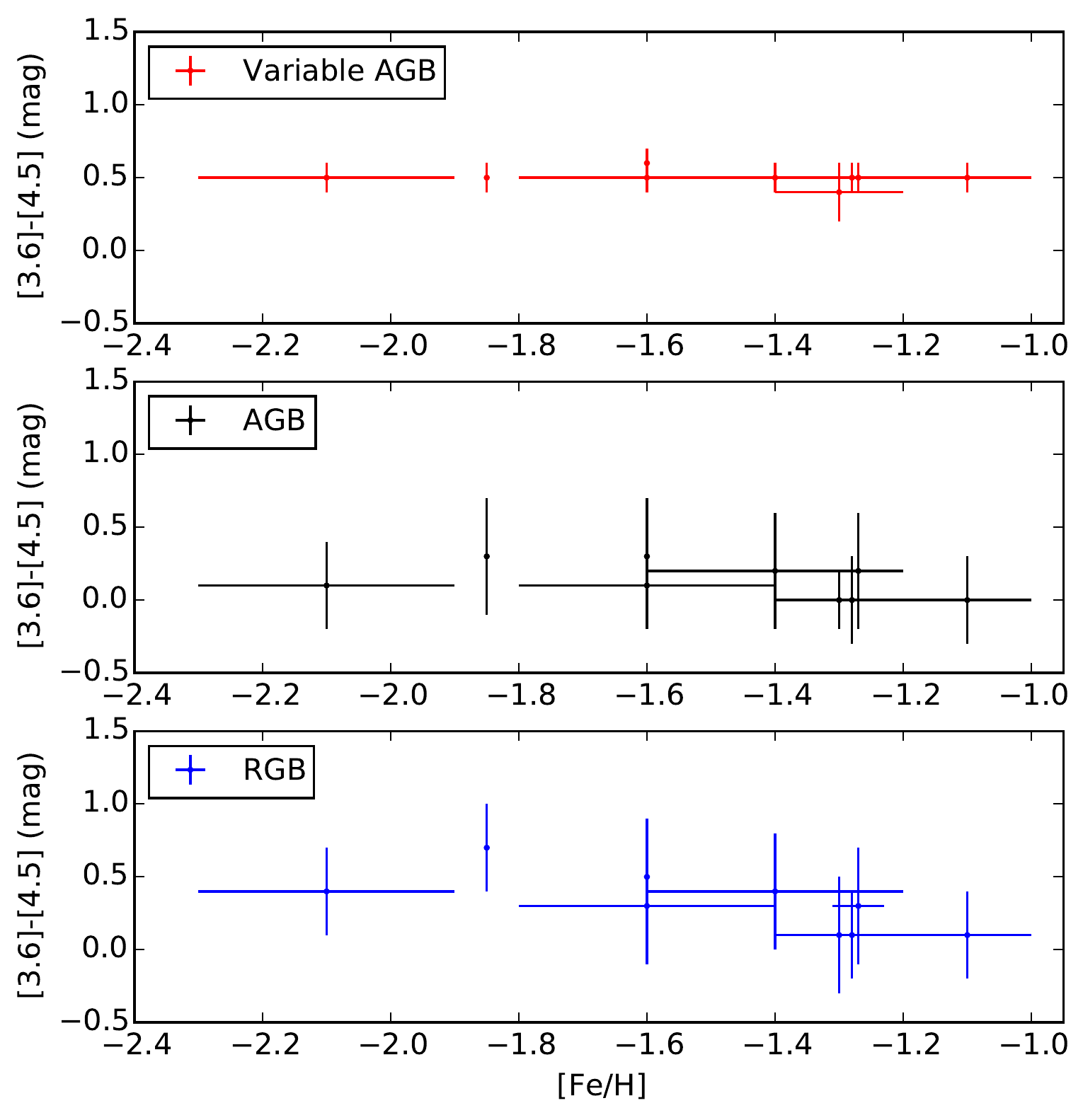}
\caption{The average [3.6]$-$[4.5] color and standard deviation of the variable AGB, AGB, and RGB populations as a function of galaxy metallicity. The variable AGB stars are redder due to their dust production. Neither the variable AGB nor the AGB colors show a correlation with metallicity; the RGB colors tend to be somewhat bluer at higher metallicity. This may be an effect of the higher luminosity and higher degree of crowding in these galaxies which reduces the impact of red background contamination.}
\label{fig:agb_colors}
\end{figure}

We examined the [3.6] $-$ [4.5] colors of the stars as a function of galactocentric radius, and we found no significant change in color across the disks. Figure~\ref{fig:agb_colors} shows the average [3.6] $-$ [4.5] color and standard deviation of the variable AGB, AGB, and RGB stars for each galaxy. The colors for the variable AGB stars are the average over two epochs and have redder colors than the AGB and RGB stars (as expected from their dust production). The AGB populations are slightly bluer than the RGB populations. This is likely due to CN and C$_2$  absorption in the 4.5 $\micron$ band present in the carbon stars \citep{Boyer2011}, as well as the contribution of background galaxies to the RGB colors. We find no correlation between the average AGB color and the gas-phase oxygen abundance or the stellar metallicity in the sample, based on measurements reported in \citet{McConnachie2012}. The more metal-rich galaxies have a slighter bluer average color than the metal-poor galaxies in our sample. This is likely the combined effect of the higher luminosity of these galaxies and higher degree of crowding that reduces the impact of red background contamination.

\section{AGB stars as Probes for Galaxy Formation and Evolution Scenarios}\label{sec:scenarios}
Galaxy formation and evolution scenarios predict star formation intrinsically creates radial age gradients in dwarf galaxies at both young and intermediate epochs, with an increasing number of stars forming at inner radii due to the availability of denser gas. However, gradients can be diluted or erased over time via both internal evolution and external events. Here, we consider various scenarios that can impact the stellar age gradients from the literature and compare them with the stellar distributions in our sample.

External events, such as recent mergers, accretion, or environmental interactions, can erase radial age gradients and mix the distribution of stellar populations \citep[e.g.,][]{Benitez-Llambay2015, Lokas2012}. Given the low predicted frequency of 20\% for dwarf-dwarf major mergers since z$=$1 within a Local Group environment \citep{Deason2014}, it is unlikely that merger events are responsible for the extended distribution of intermediate-age stars found in six of the nine galaxies in our sample. Yet, three of these galaxies (Sex~A, NGC~147, and Sag~DIG) do show signs of a gravitational interaction. Interestingly, Sex~A and Sag~DIG have a uniquely extended AGB population reaching {\em farther} than the older RGB population, showing an enhancement of intermediate-age stars at greater radii is possible. While stellar tidal features have been predicted to persist for only $\ltsimeq$1 Gyr \citep{Conselice2009}, there is still uncertainty about the longevity of such features, particularly for low density structures. Therefore, if the extended AGB populations in Sex~A, NGC~147, and Sag~DIG have a tidal origin, it is unclear whether their current location is due to tidal mixing or interaction-driven in-situ formation. Finally, the recent star-formation activity in IC~10 may be the result of a collision with an \HI\ cloud \citep{Huchtmeier1979, Sanna2010}. In this case, the more concentrated AGB stars in this system are likely a result of interaction-driven in-situ star formation. 

A number of internal processes can impact radial stellar distributions and drive radial migration through both direct and indirect means. Stars may be directly ejected to larger radii and/or into the outer extremity after forming in supernova-driven shocks. Stars may migrate as an indirect result of gas motions driven by stellar feedback. Fluctuations in a galaxy's potential due to the outward flow of gas driven by supernovae and radiation pressure can subsequently add energy to stellar orbits, changing the kinematics of the stars and the distribution of dark matter. Finally, galactic dynamics in the form of star-star or star-cloud encounters can disperse stellar clusters and relax stellar associations. Observationally, radial migration has been measured to be of lesser importance in dwarfs on timescales of $\sim100-300$ Myr \citep{Gieles2008, Bastian2009, Bastian2011}. This constrains significant changes in stellar kinematics and large-scale radial migrations to timescales longer than a few 100 Myr. This is not surprising as young stars ($\ltsimeq$ few 100 Myr) are often centrally concentrated in star-forming dwarfs \citep[e.g.,][]{McQuinn2012a}. 

Recent simulations have studied the relative impact a few of these internal processes have on the age distributions of stars in low-mass galaxies, making specific predictions to which we can compare our results. The models from \citet{El-Badry2016} suggest that star formation is always centrally concentrated in dwarfs, particularly at early times, but stars are pushed outward by strong fluctuations in the global potential driven by gas motions. In this scenario, old stars are influenced over longer timescales and therefore move the greatest distances into the outer disks and extremities. Intermediate-age stars are predicted to be less extended in the simulation, as stars migrate $\ltsimeq 2 \times R_e$ over timescales of 8 Gyr. In contrast to this prediction, we find a roughly constant distribution of AGB stars relative to the RGB population out to $3 \times R_e$, with variable AGB star candidates in their extremities in six galaxies. As mentioned, the distributions in two of these galaxies may have, at least in part, a tidal origin. For the remaining four systems, the radial extents of the intermediate-age stars suggest that either there are additional factors to consider, or the timescales for migration are shorter than predicted.

Other simulations by \citet{Stinson2009} show that population gradients in dwarf galaxies can be explained by a combination of in-situ star formation in a contracting gaseous disk and stellar migration, again without invoking external merger/interaction events to mix the stellar populations. Cumulatively, stars formed and ejected via supernova shocks can create a spherically distributed halo. Stars relocated by the sloshing of a changing potential due to random bulk gas motions contribute to a puffy, thick disk of intermediate-age and old stars. The similar scale lengths of the intemediate-age and old stars found in our sample and the presence of variable AGB star candidates at large radii along both the major and minor axes support this description of radial migration. 

Fluctuations in the potential have also been discussed as a possible explanation for the flatter, cored central density profiles identified in many dwarf galaxies \citep[e.g.,][]{Navarro1996, Mashchenko2008, Pontzen2012}. In high-resolution, cosmological hydrodynamical simulations, \citet{Onorbe2015} show that stellar feedback is effective in creating cored mass-density profiles observable at $z$$\sim$0 in dwarfs more massive than $M_* \gtsimeq10^6$ \msun\ if they have undergone star formation at late times. If so, stellar feedback responsible for the flatter inner density profiles of dwarf galaxies could also explain the presence of intermediate-age stars in the outer reaches of dwarf galaxies. In this case, one would expect the systems with flatter inner profiles to also have a more extended AGB population, which could be tested with mass models of the dwarf galaxies in our sample. 

\section{Conclusion}\label{sec:conclusion}
The DUSTiNGS survey cataloged the evolved stellar populations in 50 Local Group dwarf galaxies with {\it Spitzer}. As part of the DUSTiNGS survey, we have presented the radial distribution of the intermediate-age TP-AGB populations and the older RGB populations within the disks of nine dwarf galaxies, and the distribution of intermediate-age variable AGBs stars within the disks and outer extremities of the systems. Here, we summarize our findings.

\begin{enumerate}
\item We uniquely identify the TRGB at 3.6 $\micron$ using photometry matched to the F814W photometry from HST optical imaging. Within our sample, the absolute magnitude of the TRGB at 3.6 $\micron$ varies from $-5.94^{+0.25}_{0.17}$ to $-6.65^{+0.22}_{0.14}$ mag. The $\sim0.7$ mag variation in TRGB magnitudes shows a loose correlation with metallicity but no clear trend with the age of the population. While both metallicity and age are predicted to impact the TRGB luminosity at 3.6 $\micron$ (and add to the scatter), these factors seem unlikely to account for the full difference we find. Photometric crowding, incompleteness also likely contribute to the offset and scatter in the measurement. Without a clear understanding of the dominant driver for the variability of the TRGB luminosity at 3.6 $\micron$ (and a well-calibrated correction), the $\sim0.7$ mag variability makes the TRGB magnitude an unreliable distance indicator at this wavelength.

\item We measure the ellipticities and position angles of the galaxies by fitting surface-brightness isophotes to the 3.6 $\micron$ images. Generally, the measurements in the infrared agree with previous values measured at optical wavelengths. 

\item We measure the effective radii from S\'ersic fits to the radial profiles of AGB and RGB stars and from the cumulative stellar distributions. Our values from the two approaches agree within the uncertainties and also agree with previously measured scales, except for two galaxies (IC~10, NGC~147) where we find larger values from the S\'ersic fits. In both cases, the inner regions have significant crowding which may be impacting the S\'ersic profiles.

\item The intermediate-age and old stellar populations are well-mixed in six of the nine dwarf galaxies studied. Variable AGB stars are identified in the outer extremities of these dwarfs galaxies out to distances of nearly $6 \times R_e$. These AGB stars have dust-driven winds that are chemically enriched with dredged-up, newly-synthesized material. The presence of these stars in the outer regions of the galaxies indicates that chemical enrichment may occur in the outer regions of dwarf galaxies with frequency. Previous tidal interactions may have played a role in the extended spatial distributions of the AGB and variable AGB stars in two of these galaxies, whereas internal evolution is likely responsible for the mixing of the intermediate-age and old populations in the other four. Simulations show a variety of physical internal drivers are possible, as discussed above, including star formation occurring in supernovae-driven gas, fluctuations in a galaxy potential from stellar feedback driving gas motions that subsequently affect stellar orbits, and the heating of stellar orbits from ongoing star formation. 

\item We find evidence of a positive radial intermediate-age gradient in one galaxy (IC~10), which is likely biased by a population of younger AGB stars formed in the on-going starburst event in this system. We also find evidence of an intermediate-age radial gradient at small radii in the two dSphs of the sample (NGC~185 and NGC~147), with a relatively constant AGB / RGB ratio out to larger radii. 

\item Variable AGB star candidates are identified in an extended stellar feature in Sex~A that has been previously identified as a stellar tidal tail \citep{Bellazzini2014} and a similarly extended stellar feature outside of the stellar disk in Sag~DIG. Both galaxies show a {\it longer} scale length of TP-AGB stars compared to RGB stars; the distribution of the extended AGB stars may have a tidal origin.

\end{enumerate}

\acknowledgements
This work is supported by {\it Spitzer} via grant GO80063 and by the NASA Astrophysics Data Analysis Program grant number N3$-$ADAP13$-$0058. RDG was supported, in part, by the United States Air Force. AZ and IM acknowledge support from the UK Science and Technology Facility Council under grant ST/L000768/1. PAW thanks the South African National Research Foundation for a research grant. Many thanks to Andy Dolphin for assistance with HST/WFPC2 photometry. This research made use of NASA's Astrophysical Data System, the NASA/IPAC Extragalactic Database which is operated by the Jet Propulsion Laboratory, California Institute of Technology, under contract with the National Aeronautics and Space Administration. We acknowledge the use of the HyperLeda database (http://leda.univ-lyon1.fr).

{\it Facilities:} \facility{{\it Spitzer Space Telescope}, \facility{Hubble Space Telescope}}

\appendix \label{sec:appendix}
\section{Measuring the TRGB luminosity at 3.6$\micron$}\label{sec:append_trgb}
We used four strategies to measure the TRGB luminosity at 3.6$\micron$ from the HST$+${\it Spitzer} photometry, each with their own strengths and weaknesses. All four estimates should agree for galaxies with large stellar populations unaffected by photometric incompleteness in the {\it Spitzer} data:

\begin{enumerate}
  \setlength\itemsep{0.1em}
\item Select stars from the HST data within $\pm$0.1~mag of the F814W TRGB, and compute the mean of their  3.6~\micron\ magnitudes.
\item Measure the response of an edge-detection filter on a 3.6~\micron\ luminosity function constructed only from those stars with a match in HST.
\item Measure the response of an edge-detection filter on a 3.6~\micron\ luminosity function constructed only from RGB stars identified in the HST data.
\item Measure the response of an edge-detection filter on a 3.6~\micron\ luminosity function constructed from all {\it Spitzer} sources, but  restricted to an area that minimizes crowding and background galaxy contamination. 
\end{enumerate}

\input{tab6}

\begin{figure}
\begin{center}
\includegraphics[width=0.5\columnwidth]{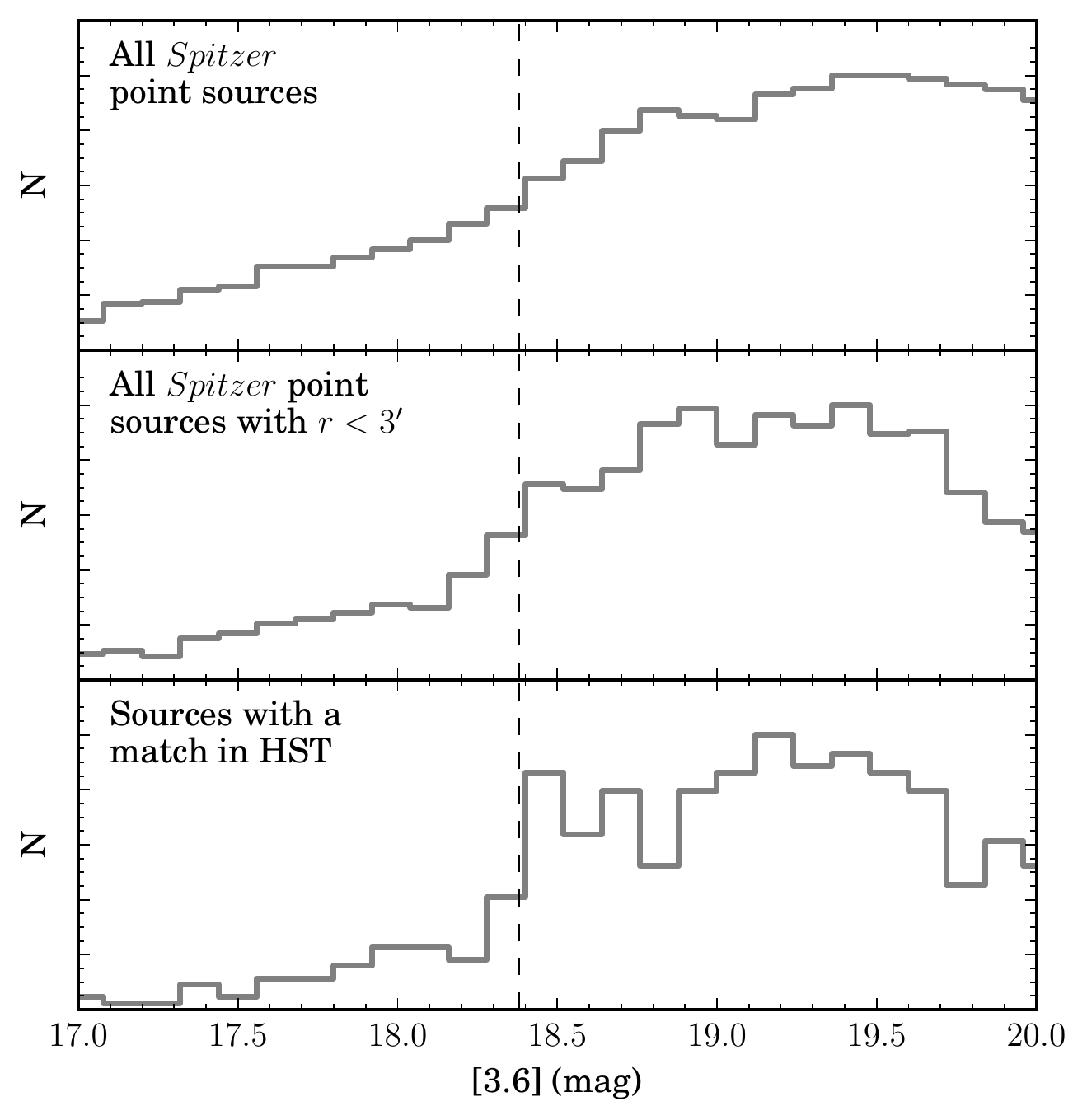}
\caption{The 3.6~\micron\ luminosity function of IC~1613. The TRGB is not evident in the full {\it Spitzer} catalog (upper panel), but becomes more prominent when including only the central region ($r<3$\arcmin) and when including sources only if they have a match in HST. The dashed line marks the TRGB found by averaging the four techniques listed in the text. \label{fig:lfunc}}
\end{center}
\end{figure}

The results of the four strategies are listed in Table~\ref{tab:trgb_spitzer} for each field studied, including the F814W TRGB measurements for comparison. For nearby galaxies with large stellar populations, minimal crowding, and sufficient photometric depth, all four estimates agree within the uncertainties as illustrated for IC~1613 in Figure~\ref{fig:lfunc}. The first strategy is less reliable in galaxies with small stellar populations, resulting in few stars around the optical TRGB from which to estimate the 3.6~\micron\ mean magnitude. This method will also bias TRGB estimates to slightly fainter magnitudes, since it will always include more stars at $+0.1$~mag than at $-0.1$~mag. To obtain the most robust estimate using this strategy, we use 1000 Monte Carlo Bootstrap trials, each time adding Gaussian random errors (over the range of $-4$ to $+4\,\sigma$) to both the HST and {\it Spitzer} photometry, based on the measured photometric uncertainties of individual point sources. The resulting TRGB estimates are fit with a Gaussian, adopting the mean and width of the distribution as the TRGB and its uncertainty, combined in quadrature with 0.1 mag to account for the magnitude range in selected stars. 

For strategies 2$-$4, we employ the same procedure used to find the optical TRGB, described above. The difference lies in the initial input luminosity function of the {\it Spitzer} data. Strategies 2 and 3 use sources matched with the HST data, reducing contamination from background galaxies that are unresolved in {\it Spitzer}, and thereby significantly improving the faint end of the luminosity function. Strategy 4 uses only the {\it Spitzer} data, but is restricted spatially to avoid crowding and background contamination. This strategy has the advantage of including more stars by not restricting itself to the small HST field. It is also a better global estimate of the TRGB, being less affected by regional variations that might be reflected in the small HST field.

Strategies 2 and 4 are unreliable in galaxies where the infrared TRGB is near the photometric sensitivity limit (Sextans\,A, Sextans\,B, and Sag\,DIG). This is because stars near the TRGB in these galaxies have large photometric uncertainties and because photometric incompleteness severely affects the luminosity function below the TRGB. For such cases limited by photometric depth, strategy 3 has the advantage. By selecting only RGB stars matched from the HST data, the luminosity function drops to near zero at the TRGB (within photometric uncertainties), allowing us to identify the TRGB luminosity at 3.6 $\micron$ even when it lies fainter than where the luminosity function turns over due to incompleteness. For these more difficult fields, we take the weighted mean of only strategies 1 and 3 for our final TRGB 3.6$\micron$ measurement. Note that, even for the more difficult fields, the TRGB luminosities at 3.6$\micron$ identified using these strategies are sufficient for our purposes of separating the more luminous TP-AGB stars from the bulk of the RGB population. The final TRGB luminosities at 3.6 $\micron$ and their absolute magnitudes are listed in Table~\ref{tab:trgb}. 

\section{Calculating Areas of Elliptical Sectors for Partial Radial Annuli}\label{sec:append_sectors}
The area of an elliptical sector for an axis-aligned ellipse centered at the origin in polar coordinates and delimited by angles $\theta_0$ and $\theta_1$ is described by:
\begin{align}
A_s(\theta_0, \theta_1)=\int_{\theta_0}^{\theta_1}\tfrac{1}{2}r^2d\theta=\int_{\theta_0}^{\theta_1} \frac{(a^2b^2/2)}{b^2\cos^2\theta+a^2\sin^2\theta}d\theta
\end{align}

\noindent where $r$ is the radius of the ellipse, and $a$ and $b$ are the semi-major and semi-minor axes. A solution of this integral is the function:
\begin{align}
F(\theta)\Big|_{\theta_0}^{\theta_1}=\tfrac{ab}{2}\Big[\theta-\tan^{-1}\Big(\frac{(b-a)\sin{2\theta}}{(b+a)+(b-a)\cos{2\theta}}\Big)\Big]\Big|_{\theta_0}^{\theta_1}
\end{align}

\noindent where $\tan^{-1}(\alpha)$ is the principle branch of the inverse tangent function, which has a range of $(-\tfrac{\pi}{2}, \tfrac{\pi}{2})$. To obtain the area of the elliptical segment delimited by two angles, $\theta_0$ and $\theta_1$, we subtract the area of the triangle created by the origin and the arc endpoints, which are $(x_0, y_0)=(r_0\cos\theta_0, r_0\sin\theta_0)$ and $(x_1, y_1)=(r_1\cos\theta_1, r_1\sin\theta_1)$. This area is given by:

\begin{align}
A_{\vartriangle}=\frac{r_0r_1}{2}sin(\theta_1-\theta_0)
\end{align}

The final area of an elliptical segment is then:

\begin{align}
A=A_s-A_{\vartriangle}=F(\theta_1)-F(\theta_0)-\tfrac{r_0r_1}{2}sin(\theta_1-\theta_0)
\end{align}

\renewcommand\bibname{{References}}
\bibliography{../../bibliography.bib}

\end{document}

%% file: tab1.tex
\begin{table*}
\begin{center}
\caption{Basic Properties and Observational Summary of Targets}
\label{tab:properties}
\end{center}
\begin{center}
\vspace{-15pt}
\begin{tabular}{lcccccccc}
\hline 
\hline 
Galaxy  &	RA, Dec  		& $M_V$	& M$_*$	& AOR Key 	& AOR Key     	& AOR Key 	& Coverage	& Refs. \\
            &  (J2000)   		& (mag) & ($\times10^6$ \msun)  & Epoch 0	&  Epoch 1	&  Epoch 2   	& (arcmin$^2$)&                    \\
\hline
IC 1613	& 01 04 47.8 $+02$ 07 04	&  $-15.2 \pm 0.2$  	& 100& 15892736 & 42321664   &  42321920   	&  356.3  		&1,2,6  \\
IC 10	& 00 20 17.3 $+59$ 18 14	&  $-15.0 \pm 0.2$  	& 86	& \nodata	& 42321152   &  42321408 	&  195.9  		&1,4,5  \\
NGC 185 	& 00 38 58.0 $+48$ 20 15	&  $-14.8 \pm 0.1$	& 68	& \nodata	& 42324736   &  42324992  	&  192.5   		& 1,4  \\
NGC 147	& 00 33 12.1 $+48$ 30 32	&  $-14.6 \pm 0.1$  	& 62	& \nodata	& 42324224   &  42324480  	&  201.3   		&1  \\
Sextans B	& 10 00 00.1 $+05$ 19 56	&  $-14.5 \pm 0.2$	& 52 	& \nodata	& 42327552   &  42327296	&  195.7 		&1,2,3,4  \\
Sextans A	& 10 11 00.8 $-04$ 41 34 	&  $-14.3 \pm 0.1$	& 44	&  15892224 & 42327040   &  42326784 	&  196.8  		&1,2,3  \\
WLM		& 00 01 58.2 $-15$ 27 39	&  $-14.2 \pm 0.1$	& 43 	&  15892480 &  42328832  &  42329088	&  185.5 		&1,2,3  \\
DDO 216	(Pegasus) & 23 28 58.0 $+14$ 44 35	&  $-12.2 \pm 0.2$ & 6.6	 & 23043840 & 42320128   &  42320384 	&  179.8   		& 1,2,3,7  \\
Sag DIG  	& 19 29 59.0 $-17$ 40 41	&  $-11.5 \pm 0.3$	& 3.5	& \nodata	& 42326016   &  42325760  	&  85.8     		&1,4  \\ 
\hline
\\
\end{tabular}
\end{center}
\vspace{-10pt}
\tablecomments{Summary of the fundamental properties and observational data of the nine DUSTiNGS galaxies in this study, ordered by absolute visual magnitude. Values for coordinates and magnitudes are adopted from (1) \citet{McConnachie2012}; original references include (2) \cite{Lee2006}, (3) \citet{Tammann2011},  (4) \citet{Mateo1998}, (5) \cite{Kim2009}, (6) \citet{Bernard2010}, and (7) \citet{McConnachie2005}. All observations that are part of the DUSTiNGS program had a combined exposure time from epochs 1 and 2 of 1080 sec with a 5$\sigma$ sensitivity limit of 1.6 $\mu$Jy. Epoch 0 observations consist of archival data for three of our selected sample which were processed in a uniform manner as Epochs 1 and 2. The total areal coverage is that which includes all epochs and all wavelengths. This is the total coverage within which we can identify variable star candidates, which is smaller than the total map size for each wavelength at a given epoch. See \citet{Boyer2015a} for details.}

\end{table*}

%% file: tab2.tex
\begin{table*}
\begin{center}
\caption{HST and {\it Spitzer} TRGB Magnitudes}
\label{tab:trgb}
\vspace{-10pt}
\end{center}
\begin{center}
\begin{tabular}{lccccccc}
\hline 
\hline 
Galaxy	  	& HST  		    	& $A_V$ & [Fe/H] & $(m-M)_0$ &(TRGB$_{\rm F814W})_0$	& TRGB$_{\rm 3.6 \micron}$ & M$_{3.6}$ TRGB \\
	        & Field(s)			& (mag) &        & (mag)     &(mag) 	    	 	& (mag)			           & (mag)  \\
\hline
IC~1613     & u40401/u56750 		& 0.07 & $-1.60 \pm 0.20$ & 24.39$\pm0.03$ & 20.342$\pm$0.013 & 18.38$^{+0.17}_{-0.12}$& $-6.01^{+0.17}_{-0.12}$ \\
IC~10 		& 10242			& 2.41 & $-1.28$          & 24.43$\pm0.03$ & 20.447$\pm$0.016 & 17.87$^{+0.19}_{-0.15}$	& $-6.56^{+0.19}_{-0.15}$\\
NGC~185 	& u3kl01/u3kl04 	& 0.41 & $-1.30 \pm 0.10$ & 24.17$\pm0.03$ & 20.208$\pm$0.015 & 17.73$^{+0.15}_{-0.10}$ & $-6.44^{+0.15}_{-0.10}$ \\
NGC~147 	& u2ob01 			& 0.37 & $-1.10 \pm 0.10$ & 24.39$\pm0.06$ & 20.429$\pm$0.056 & 18.04$^{+0.21}_{-0.16}$ & $-6.35^{+0.21}_{-0.17}$\\
Sextans~B 	& u67204 			& 0.09 & $-1.60 \pm 0.10$ & 25.77$\pm0.03$ & 21.735$\pm$0.021 & 19.12$^{+0.22}_{-0.14}$ & $-6.65^{+0.22}_{-0.14}$ \\
Sextans~A	& u2x502/u56710 	& 0.12 & $-1.85$          & 25.82$\pm0.03$ & 21.760$\pm$0.023 & 19.54$^{+0.21}_{-0.12}$ & $-6.28^{+0.21}_{-0.12}$\\
WLM 		& u37h01/u48i01 	& 0.10 & $-1.27 \pm 0.04$ & 24.94$\pm0.03$ & 20.909$\pm$0.024 & 18.78$^{+0.17}_{-0.11}$ & $-6.16^{+0.17}_{-0.11}$\\
DDO~216		& u2x504	 		& 0.19 & $-1.40 \pm 0.20$ & 24.96$\pm0.04$ & 20.960$\pm$0.038 & 18.63$^{+0.22}_{-0.16}$ & $-6.33^{+0.22}_{-0.16}$\\
Sag~DIG     & u67202 			& 0.34 & $-2.10 \pm 0.20$ & 25.18$\pm0.04$ & 21.130$\pm$0.037 & 19.24$^{+0.25}_{-0.17}$ & $-5.94^{+0.25}_{-0.17}$\\
\hline 
\end{tabular}
\end{center}
\tablecomments{The TRGBs are the mean values for all HST fields. The distance moduli are the values we derive from the HST F814W data corrected for extinction; uncertainties include our measured uncertainties and the uncertainties in the zero-point and metallicity correction from the calibration of \citet{Rizzi2007a}. With the exception of IC\,10, photometry is from the \citet{Holtzman2006} archive. For IC\,10, we perform photometry on data from HST GO program 10242 from the HST archive. The $A_V$ values were taken from \citet{Weisz2014a}. [Fe/H] measurements were taken from \citet[][and references therein]{McConnachie2012} with the exception of Sextans~B, which was reported in \citet{Bellazzini2014}. The higher upper uncertainties on the TRGB 3.6~$\micron$ magnitudes take into account the $\sim0.05-0.10$ mag bias towards brighter magnitudes estimated by artificial star tests. See text for details.}
\end{table*}

%% file: tab3.tex
\begin{table*}
\begin{center}
\caption{Star Counts and Contamination Estimates}
\label{tab:photometry}
\end{center}
\begin{center}
\vspace{-10pt}
\begin{tabular}{lccccccccc}
\hline 
\hline 
Galaxy     	& $N_{RGB}$ & $N_{AGB}$ &  $N_{AGB}$& $N_{x-AGB}$	& \multicolumn{3}{c}{Foreground Density}	& \multicolumn{2}{c}{Background Density}	\\
	        	&                     &                     &      (var)     	&     (var)                 	& AGB			& RGB			& Radius 	& AGB 			& RGB \\
                	&                     &                     &                             	&              	& (arcmin$^{-2}$)	& (arcmin$^{-2}$)     	& (arcmin)	& (arcmin$^{-2}$)     & (arcmin$^{-2}$) \\
\hline
IC 1613	&   16544  	&  4880    	   &     38       	&   30            	    	&   ...                      	&  ...    	             	& ...		& 11.3                    	& 47.6    \\
IC 10	&    31536      	&   16316     &   257       	&  235       	       	&  13.2                 	&  1.2      	  		& 3		&  25.0                    	& 51.8 \\ 
NGC 185	&   25996       	&   4783       &   69          	&  58              	     	&  11.2                 	&  3.1       		   	& 2		&  8.9                    	& 83.5          \\ 
NGC 147	&    25428      	&   6191   	   &    89          	&   76            	     	&  13.4                	&  0.7	      		& 2		&  9.0                     	& 38.0       \\
Sextans B	&    3870  	    	&   3782   	   &    35         	&  32           	   	&   ...                     	&     ...   	           	& ...		& 15.2                  	& 22.6	        \\
Sextans A &    2294  	    	&   5678   	   &    25        	&   24	      	    	&   ...                     	&     ...	               	& ...		& 14.5	                & 7.2	       \\
WLM       	&  8799    	    	&   3566   	   &    27          	&   22          	    	&   ...                     	&       ...	                	& ...		& 14.6 	                & 39.9	       \\    
DDO 216	&   8101   	    	&   2195   	   &    6           	&    6                     	&   ...                     	&       ...                   	& ...		&  8.8 	                & 36.0	       \\
Sag DIG  	&    2535  	    	&    7233  	   &   7            	&    6               	    	&   ...                     	&        ...                  	& ...		& 42.0 	                & 14.6	      \\
\hline
\\
\end{tabular}
\end{center}
\vspace{-15pt}
\tablecomments{Star counts for each galaxy and estimated contamination. The numbers of RGB stars (Col. 2) and AGB stars (Col. 3) are based on the numbers of stars below and above the TRGB luminosity at 3.6$\micron$ for all sources in the full DUSTiNGS maps, regardless of epoch. The total number of variable AGB (Col 4.), and the number  of variable x-AGB stars (Col. 5) are from the DUSTiNGS catalogs (see Paper~II). The foreground density of Galactic stars was estimated for galaxies with significant crowding in their inner regions using TRILEGAL simulations \citep{Girardi2012}. The background density of contamination was estimated using the DUSTiNGS data set. See text for details.}

\end{table*}

%% file: tab4.tex
\begin{table*}
\begin{center}
\caption{Structural Parameters}
\label{tab:structure_params}
\end{center}
\begin{center}
\vspace{-10pt}
\begin{tabular}{lccccl}
\hline 
\hline 
Galaxy             & Ellipticity from		    & Ellipticity from 	& PA (\degree) from 	 	& PA (\degree)	from	&	Refs. \\
                        & 3.6$\micron$ images  &  literature		& 3.6$\micron$ images   	& literature		&		 \\
\hline
IC~1613		&    \nodata               	&  0.11 $\pm$ 0.03   	&    \nodata      			& 50$\pm2$		& M12 \\
                         &                              	&  0.19 $\pm$ 0.02   	&            				&    87		        	& BDE07 \\
                         &                              	&  0.15                  	&         		 		&    80		      	& B$+$07 \\
                         &                              	&  0.18              	&             				&    81		     	& A72 \\
\\
IC~10		& 0.27 $\pm$  0.06  	&  0.19 $\pm$ 0.02 	& 121$\pm$4    		&   \nodata             	& M12 \\
\\
NGC~185		& 0.31 $\pm$  0.02   &  0.15 $\pm$ 0.01 	& 57 $\pm$1			& 35$\pm$3		& M12 \\
                         &                             	&  0.22 $\pm$ 0.01   	&                   			& 45.9$\pm1.2$	& C$+$14 \\
                         &                              	&  0.23 		        	&                     			& 42.9  		      	& G$+$10 \\
\\
NGC~147		& 0.47 $\pm$  0.03 	&  0.41 $\pm$ 0.02 	& 30 $\pm$1			& 25$\pm3$	   	& M12 \\
                         &                              	&  0.46 $\pm$ 0.02   	&                     			& 34.2$\pm3.6$	& C$+$14 \\
                         &                             	&  0.44    		        	&                     			&    28.4  		  	& G$+$10 \\
\\
Sex~B		& 0.38 $\pm$ 0.04     &  0.31 $\pm$ 0.03	& 105 $\pm$2			& 110$\pm2$	 	& M12 \\
                         &                                &  0.41    			&                  			& 95 $\pm15$		& H$+$06, B$+$14 \\
                         &                                &           			&                 			& 88$\pm15$		& H$+$06  \\
\\
Sex~A	        & 0.27 $\pm$ 0.03    	&  0.17 $\pm$ 0.02	& 85 $\pm$ 3			& 0 $\pm$		 	& M12 \\
                         &                              	&  0.15                  	&                  			& 41.8  		   	& H$+$06 \\
\\
WLM		        & 0.58 $\pm$  0.03  	&   0.65 $\pm$ 0.01	& 2.2 $\pm$2.9			& 4 $\pm2$	   	& M12 \\
                         &                             	&   0.55            		&                   			& 179 $\pm2$		& L$+$12 \\
\\
DDO~216		& 0.58 $\pm$  0.04 	&  0.46 $\pm$ 0.02	& 118 $\pm$ 2			& 120$\pm2$		& M12, deV$+$91\\
                         &                        	& $0.6\pm0.1$		&                    			& $-55.7\pm0.1$	& K$+$09 \\
\\
Sag~DIG 		& \nodata        		&  0.53  	           	&    \nodata     			& $-0.5$ 			& H$+$16 \\ 
		  	& 			      	&  0.5  	           	& 		   			& 90 			    	& L$+$00 \\
\hline
\\
\end{tabular}
\end{center}
\vspace{-10pt}
\tablecomments{The ellipticity and position angle (PA) of each galaxy measured by fitting isophotes to the 3.6$\micron$ images after significant foreground stars were removed. Ellipticity is defined as $1 - b/a$ where $b$ is the semiminor axis and $a$ is the semimajor axis. IC~1613 and Sag~DIG were not well fit by the IRAF task {\sc ellipse} in the images. For these galaxies, we adopt the listed elliptical parameters from \citet{Bernard2007} and \citet{Higgs2016} respectively.}
\vspace{5pt}

{\bf References:} M12 $-$ \citet[][and references therein]{McConnachie2012}, BDE07 $-$ \citet{Battinelli2007}, B$+$07 $-$ \citet{Bernard2007}, A72 $-$ \citet{Ables1972}, C$+$14 $-$ \citet{Crnojevic2014}, G$+$10 $-$ \citet{Geha2010}, H$+$06 $-$ \citet{Hunter2006}, B$+$14 $-$ \citet{Beccari2014}, L$+$12 $-$ \citet{Leaman2012}, deV+91 $-$ \citet{deVaucouleurs1991}, K$+$09 $-$ \citet{Kniazev2009}, H$+$16 $-$ \citet{Higgs2016}, L$+$00 $-$ \citet{Lee2000}, . 

\end{table*}

%% file: tab5.tex
\begin{table*}
\begin{center}
\caption{Effective Radii}
\label{tab:radial_extent}
\end{center}
\begin{center}
\begin{tabular}{l | ccccc | ccccc | c}
\hline 
\hline 
		& \multicolumn{5}{c}{AGB}																& \multicolumn{5}{c}{RGB}																& 	\\
		& \multicolumn{4}{c}{S\'ersic}												& Cumulative		& \multicolumn{4}{c}{S\'ersic}												& Cumulative		& \\
Galaxy	& Index		& $R_e$		& $3\times~R_e$ 		& $3 \times~R_e$ 	& $R_e$ 			& Index	& $R_e$ 			& $3\times~R_e$ 			& $3 \times~R_e$ 		& $R_e$ 			& $r_{25}$  \\
		& 			& (\arcmin)	& (\arcmin)			& (kpc)			& (\arcmin)		& 		& (\arcmin)		&  (\arcmin)				& (kpc)				& (\arcmin)		& (\arcmin) \\
\hline
IC~1613 	& $0.35\pm0.07$	& $3.0\pm0.2$		& 9.0 			& 2.0					& $3.5\pm0.5$		& 0.96$\pm0.07$	& $3.1\pm0.2$			& 9.3				& 2.0					& $2.5\pm0.5$		& 9.1		\\
IC~10	& $0.45\pm0.05$	& $2.4\pm0.1$		& 7.2	 			& 1.6					& $3.5\pm0.5$		& 1.00			& $7.0\pm0.5$			& 21.0 			& 4.7					& $5.5\pm0.5$		& 3.4		\\
NGC~185 & 1.00			& $1.7\pm0.1$		& 5.1				& 0.9					& $2.5\pm0.5$		& 1.00			& $3.7\pm	0.1$			& 11.1			& 2.2					& $3.5\pm0.5$		& 6.4		\\
NGC~147 & $1.01\pm0.07$	& $2.7\pm0.2$		& 8.1				& 1.8					& $2.5\pm0.5$		& 1.10$\pm0.12$	& $7.5\pm0.8$			& 22.5			& 4.9					& $5.5\pm0.5$		& 4.7		\\
Sex~B	& $1.05\pm0.07$	& $1.5\pm0.1$		& 4.5				& 1.9					& $1.5\pm0.5$		& 0.29$\pm0.03$	& $1.5\pm	0.05$		& 4.5				& 1.9					& $1.5\pm0.5$		& 2.4		\\
Sex~A	& $0.38\pm0.05$	& $1.6\pm0.1$		& 4.8				& 2.0					& $2.5\pm0.5$		& 0.71$\pm0.18$	& $2.1\pm0.4$			& 6.3				& 2.7					& $2.5\pm0.5$		& 2.7		\\
WLM		& $0.45\pm0.08$	& $2.2\pm0.2$		& 6.6				& 1.9					& $2.5\pm0.5$		& 0.61$\pm0.03$	& $3.1\pm0.1$			& 9.3				& 2.6					& $2.5\pm0.5$		& 5.2 	\\
DDO~216 & $0.49\pm0.08$	& $1.7\pm0.2$		& 5.1		 		& 1.5					& $1.5\pm0.5$		& 1.05$\pm0.07$	& $2.1\pm0.03$		& 6.3				& 1.8					& $2.5\pm0.5$		& 2.3 	\\
Sag~DIG 	& 1.00			& $1.3\pm0.2$		& 3.9				& 1.3					& $4.5\pm0.5$		& 0.21			& $1.5 $				& 4.5 			& 1.4					& $4.5\pm0.5$		& 1.4		\\
\hline
\end{tabular}
\end{center}
\tablecomments{In Cols. $2-6$, we list the indices and effective radii from the best-fitting S\'ersic profiles to the AGB surface densities and the radial bin that encompasses 50\% of the AGB stars for each galaxy. The S\'ersic profile values for Sag~DIG are uncertain due to the small number of points used in the fit. Uncertainties on the cumulative distributions are based on the 1$\arcmin$ width of the annuli used to bin the stars. Cols. $7-11$ list the same measured from the RGB stars. We adopt the RGB $3\times R_e$ as the transition between the disks and the outer extremities of the galaxies. The effective radii of the RGB stars measured by the cumulative distributions are consistent with those measured by the best-fitting S\'ersic profiles with the exception of IC~10 and NGC~147 (see Section~\ref{sec:galaxies} for discussion). Col. 12 lists a separate measurement of the radial extent of the galaxies, based on the distance along the major axis of the 25 mag arcsec$^{-2}$ B-band isophote taken from the HyperLeda database \citep{Makarov2014}.}

\end{table*}

%% file: tab6.tex
\begin{table*}
\begin{center}
\caption{{\it Spitzer} TRGBs}
\label{tab:trgb_spitzer}
\end{center}
\begin{center}
\begin{tabular}{llcccccc}
\hline
\hline
		& 		& F814W	& \multicolumn{5}{l}{3.6~$\mu$m\dotfill}\\
Galaxy  	& HST  	&TRGB  	& TRGB 1 & N$^{a}$&TRGB 2 & TRGB 3 & TRGB 4 \\
		& Field  	&(mag)	& (mag) 	& 		& (mag) 	& (mag) 	& (mag)	\\
\hline
    IC\,1613 & u40401 & 20.397$\pm$0.024 & 18.34$\pm$0.11 & 30 & 18.40$\pm$0.22 & 18.37$\pm$0.21 & 18.38$\pm$0.17 \\
    IC\,1613 & u56750 & 20.321$\pm$0.015 & 18.40$\pm$0.29 & 3  & 18.45$\pm$0.27 & 18.47$\pm$0.22 & 18.37$\pm$0.17 \\
      IC\,10 & 10242$^{b}$ & 20.447$\pm$0.016 & 17.86$\pm$0.11 & 38 & 17.85$\pm$0.23& 17.90$\pm$0.21& 17.90$\pm$0.14 \\
    NGC\,185 & u3kl01 & 20.174$\pm$0.022 & 17.76$\pm$0.11 & 45 & 17.74$\pm$0.29 & 17.95$\pm$0.27 & 17.76$\pm$0.15 \\
    NGC\,185 & u3kl04 & 20.238$\pm$0.021 & 17.51$\pm$0.11 & 24 & 17.98$\pm$0.26 & 18.02$\pm$0.23 & 17.75$\pm$0.15 \\
    NGC\,147 & u2ob01 & 20.429$\pm$0.056 & 18.04$\pm$0.11 & 29 & 18.03$\pm$0.26 & 18.11$\pm$0.22 & 18.01$\pm$0.20 \\
 Sextans\,B & u67204 & 21.735$\pm$0.021 & 19.14$\pm$0.12 & 56 & 18.59$\pm$0.29 & 18.91$\pm$0.38 & 18.61$\pm$0.27 \\
  Sextans\,A & u2x502 & 21.794$\pm$0.055 & 19.66$\pm$0.13 & 25 & 18.56$\pm$0.24 & 19.31$\pm$0.40 & 18.60$\pm$0.26 \\
  Sextans\,A & u56710 & 21.753$\pm$0.025 & 19.46$\pm$0.14 & 18 & 18.68$\pm$0.30 & 19.33$\pm$0.41 & 18.59$\pm$0.25 \\
          WLM & u37h01 & 20.909$\pm$0.024 & 18.94$\pm$0.12 & 15 & 18.73$\pm$0.24 & 18.81$\pm$0.24 & 18.68$\pm$0.17 \\
         WLM & u48i01 & 20.909$\pm$0.023 & 18.79$\pm$0.11 & 39 & 18.68$\pm$0.20 & 18.71$\pm$0.20 & 18.68$\pm$0.17 \\
DDO~216 & u2x504 & 20.960$\pm$0.038 & 18.64$\pm$0.11 & 41 & 18.56$\pm$0.24 & 18.68$\pm$0.23 & 18.59$\pm$0.21 \\
    Sag\,DIG & u67202 & 21.130$\pm$0.037 & 19.28$\pm$0.14 & 11 & 18.51$\pm$0.42 & 19.11$\pm$0.30 & 18.53$\pm$0.40 \\
\hline
\end{tabular}
\end{center}
\tablecomments{Uncertainties are derived from the Monte Carlo trials, with 4\,$\sigma$ Gaussian random deviations in the photometric uncertainties. For this work, we adopt a 3.6$\micron$ TRGB that is a weighted mean of 2--4 of the TRGB methods, depending on the level of crowding and photometric depth.}
$^{a}${Number of stars within $\pm$0.1~mag of the F814W TRGB; their mean magnitude provides one estimate of 3.6$\micron$ TRGB.}\\
$^{b}${With the exception of IC\,10, photometry is from the \citet{Holtzman2006} archive. For IC\,10, we perform photometry on data from HST GO program 10242 from the HST archive.}
\end{table*}

%% file: ms_final.bbl
\begin{thebibliography}{}
\expandafter\ifx\csname natexlab\endcsname\relax\def\natexlab#1{#1}\fi

\bibitem[{{Ables}(1972)}]{Ables1972}
{Ables}, H.~D. 1972, Publications of the U.S.~Naval Observatory Second Series,
  20, 1

\bibitem[{{Albert} {et~al.}(2000){Albert}, {Demers}, \& {Kunkel}}]{Albert2000}
{Albert}, L., {Demers}, S., \& {Kunkel}, W.~E. 2000, \aj, 119, 2780

\bibitem[{{Aparicio} \& {Tikhonov}(2000)}]{Aparicio2000a}
{Aparicio}, A., \& {Tikhonov}, N. 2000, \aj, 119, 2183

\bibitem[{{Aparicio} {et~al.}(2000){Aparicio}, {Tikhonov}, \&
  {Karachentsev}}]{Aparicio2000b}
{Aparicio}, A., {Tikhonov}, N., \& {Karachentsev}, I. 2000, \aj, 119, 177

\bibitem[{{Arias} {et~al.}(2016){Arias}, {Guglielmo}, {Fernando}, {Lewis},
  {Bland-Hawthorn}, {Bate}, {Conn}, {Irwin}, {Ferguson}, {Ibata},
  {McConnachie}, \& {Martin}}]{Arias2016}
{Arias}, V., {Guglielmo}, M., {Fernando}, N., {et~al.} 2016, \mnras, 456, 1654

\bibitem[{{Bastian} {et~al.}(2009){Bastian}, {Gieles}, {Ercolano}, \&
  {Gutermuth}}]{Bastian2009}
{Bastian}, N., {Gieles}, M., {Ercolano}, B., \& {Gutermuth}, R. 2009, \mnras,
  392, 868

\bibitem[{{Bastian} {et~al.}(2011){Bastian}, {Weisz}, {Skillman}, {McQuinn},
  {Dolphin}, {Gutermuth}, {Cannon}, {Ercolano}, {Gieles}, {Kennicutt}, \&
  {Walter}}]{Bastian2011}
{Bastian}, N., {Weisz}, D.~R., {Skillman}, E.~D., {et~al.} 2011, \mnras, 412,
  1539

\bibitem[{{Battinelli} \& {Demers}(2000)}]{Battinelli2000}
{Battinelli}, P., \& {Demers}, S. 2000, \aj, 120, 1801

\bibitem[{{Battinelli} \& {Demers}(2004{\natexlab{a}})}]{Battinelli2004c}
---. 2004{\natexlab{a}}, \aap, 418, 33

\bibitem[{{Battinelli} \& {Demers}(2004{\natexlab{b}})}]{Battinelli2004b}
---. 2004{\natexlab{b}}, \aap, 417, 479

\bibitem[{{Battinelli} \& {Demers}(2004{\natexlab{c}})}]{Battinelli2004a}
---. 2004{\natexlab{c}}, \aap, 416, 111

\bibitem[{{Battinelli} {et~al.}(2007){Battinelli}, {Demers}, \&
  {Artigau}}]{Battinelli2007}
{Battinelli}, P., {Demers}, S., \& {Artigau}, {\'E}. 2007, \aap, 466, 875

\bibitem[{{Beccari} {et~al.}(2014){Beccari}, {Bellazzini}, {Fraternali},
  {Battaglia}, {Perina}, {Sollima}, {Oosterloo}, {Testa}, \&
  {Galleti}}]{Beccari2014}
{Beccari}, G., {Bellazzini}, M., {Fraternali}, F., {et~al.} 2014, \aap, 570,
  A78

\bibitem[{{Bellazzini} {et~al.}(2014){Bellazzini}, {Beccari}, {Fraternali},
  {Oosterloo}, {Sollima}, {Testa}, {Galleti}, {Perina}, {Faccini}, \&
  {Cusano}}]{Bellazzini2014}
{Bellazzini}, M., {Beccari}, G., {Fraternali}, F., {et~al.} 2014, \aap, 566,
  A44

\bibitem[{{Ben{\'{\i}}tez-Llambay} {et~al.}(2015){Ben{\'{\i}}tez-Llambay},
  {Navarro}, {Abadi}, {Gottloeber}, {Yepes}, {Hoffman}, \&
  {Steinmetz}}]{Benitez-Llambay2015}
{Ben{\'{\i}}tez-Llambay}, A., {Navarro}, J.~F., {Abadi}, M.~G., {et~al.} 2015,
  ArXiv e-prints, arXiv:1511.06188

\bibitem[{{Bernard} {et~al.}(2007){Bernard}, {Aparicio}, {Gallart},
  {Padilla-Torres}, \& {Panniello}}]{Bernard2007}
{Bernard}, E.~J., {Aparicio}, A., {Gallart}, C., {Padilla-Torres}, C.~P., \&
  {Panniello}, M. 2007, \aj, 134, 1124

\bibitem[{{Bernard} {et~al.}(2010){Bernard}, {Monelli}, {Gallart}, {Aparicio},
  {Cassisi}, {Drozdovsky}, {Hidalgo}, {Skillman}, \& {Stetson}}]{Bernard2010}
{Bernard}, E.~J., {Monelli}, M., {Gallart}, C., {et~al.} 2010, \apj, 712, 1259

\bibitem[{{Blum} {et~al.}(2006){Blum}, {Mould}, {Olsen}, {Frogel}, {Werner},
  {Meixner}, {Markwick-Kemper}, {Indebetouw}, {Whitney}, {Meade}, {Babler},
  {Churchwell}, {Gordon}, {Engelbracht}, {For}, {Misselt}, {Vijh}, {Leitherer},
  {Volk}, {Points}, {Reach}, {Hora}, {Bernard}, {Boulanger}, {Bracker},
  {Cohen}, {Fukui}, {Gallagher}, {Gorjian}, {Harris}, {Kelly}, {Kawamura},
  {Latter}, {Madden}, {Mizuno}, {Mizuno}, {Nota}, {Oey}, {Onishi}, {Paladini},
  {Panagia}, {Perez-Gonzalez}, {Shibai}, {Sato}, {Smith}, {Staveley-Smith},
  {Tielens}, {Ueta}, {Van Dyk}, \& {Zaritsky}}]{Blum2006}
{Blum}, R.~D., {Mould}, J.~R., {Olsen}, K.~A., {et~al.} 2006, \aj, 132, 2034

\bibitem[{{Boyer} {et~al.}(2015{\natexlab{a}}){Boyer}, {McDonald},
  {Srinivasan}, {Zijlstra}, {van Loon}, {Olsen}, \& {Sonneborn}}]{Boyer2015c}
{Boyer}, M.~L., {McDonald}, I., {Srinivasan}, S., {et~al.} 2015{\natexlab{a}},
  \apj, 810, 116

\bibitem[{{Boyer} {et~al.}(2009){Boyer}, {Skillman}, {van Loon}, {Gehrz}, \&
  {Woodward}}]{Boyer2009}
{Boyer}, M.~L., {Skillman}, E.~D., {van Loon}, J.~T., {Gehrz}, R.~D., \&
  {Woodward}, C.~E. 2009, \apj, 697, 1993

\bibitem[{{Boyer} {et~al.}(2011){Boyer}, {Srinivasan}, {van Loon}, {McDonald},
  {Meixner}, {Zaritsky}, {Gordon}, {Kemper}, {Babler}, {Block}, {Bracker},
  {Engelbracht}, {Hora}, {Indebetouw}, {Meade}, {Misselt}, {Robitaille},
  {Sewi{\l}o}, {Shiao}, \& {Whitney}}]{Boyer2011}
{Boyer}, M.~L., {Srinivasan}, S., {van Loon}, J.~T., {et~al.} 2011, \aj, 142,
  103

\bibitem[{{Boyer} {et~al.}(2015{\natexlab{b}}){Boyer}, {McQuinn}, {Barmby},
  {Bonanos}, {Gehrz}, {Gordon}, {Groenewegen}, {Lagadec}, {Lennon}, {Marengo},
  {Meixner}, {Skillman}, {Sloan}, {Sonneborn}, {van Loon}, \&
  {Zijlstra}}]{Boyer2015a}
{Boyer}, M.~L., {McQuinn}, K.~B.~W., {Barmby}, P., {et~al.} 2015{\natexlab{b}},
  \apjs, 216, 10

\bibitem[{{Boyer} {et~al.}(2015{\natexlab{c}}){Boyer}, {McQuinn}, {Barmby},
  {Bonanos}, {Gehrz}, {Gordon}, {Groenewegen}, {Lagadec}, {Lennon}, {Marengo},
  {McDonald}, {Meixner}, {Skillman}, {Sloan}, {Sonneborn}, {van Loon}, \&
  {Zijlstra}}]{Boyer2015b}
---. 2015{\natexlab{c}}, \apj, 800, 51

\bibitem[{{Bressan} {et~al.}(2012){Bressan}, {Marigo}, {Girardi}, {Salasnich},
  {Dal Cero}, {Rubele}, \& {Nanni}}]{Bressan2012}
{Bressan}, A., {Marigo}, P., {Girardi}, L., {et~al.} 2012, \mnras, 427, 127

\bibitem[{{Capaccioli}(1989)}]{Capaccioli1989}
{Capaccioli}, M. 1989, in World of Galaxies (Le Monde des Galaxies), ed. H.~G.
  {Corwin}, Jr. \& L.~{Bottinelli}, 208--227

\bibitem[{{Choudhury} {et~al.}(2016){Choudhury}, {Subramaniam}, \&
  {Cole}}]{Choudhury2016}
{Choudhury}, S., {Subramaniam}, A., \& {Cole}, A.~A. 2016, \mnras, 455, 1855

\bibitem[{{Conn} {et~al.}(2012){Conn}, {Ibata}, {Lewis}, {Parker}, {Zucker},
  {Martin}, {McConnachie}, {Irwin}, {Tanvir}, {Fardal}, {Ferguson}, {Chapman},
  \& {Valls-Gabaud}}]{Conn2012}
{Conn}, A.~R., {Ibata}, R.~A., {Lewis}, G.~F., {et~al.} 2012, \apj, 758, 11

\bibitem[{{Conselice}(2009)}]{Conselice2009}
{Conselice}, C.~J. 2009, \mnras, 399, L16

\bibitem[{{Crnojevi{\'c}} {et~al.}(2014){Crnojevi{\'c}}, {Ferguson}, {Irwin},
  {McConnachie}, {Bernard}, {Fardal}, {Ibata}, {Lewis}, {Martin}, {Navarro},
  {No{\"e}l}, \& {Pasetto}}]{Crnojevic2014}
{Crnojevi{\'c}}, D., {Ferguson}, A.~M.~N., {Irwin}, M.~J., {et~al.} 2014,
  \mnras, 445, 3862

\bibitem[{{Dalcanton} {et~al.}(2012){Dalcanton}, {Williams}, {Melbourne},
  {Girardi}, {Dolphin}, {Rosenfield}, {Boyer}, {de Jong}, {Gilbert}, {Marigo},
  {Olsen}, {Seth}, \& {Skillman}}]{Dalcanton2012}
{Dalcanton}, J.~J., {Williams}, B.~F., {Melbourne}, J.~L., {et~al.} 2012,
  \apjs, 198, 6

\bibitem[{{Davidge}(2005)}]{Davidge2005}
{Davidge}, T.~J. 2005, \aj, 130, 2087

\bibitem[{{de Vaucouleurs} {et~al.}(1991){de Vaucouleurs}, {de Vaucouleurs},
  {Corwin}, {Buta}, {Paturel}, \& {Fouqu{\'e}}}]{deVaucouleurs1991}
{de Vaucouleurs}, G., {de Vaucouleurs}, A., {Corwin}, Jr., H.~G., {et~al.}
  1991, {Third Reference Catalogue of Bright Galaxies. Volume I: Explanations
  and references. Volume II: Data for galaxies between 0$^{h}$ and 12$^{h}$.
  Volume III: Data for galaxies between 12$^{h}$ and 24$^{h}$.}

\bibitem[{{Deason} {et~al.}(2014){Deason}, {Wetzel}, \&
  {Garrison-Kimmel}}]{Deason2014}
{Deason}, A., {Wetzel}, A., \& {Garrison-Kimmel}, S. 2014, \apj, 794, 115

\bibitem[{{Dell'Agli} {et~al.}(2015{\natexlab{a}}){Dell'Agli},
  {Garc{\'{\i}}a-Hern{\'a}ndez}, {Ventura}, {Schneider}, {Di Criscienzo}, \&
  {Rossi}}]{Dell'Agli2015b}
{Dell'Agli}, F., {Garc{\'{\i}}a-Hern{\'a}ndez}, D.~A., {Ventura}, P., {et~al.}
  2015{\natexlab{a}}, \mnras, 454, 4235

\bibitem[{{Dell'Agli} {et~al.}(2015{\natexlab{b}}){Dell'Agli}, {Ventura},
  {Schneider}, {Di Criscienzo}, {Garc{\'{\i}}a-Hern{\'a}ndez}, {Rossi}, \&
  {Brocato}}]{Dell'Agli2015a}
{Dell'Agli}, F., {Ventura}, P., {Schneider}, R., {et~al.} 2015{\natexlab{b}},
  \mnras, 447, 2992

\bibitem[{{Demers} \& {Battinelli}(2002)}]{Demers2002}
{Demers}, S., \& {Battinelli}, P. 2002, \aj, 123, 238

\bibitem[{{Demers} {et~al.}(2004){Demers}, {Battinelli}, \&
  {Letarte}}]{Demers2004}
{Demers}, S., {Battinelli}, P., \& {Letarte}, B. 2004, \aap, 424, 125

\bibitem[{{Dobbie} {et~al.}(2014){Dobbie}, {Cole}, {Subramaniam}, \&
  {Keller}}]{Dobbie2014}
{Dobbie}, P.~D., {Cole}, A.~A., {Subramaniam}, A., \& {Keller}, S. 2014,
  \mnras, 442, 1680

\bibitem[{{Dohm-Palmer} {et~al.}(1997){Dohm-Palmer}, {Skillman}, {Saha},
  {Tolstoy}, {Mateo}, {Gallagher}, {Hoessel}, {Chiosi}, \&
  {Dufour}}]{Dohm-Palmer1997}
{Dohm-Palmer}, R.~C., {Skillman}, E.~D., {Saha}, A., {et~al.} 1997, \aj, 114,
  2527

\bibitem[{{Dohm-Palmer} {et~al.}(1998){Dohm-Palmer}, {Skillman}, {Gallagher},
  {Tolstoy}, {Mateo}, {Dufour}, {Saha}, {Hoessel}, \&
  {Chiosi}}]{Dohm-Palmer1998}
{Dohm-Palmer}, R.~C., {Skillman}, E.~D., {Gallagher}, J., {et~al.} 1998, \aj,
  116, 1227

\bibitem[{{Dolphin}(2000)}]{Dolphin2000}
{Dolphin}, A.~E. 2000, \pasp, 112, 1383

\bibitem[{{Dotter} {et~al.}(2008){Dotter}, {Chaboyer}, {Jevremovi{\'c}},
  {Kostov}, {Baron}, \& {Ferguson}}]{Dotter2008}
{Dotter}, A., {Chaboyer}, B., {Jevremovi{\'c}}, D., {et~al.} 2008, \apjs, 178,
  89

\bibitem[{{El-Badry} {et~al.}(2016){El-Badry}, {Wetzel}, {Geha}, {Hopkins},
  {Kere{\v s}}, {Chan}, \& {Faucher-Gigu{\`e}re}}]{El-Badry2016}
{El-Badry}, K., {Wetzel}, A., {Geha}, M., {et~al.} 2016, \apj, 820, 131

\bibitem[{{Fazio} {et~al.}(2004){Fazio}, {Hora}, {Allen}, {Ashby}, {Barmby},
  {Deutsch}, {Huang}, {Kleiner}, {Marengo}, {Megeath}, {Melnick}, {Pahre},
  {Patten}, {Polizotti}, {Smith}, {Taylor}, {Wang}, {Willner}, {Hoffmann},
  {Pipher}, {Forrest}, {McMurty}, {McCreight}, {McKelvey}, {McMurray}, {Koch},
  {Moseley}, {Arendt}, {Mentzell}, {Marx}, {Losch}, {Mayman}, {Eichhorn},
  {Krebs}, {Jhabvala}, {Gezari}, {Fixsen}, {Flores}, {Shakoorzadeh}, {Jungo},
  {Hakun}, {Workman}, {Karpati}, {Kichak}, {Whitley}, {Mann}, {Tollestrup},
  {Eisenhardt}, {Stern}, {Gorjian}, {Bhattacharya}, {Carey}, {Nelson},
  {Glaccum}, {Lacy}, {Lowrance}, {Laine}, {Reach}, {Stauffer}, {Surace},
  {Wilson}, {Wright}, {Hoffman}, {Domingo}, \& {Cohen}}]{Fazio2004}
{Fazio}, G.~G., {Hora}, J.~L., {Allen}, L.~E., {et~al.} 2004, \apjs, 154, 10

\bibitem[{{Feast} {et~al.}(2006){Feast}, {Whitelock}, \& {Menzies}}]{Feast2006}
{Feast}, M.~W., {Whitelock}, P.~A., \& {Menzies}, J.~W. 2006, \mnras, 369, 791

\bibitem[{{Gallagher} {et~al.}(1998){Gallagher}, {Tolstoy}, {Dohm-Palmer},
  {Skillman}, {Cole}, {Hoessel}, {Saha}, \& {Mateo}}]{Gallagher1998}
{Gallagher}, J.~S., {Tolstoy}, E., {Dohm-Palmer}, R.~C., {et~al.} 1998, \aj,
  115, 1869

\bibitem[{{Geha} {et~al.}(2010){Geha}, {van der Marel}, {Guhathakurta},
  {Gilbert}, {Kalirai}, \& {Kirby}}]{Geha2010}
{Geha}, M., {van der Marel}, R.~P., {Guhathakurta}, P., {et~al.} 2010, \apj,
  711, 361

\bibitem[{{Geha} {et~al.}(2015){Geha}, {Weisz}, {Grocholski}, {Dolphin}, {van
  der Marel}, \& {Guhathakurta}}]{Geha2015}
{Geha}, M., {Weisz}, D., {Grocholski}, A., {et~al.} 2015, \apj, 811, 114

\bibitem[{{Gehrz} {et~al.}(2007){Gehrz}, {Roellig}, {Werner}, {Fazio}, {Houck},
  {Low}, {Rieke}, {Soifer}, {Levine}, \& {Romana}}]{Gehrz2007}
{Gehrz}, R.~D., {Roellig}, T.~L., {Werner}, M.~W., {et~al.} 2007, Review of
  Scientific Instruments, 78, 011302

\bibitem[{{Gerbrandt} {et~al.}(2015){Gerbrandt}, {McConnachie}, \&
  {Irwin}}]{Gerbrandt2015}
{Gerbrandt}, S.~A.~N., {McConnachie}, A.~W., \& {Irwin}, M. 2015, \mnras, 454,
  1000

\bibitem[{{Gieles} {et~al.}(2008){Gieles}, {Bastian}, \&
  {Ercolano}}]{Gieles2008}
{Gieles}, M., {Bastian}, N., \& {Ercolano}, B. 2008, \mnras, 391, L93

\bibitem[{{Girardi} \& {Marigo}(2007)}]{Girardi2007}
{Girardi}, L., \& {Marigo}, P. 2007, \aap, 462, 237

\bibitem[{{Girardi} {et~al.}(2008){Girardi}, {Dalcanton}, {Williams}, {de
  Jong}, {Gallart}, {Monelli}, {Groenewegen}, {Holtzman}, {Olsen}, {Seth},
  {Weisz}, \& {ANGST/ANGRRR Collaboration}}]{Girardi2008}
{Girardi}, L., {Dalcanton}, J., {Williams}, B., {et~al.} 2008, \pasp, 120, 583

\bibitem[{{Girardi} {et~al.}(2012){Girardi}, {Barbieri}, {Groenewegen},
  {Marigo}, {Bressan}, {Rocha-Pinto}, {Santiago}, {Camargo}, \& {da
  Costa}}]{Girardi2012}
{Girardi}, L., {Barbieri}, M., {Groenewegen}, M.~A.~T., {et~al.} 2012,
  {TRILEGAL, a TRIdimensional modeL of thE GALaxy: Status and Future}, ed.
  A.~{Miglio}, J.~{Montalb{\'a}n}, \& A.~{Noels}, 165

\bibitem[{{Gordon} {et~al.}(2011){Gordon}, {Meixner}, {Meade}, {Whitney},
  {Engelbracht}, {Bot}, {Boyer}, {Lawton}, {Sewi{\l}o}, {Babler}, {Bernard},
  {Bracker}, {Block}, {Blum}, {Bolatto}, {Bonanos}, {Harris}, {Hora},
  {Indebetouw}, {Misselt}, {Reach}, {Shiao}, {Tielens}, {Carlson},
  {Churchwell}, {Clayton}, {Chen}, {Cohen}, {Fukui}, {Gorjian}, {Hony},
  {Israel}, {Kawamura}, {Kemper}, {Leroy}, {Li}, {Madden}, {Marble},
  {McDonald}, {Mizuno}, {Mizuno}, {Muller}, {Oliveira}, {Olsen}, {Onishi},
  {Paladini}, {Paradis}, {Points}, {Robitaille}, {Rubin}, {Sandstrom}, {Sato},
  {Shibai}, {Simon}, {Smith}, {Srinivasan}, {Vijh}, {Van Dyk}, {van Loon}, \&
  {Zaritsky}}]{Gordon2011}
{Gordon}, K.~D., {Meixner}, M., {Meade}, M.~R., {et~al.} 2011, \aj, 142, 102

\bibitem[{{Held} {et~al.}(2010){Held}, {Gullieuszik}, {Rizzi}, {Girardi},
  {Marigo}, \& {Saviane}}]{Held2010}
{Held}, E.~V., {Gullieuszik}, M., {Rizzi}, L., {et~al.} 2010, \mnras, 404, 1475

\bibitem[{{Higgs} {et~al.}(2016){Higgs}, {McConnachie}, {Irwin}, {Bate},
  {Lewis}, {Walker}, {C{\^o}t{\'e}}, {Venn}, \& {Battaglia}}]{Higgs2016}
{Higgs}, C.~R., {McConnachie}, A.~W., {Irwin}, M., {et~al.} 2016, \mnras, 458,
  1678

\bibitem[{{Ho} {et~al.}(2015){Ho}, {Geha}, {Tollerud}, {Zinn}, {Guhathakurta},
  \& {Vargas}}]{Ho2015}
{Ho}, N., {Geha}, M., {Tollerud}, E.~J., {et~al.} 2015, \apj, 798, 77

\bibitem[{{Holtzman} {et~al.}(2006){Holtzman}, {Afonso}, \&
  {Dolphin}}]{Holtzman2006}
{Holtzman}, J.~A., {Afonso}, C., \& {Dolphin}, A. 2006, \apjs, 166, 534

\bibitem[{{Huchtmeier}(1979)}]{Huchtmeier1979}
{Huchtmeier}, W.~K. 1979, \aap, 75, 170

\bibitem[{{Hunter} \& {Elmegreen}(2006)}]{Hunter2006}
{Hunter}, D.~A., \& {Elmegreen}, B.~G. 2006, \apjs, 162, 49

\bibitem[{{Hunter} {et~al.}(2012){Hunter}, {Ficut-Vicas}, {Ashley}, {Brinks},
  {Cigan}, {Elmegreen}, {Heesen}, {Herrmann}, {Johnson}, {Oh}, {Rupen},
  {Schruba}, {Simpson}, {Walter}, {Westpfahl}, {Young}, \&
  {Zhang}}]{Hunter2012}
{Hunter}, D.~A., {Ficut-Vicas}, D., {Ashley}, T., {et~al.} 2012, \aj, 144, 134

\bibitem[{{Ita} \& {Matsunaga}(2011)}]{Ita2011}
{Ita}, Y., \& {Matsunaga}, N. 2011, \mnras, 412, 2345

\bibitem[{{Jackson} {et~al.}(2007{\natexlab{a}}){Jackson}, {Skillman}, {Gehrz},
  {Polomski}, \& {Woodward}}]{Jackson2007a}
{Jackson}, D.~C., {Skillman}, E.~D., {Gehrz}, R.~D., {Polomski}, E., \&
  {Woodward}, C.~E. 2007{\natexlab{a}}, \apj, 656, 818

\bibitem[{{Jackson} {et~al.}(2007{\natexlab{b}}){Jackson}, {Skillman}, {Gehrz},
  {Polomski}, \& {Woodward}}]{Jackson2007b}
---. 2007{\natexlab{b}}, \apj, 667, 891

\bibitem[{{Javadi} {et~al.}(2013){Javadi}, {van Loon}, {Khosroshahi}, \&
  {Mirtorabi}}]{Javadi2013}
{Javadi}, A., {van Loon}, J.~T., {Khosroshahi}, H., \& {Mirtorabi}, M.~T. 2013,
  \mnras, 432, 2824

\bibitem[{{Javadi} {et~al.}(2011){Javadi}, {van Loon}, \&
  {Mirtorabi}}]{Javadi2011}
{Javadi}, A., {van Loon}, J.~T., \& {Mirtorabi}, M.~T. 2011, \mnras, 414, 3394

\bibitem[{{Kim} {et~al.}(2009){Kim}, {Kim}, {Hwang}, {Lee}, {Im}, {Karoji},
  {Noumaru}, \& {Tanaka}}]{Kim2009}
{Kim}, M., {Kim}, E., {Hwang}, N., {et~al.} 2009, \apj, 703, 816

\bibitem[{{Kniazev} {et~al.}(2009){Kniazev}, {Brosch}, {Hoffman}, {Grebel},
  {Zucker}, \& {Pustilnik}}]{Kniazev2009}
{Kniazev}, A.~Y., {Brosch}, N., {Hoffman}, G.~L., {et~al.} 2009, \mnras, 400,
  2054

\bibitem[{{Le Bertre}(1992)}]{LeBetre1992}
{Le Bertre}, T. 1992, \aaps, 94, 377

\bibitem[{{Le Bertre}(1993)}]{LeBetre1993}
---. 1993, \aaps, 97, 729

\bibitem[{{Leaman} {et~al.}(2012){Leaman}, {Venn}, {Brooks}, {Battaglia},
  {Cole}, {Ibata}, {Irwin}, {McConnachie}, {Mendel}, \& {Tolstoy}}]{Leaman2012}
{Leaman}, R., {Venn}, K.~A., {Brooks}, A.~M., {et~al.} 2012, \apj, 750, 33

\bibitem[{{Lee} {et~al.}(2006){Lee}, {Skillman}, \& {Venn}}]{Lee2006}
{Lee}, H., {Skillman}, E.~D., \& {Venn}, K.~A. 2006, \apj, 642, 813

\bibitem[{{Lee} {et~al.}(2000){Lee}, {Salzer}, {Law}, \& {Rosenberg}}]{Lee2000}
{Lee}, J.~C., {Salzer}, J.~J., {Law}, D.~A., \& {Rosenberg}, J.~L. 2000, \apj,
  536, 606

\bibitem[{{Letarte} {et~al.}(2002){Letarte}, {Demers}, {Battinelli}, \&
  {Kunkel}}]{Letarte2002}
{Letarte}, B., {Demers}, S., {Battinelli}, P., \& {Kunkel}, W.~E. 2002, \aj,
  123, 832

\bibitem[{{{\L}okas} {et~al.}(2012){{\L}okas}, {Kazantzidis}, \&
  {Mayer}}]{Lokas2012}
{{\L}okas}, E.~L., {Kazantzidis}, S., \& {Mayer}, L. 2012, \apjl, 751, L15

\bibitem[{{Makarov} {et~al.}(2006){Makarov}, {Makarova}, {Rizzi}, {Tully},
  {Dolphin}, {Sakai}, \& {Shaya}}]{Makarov2006}
{Makarov}, D., {Makarova}, L., {Rizzi}, L., {et~al.} 2006, \aj, 132, 2729

\bibitem[{{Makarov} {et~al.}(2014){Makarov}, {Prugniel}, {Terekhova},
  {Courtois}, \& {Vauglin}}]{Makarov2014}
{Makarov}, D., {Prugniel}, P., {Terekhova}, N., {Courtois}, H., \& {Vauglin},
  I. 2014, \aap, 570, A13

\bibitem[{{Marigo} {et~al.}(2013){Marigo}, {Bressan}, {Nanni}, {Girardi}, \&
  {Pumo}}]{Marigo2013}
{Marigo}, P., {Bressan}, A., {Nanni}, A., {Girardi}, L., \& {Pumo}, M.~L. 2013,
  \mnras, 434, 488

\bibitem[{{Marleau} {et~al.}(2010){Marleau}, {Noriega-Crespo}, \&
  {Misselt}}]{Marleau2010}
{Marleau}, F.~R., {Noriega-Crespo}, A., \& {Misselt}, K.~A. 2010, \apj, 713,
  992

\bibitem[{{Mart{\'{\i}}nez-Delgado} {et~al.}(1999){Mart{\'{\i}}nez-Delgado},
  {Aparicio}, \& {Gallart}}]{Martinez-Delgado1999}
{Mart{\'{\i}}nez-Delgado}, D., {Aparicio}, A., \& {Gallart}, C. 1999, \aj, 118,
  2229

\bibitem[{{Mashchenko} {et~al.}(2008){Mashchenko}, {Wadsley}, \&
  {Couchman}}]{Mashchenko2008}
{Mashchenko}, S., {Wadsley}, J., \& {Couchman}, H.~M.~P. 2008, Science, 319,
  174

\bibitem[{{Mateo}(1998)}]{Mateo1998}
{Mateo}, M.~L. 1998, \araa, 36, 435

\bibitem[{{McConnachie}(2012)}]{McConnachie2012}
{McConnachie}, A.~W. 2012, \aj, 144, 4

\bibitem[{{McConnachie} {et~al.}(2007){McConnachie}, {Arimoto}, \&
  {Irwin}}]{McConnachie2007}
{McConnachie}, A.~W., {Arimoto}, N., \& {Irwin}, M. 2007, \mnras, 379, 379

\bibitem[{{McConnachie} {et~al.}(2005){McConnachie}, {Irwin}, {Ferguson},
  {Ibata}, {Lewis}, \& {Tanvir}}]{McConnachie2005}
{McConnachie}, A.~W., {Irwin}, M.~J., {Ferguson}, A.~M.~N., {et~al.} 2005,
  \mnras, 356, 979

\bibitem[{{McDonald} \& {Zijlstra}(2016)}]{McDonald2016}
{McDonald}, I., \& {Zijlstra}, A.~A. 2016, \apjl, 823, L38

\bibitem[{{McQuinn} {et~al.}(2012){McQuinn}, {Skillman}, {Dalcanton}, {Cannon},
  {Dolphin}, {Holtzman}, {Weisz}, \& {Williams}}]{McQuinn2012a}
{McQuinn}, K.~B.~W., {Skillman}, E.~D., {Dalcanton}, J.~J., {et~al.} 2012,
  \apj, 759, 77

\bibitem[{{McQuinn} {et~al.}(2016){McQuinn}, {Skillman}, {Dolphin}, {Berg}, \&
  {Kennicutt}}]{McQuinn2016a}
{McQuinn}, K.~B.~W., {Skillman}, E.~D., {Dolphin}, A.~E., {Berg}, D., \&
  {Kennicutt}, R. 2016, \apj, 826, 21

\bibitem[{{McQuinn} {et~al.}(2007){McQuinn}, {Woodward}, {Willner}, {Polomski},
  {Gehrz}, {Humphreys}, {van Loon}, {Ashby}, {Eicher}, \&
  {Fazio}}]{McQuinn2007}
{McQuinn}, K.~B.~W., {Woodward}, C.~E., {Willner}, S.~P., {et~al.} 2007, \apj,
  664, 850

\bibitem[{{McQuinn} {et~al.}(2013){McQuinn}, {Skillman}, {Berg}, {Cannon},
  {Salzer}, {Adams}, {Dolphin}, {Giovanelli}, {Haynes}, \&
  {Rhode}}]{McQuinn2013}
{McQuinn}, K.~B.~W., {Skillman}, E.~D., {Berg}, D., {et~al.} 2013, \aj, 146,
  145

\bibitem[{{Meixner} {et~al.}(2006){Meixner}, {Gordon}, {Indebetouw}, {Hora},
  {Whitney}, {Blum}, {Reach}, {Bernard}, {Meade}, {Babler}, {Engelbracht},
  {For}, {Misselt}, {Vijh}, {Leitherer}, {Cohen}, {Churchwell}, {Boulanger},
  {Frogel}, {Fukui}, {Gallagher}, {Gorjian}, {Harris}, {Kelly}, {Kawamura},
  {Kim}, {Latter}, {Madden}, {Markwick-Kemper}, {Mizuno}, {Mizuno}, {Mould},
  {Nota}, {Oey}, {Olsen}, {Onishi}, {Paladini}, {Panagia}, {Perez-Gonzalez},
  {Shibai}, {Sato}, {Smith}, {Staveley-Smith}, {Tielens}, {Ueta}, {van Dyk},
  {Volk}, {Werner}, \& {Zaritsky}}]{Meixner2006}
{Meixner}, M., {Gordon}, K.~D., {Indebetouw}, R., {et~al.} 2006, \aj, 132, 2268

\bibitem[{{M{\'e}ndez} {et~al.}(2002){M{\'e}ndez}, {Davis}, {Moustakas},
  {Newman}, {Madore}, \& {Freedman}}]{Mendez2002}
{M{\'e}ndez}, B., {Davis}, M., {Moustakas}, J., {et~al.} 2002, \aj, 124, 213

\bibitem[{{Menzies} {et~al.}(2015){Menzies}, {Whitelock}, \&
  {Feast}}]{Menzies2015}
{Menzies}, J.~W., {Whitelock}, P.~A., \& {Feast}, M.~W. 2015, \mnras, 452, 910

\bibitem[{{Minniti} \& {Zijlstra}(1996)}]{Minniti1996}
{Minniti}, D., \& {Zijlstra}, A.~A. 1996, \apjl, 467, L13

\bibitem[{{Mouhcine} \& {Lan{\c c}on}(2002)}]{Mouhcine2002}
{Mouhcine}, M., \& {Lan{\c c}on}, A. 2002, \aap, 393, 149

\bibitem[{{Navarro} {et~al.}(1996){Navarro}, {Eke}, \& {Frenk}}]{Navarro1996}
{Navarro}, J.~F., {Eke}, V.~R., \& {Frenk}, C.~S. 1996, \mnras, 283, L72

\bibitem[{{Nowotny} {et~al.}(2003){Nowotny}, {Kerschbaum}, {Olofsson}, \&
  {Schwarz}}]{Nowotny2003}
{Nowotny}, W., {Kerschbaum}, F., {Olofsson}, H., \& {Schwarz}, H.~E. 2003,
  \aap, 403, 93

\bibitem[{{O{\~n}orbe} {et~al.}(2015){O{\~n}orbe}, {Boylan-Kolchin}, {Bullock},
  {Hopkins}, {Ker{\v e}s}, {Faucher-Gigu{\`e}re}, {Quataert}, \&
  {Murray}}]{Onorbe2015}
{O{\~n}orbe}, J., {Boylan-Kolchin}, M., {Bullock}, J.~S., {et~al.} 2015, ArXiv
  e-prints, arXiv:1502.02036

\bibitem[{{Pontzen} \& {Governato}(2012)}]{Pontzen2012}
{Pontzen}, A., \& {Governato}, F. 2012, \mnras, 421, 3464

\bibitem[{{Reimers}(1975)}]{Reimers1975}
{Reimers}, D. 1975, Memoires of the Societe Royale des Sciences de Liege, 8,
  369

\bibitem[{{Rizzi} {et~al.}(2007){Rizzi}, {Tully}, {Makarov}, {Makarova},
  {Dolphin}, {Sakai}, \& {Shaya}}]{Rizzi2007a}
{Rizzi}, L., {Tully}, R.~B., {Makarov}, D., {et~al.} 2007, \apj, 661, 815

\bibitem[{{Rosenfield} {et~al.}(2014){Rosenfield}, {Marigo}, {Girardi},
  {Dalcanton}, {Bressan}, {Gullieuszik}, {Weisz}, {Williams}, {Dolphin}, \&
  {Aringer}}]{Rosenfield2014}
{Rosenfield}, P., {Marigo}, P., {Girardi}, L., {et~al.} 2014, \apj, 790, 22

\bibitem[{{Sage} {et~al.}(1998){Sage}, {Welch}, \& {Mitchell}}]{Sage1998}
{Sage}, L.~J., {Welch}, G.~A., \& {Mitchell}, G.~F. 1998, \apj, 507, 726

\bibitem[{{Salaris} \& {Girardi}(2005)}]{Salaris2005}
{Salaris}, M., \& {Girardi}, L. 2005, \mnras, 357, 669

\bibitem[{{S{\'a}nchez-Salcedo} \& {Hernandez}(2007)}]{Sanchez-Salcedo2007}
{S{\'a}nchez-Salcedo}, F.~J., \& {Hernandez}, X. 2007, \apj, 667, 878

\bibitem[{{Sanna} {et~al.}(2010){Sanna}, {Bono}, {Stetson}, {Ferraro},
  {Monelli}, {Nonino}, {Prada Moroni}, {Bresolin}, {Buonanno}, {Caputo},
  {Cignoni}, {Degl'Innocenti}, {Iannicola}, {Matsunaga}, {Pietrinferni},
  {Romaniello}, {Storm}, \& {Walker}}]{Sanna2010}
{Sanna}, N., {Bono}, G., {Stetson}, P.~B., {et~al.} 2010, \apjl, 722, L244

\bibitem[{{Searle} \& {Zinn}(1978)}]{Searle1978}
{Searle}, L., \& {Zinn}, R. 1978, \apj, 225, 357

\bibitem[{{S{\'e}rsic}(1963)}]{Sersic1963}
{S{\'e}rsic}, J.~L. 1963, Boletin de la Asociacion Argentina de Astronomia La
  Plata Argentina, 6, 41

\bibitem[{{Sibbons} {et~al.}(2015){Sibbons}, {Ryan}, {Irwin}, \&
  {Napiwotzki}}]{Sibbons2015}
{Sibbons}, L.~F., {Ryan}, S.~G., {Irwin}, M., \& {Napiwotzki}, R. 2015, \aap,
  573, A84

\bibitem[{{Skillman} {et~al.}(1997){Skillman}, {Bomans}, \&
  {Kobulnicky}}]{Skillman1997}
{Skillman}, E.~D., {Bomans}, D.~J., \& {Kobulnicky}, H.~A. 1997, \apj, 474, 205

\bibitem[{{Skillman} {et~al.}(2003){Skillman}, {C{\^o}t{\'e}}, \&
  {Miller}}]{Skillman2003}
{Skillman}, E.~D., {C{\^o}t{\'e}}, S., \& {Miller}, B.~W. 2003, \aj, 125, 593

\bibitem[{{Skillman} {et~al.}(1988){Skillman}, {Terlevich}, {Teuben}, \& {van
  Woerden}}]{Skillman1988}
{Skillman}, E.~D., {Terlevich}, R., {Teuben}, P.~J., \& {van Woerden}, H. 1988,
  \aap, 198, 33

\bibitem[{{Skillman} {et~al.}(2014){Skillman}, {Hidalgo}, {Weisz}, {Monelli},
  {Gallart}, {Aparicio}, {Bernard}, {Boylan-Kolchin}, {Cassisi}, {Cole},
  {Dolphin}, {Ferguson}, {Mayer}, {Navarro}, {Stetson}, \&
  {Tolstoy}}]{Skillman2014}
{Skillman}, E.~D., {Hidalgo}, S.~L., {Weisz}, D.~R., {et~al.} 2014, \apj, 786,
  44

\bibitem[{{Stetson}(1987)}]{Stetson1987}
{Stetson}, P.~B. 1987, \pasp, 99, 191

\bibitem[{{Stinson} {et~al.}(2009){Stinson}, {Dalcanton}, {Quinn}, {Gogarten},
  {Kaufmann}, \& {Wadsley}}]{Stinson2009}
{Stinson}, G.~S., {Dalcanton}, J.~J., {Quinn}, T., {et~al.} 2009, \mnras, 395,
  1455

\bibitem[{{Tammann} {et~al.}(2011){Tammann}, {Reindl}, \&
  {Sandage}}]{Tammann2011}
{Tammann}, G.~A., {Reindl}, B., \& {Sandage}, A. 2011, \aap, 531, A134

\bibitem[{{Teyssier} {et~al.}(2012){Teyssier}, {Johnston}, \&
  {Kuhlen}}]{Teyssier2012}
{Teyssier}, M., {Johnston}, K.~V., \& {Kuhlen}, M. 2012, \mnras, 426, 1808

\bibitem[{{Tolstoy} {et~al.}(1998){Tolstoy}, {Gallagher}, {Cole}, {Hoessel},
  {Saha}, {Dohm-Palmer}, {Skillman}, {Mateo}, \& {Hurley-Keller}}]{Tolstoy1998}
{Tolstoy}, E., {Gallagher}, J.~S., {Cole}, A.~A., {et~al.} 1998, \aj, 116, 1244

\bibitem[{{Tully} {et~al.}(2006){Tully}, {Rizzi}, {Dolphin}, {Karachentsev},
  {Karachentseva}, {Makarov}, {Makarova}, {Sakai}, \& {Shaya}}]{Tully2006}
{Tully}, R.~B., {Rizzi}, L., {Dolphin}, A.~E., {et~al.} 2006, \aj, 132, 729

\bibitem[{{van Loon} {et~al.}(1999){van Loon}, {Groenewegen}, {de Koter},
  {Trams}, {Waters}, {Zijlstra}, {Whitelock}, \& {Loup}}]{vanLoon1999}
{van Loon}, J.~T., {Groenewegen}, M.~A.~T., {de Koter}, A., {et~al.} 1999,
  \aap, 351, 559

\bibitem[{{Vargas} {et~al.}(2014){Vargas}, {Geha}, \& {Tollerud}}]{Vargas2014}
{Vargas}, L.~C., {Geha}, M.~C., \& {Tollerud}, E.~J. 2014, \apj, 790, 73

\bibitem[{{Vassiliadis} \& {Wood}(1993)}]{Vassiliadis1993}
{Vassiliadis}, E., \& {Wood}, P.~R. 1993, \apj, 413, 641

\bibitem[{{Vijh} {et~al.}(2009){Vijh}, {Meixner}, {Babler}, {Block}, {Bracker},
  {Engelbracht}, {For}, {Gordon}, {Hora}, {Indebetouw}, {Leitherer}, {Meade},
  {Misselt}, {Sewilo}, {Srinivasan}, \& {Whitney}}]{Vijh2009}
{Vijh}, U.~P., {Meixner}, M., {Babler}, B., {et~al.} 2009, \aj, 137, 3139

\bibitem[{{Weisz} {et~al.}(2014){Weisz}, {Dolphin}, {Skillman}, {Holtzman},
  {Gilbert}, {Dalcanton}, \& {Williams}}]{Weisz2014a}
{Weisz}, D.~R., {Dolphin}, A.~E., {Skillman}, E.~D., {et~al.} 2014, \apj, 789,
  147

\bibitem[{{Werner} {et~al.}(2004){Werner}, {Roellig}, {Low}, {Rieke}, {Rieke},
  {Hoffmann}, {Young}, {Houck}, {Brandl}, {Fazio}, {Hora}, {Gehrz}, {Helou},
  {Soifer}, {Stauffer}, {Keene}, {Eisenhardt}, {Gallagher}, {Gautier}, {Irace},
  {Lawrence}, {Simmons}, {Van Cleve}, {Jura}, {Wright}, \&
  {Cruikshank}}]{Werner2004}
{Werner}, M.~W., {Roellig}, T.~L., {Low}, F.~J., {et~al.} 2004, \apjs, 154, 1

\bibitem[{{Whitelock} {et~al.}(1987){Whitelock}, {Feast}, \&
  {Pottasch}}]{Whitelock1987}
{Whitelock}, P.~A., {Feast}, M.~W., \& {Pottasch}, S.~R. 1987, in Astrophysics
  and Space Science Library, Vol. 132, Late Stages of Stellar Evolution, ed.
  S.~{Kwok} \& S.~R. {Pottasch}, 269--272

\bibitem[{{Young} \& {Lo}(1997)}]{Young1997}
{Young}, L.~M., \& {Lo}, K.~Y. 1997, \apj, 490, 710

\end{thebibliography}
